\documentclass[10pt,aps,prd,twocolumn,eqsecnum,amssymb,amsmath,showpacs,a4paper,nofootinbib,superscriptaddress]{revtex4-2}
\usepackage{import}
\usepackage{microtype}

\usepackage{graphicx}
\usepackage{amsfonts}
\usepackage{amsmath}
\usepackage{dsfont}
\usepackage{units}
\usepackage{color}
\usepackage{dcolumn}
\usepackage{bm} 
\usepackage{float}
\usepackage{hyperref}
\usepackage{cleveref}
\usepackage{braket}
\crefname{section}{§}{§§}
\Crefname{section}{§}{§§}
\usepackage{mathtools}
\usepackage{chngcntr}
\counterwithout{equation}{section}
\usepackage[normalem]{ulem}
\usepackage{tikz}
\usepackage{comment}
\usepackage{nicefrac}

\usepackage{ulem}

\usepackage[T1]{fontenc}
\usepackage[utf8,applemac]{inputenc}

\newcommand{\be}{\begin{equation}}
\newcommand{\ee}{\end{equation}}
\newcommand{\bsub}{\begin{subequations}}
\newcommand{\esub}{\end{subequations}}
\newcommand{\bea}{\begin{eqnarray}}
\newcommand{\eea}{\end{eqnarray}}
\newcommand{\bi} {\begin{itemize}}
\newcommand{\ei} {\end{itemize}}
\newcommand{\bmat} {\begin{pmatrix}}
\newcommand{\emat} {\end{pmatrix}} 

\usepackage[final]{showlabels} 

\newcommand{\ii}{\textrm{i}}

\DeclareMathOperator{\im}{Im}
\DeclareMathOperator{\re}{Re}
\DeclareMathOperator{\sgn}{sgn}
\DeclareMathOperator{\si}{si}

\DeclareMathOperator{\Ci}{Ci}
\DeclareMathOperator{\supp}{supp}
\DeclareMathOperator{\cov}{Cov}
\DeclareMathOperator{\tr}{Tr}

\def\Xint#1{\mathchoice
   {\XXint\displaystyle\textstyle{#1}}%
   {\XXint\textstyle\scriptstyle{#1}}%
   {\XXint\scriptstyle\scriptscriptstyle{#1}}%
   {\XXint\scriptscriptstyle\scriptscriptstyle{#1}}%
   \!\int}
\def\XXint#1#2#3{{\setbox0=\hbox{$#1{#2#3}{\int}$}
     \vcenter{\hbox{$#2#3$}}\kern-.5\wd0}}

\def\dashint{\Xint-}

\makeatletter
\newcommand*{\balancecolsandclearpage}{%
  \close@column@grid
  \clearpage
  \twocolumngrid
}
\makeatother

\numberwithin{equation}{section}

\makeatletter
\let\cat@comma@active\@empty
\makeatother

\usepackage{xcolor}
\hypersetup{  colorlinks,
              linktoc         = page, 
              urlcolor        = {green!80!blue},
              linkcolor       = {orange!60!black}, 
              urlcolor        = {cyan!80!black}, 
              citecolor       = {cyan!80!black}, 
              anchorcolor     = {yellow}
}

\usepackage{silence}
\usepackage{soul}
\usepackage[colorinlistoftodos]{todonotes}

\begin{document}

\title{Back-action from inertial and non-inertial Unruh-DeWitt detectors\\ revisited in covariant perturbation theory}

\author{Adam S. Wilkinson}\thanks{adam.wilkinson@nottingham.ac.uk}

\affiliation{School of Mathematical Sciences, University of Nottingham, Nottingham NG7 2RD, United Kingdom}

\author{Leo J. A. Parry}\thanks{leo.parry@nottingham.ac.uk}

\affiliation{School of Mathematical Sciences, University of Nottingham, Nottingham NG7 2RD, United Kingdom}

\author{Jorma Louko}\thanks{jorma.louko@nottingham.ac.uk}

\affiliation{School of Mathematical Sciences, University of Nottingham, Nottingham NG7 2RD, United Kingdom}

\author{William G. Unruh}\thanks{unruh@physics.ubc.ca}

\affiliation{Department of Physics and Astronomy, University of British Columbia, Vancouver, British Columbia V6T 1Z1, Canada}
\affiliation{Institute for Quantum Science and Engineering, Department of Physics and Astronomy,\\ 
Texas A\&M University, College
Station, TX 77843, USA}

\date{December 2025. Revised March 2026. \\ aaPublished in Phys.\ Rev.\ D \textbf{113}, 105002 (2026), doi.org/10.1103/xgv5-9ykl.}

\begin{abstract}

We investigate the back-action from a spatially pointlike particle detector on a quantum scalar field, as characterised by the expectation value of the field's stress-energy tensor, without conditioning on a measurement of the detector. First, assuming the field to be initially in a zero-mean Gaussian Hadamard state in a globally hyperbolic spacetime, we evaluate the field's two-point function in second-order perturbation theory by techniques of covariant curved spacetime quantum field
theory, which allow a full control of the time and space localisation of the interaction, and do not rely on field mode decompositions or non-local particle countings.
The detector's two-point function splits into a deterministic and a fluctuating part, and we show that this split is maintained in the back-action. 

We then specialise to a two-level Unruh-DeWitt detector, prepared in an energy eigenstate, for which the back-action is fully fluctuating. 
We compute the renormalised stress-energy tensor for a massless scalar field in $(3+1)$-dimensional Minkowski spacetime for a general detector trajectory, using the manifestly causal two-point function. We present explicit analytic and numerical results for an inertial detector and a uniformly linearly accelerated detector, switched on in the asymptotic past. 
The energy flux into and out of the accelerated detector accounts exactly for the energy gained and lost by the detector in its transitions due to the Unruh effect. 
The same holds for the outward flux associated with de-excitations of the inertial detector, which has a vanishing excitation rate and no inward flux.
A~novelty with the accelerated detector is two regions of negative energy density when
the detector is initially prepared in its ground state, one near the Rindler horizon that bounds the causal future of the trajectory, 
the other in the far future of the trajectory.

\end{abstract}

\maketitle

\setlength{\belowdisplayskip}{2pt} \setlength{\belowdisplayshortskip}{2pt}
\setlength{\abovedisplayskip}{2pt} \setlength{\abovedisplayshortskip}{2pt}

\section{Introduction}

\subsection{Background and scope}

In quantum field theory, the particle content of the state of a quantum field depends on both the background spacetime and the motion of the observer. Furthermore, while particles are often defined through \textit{global} mode decompositions, real observations are \textit{local}. 
This tension is resolved by particle detectors, quantum systems localised in space and time that couple to the field around specified trajectories. 
They respond regardless of the availability of an underlying description of the field state in terms of particles, thereby providing an operational framework for converting abstract field-theoretic statements into measurable predictions without the need for the notion of a particle. Particle detectors have been used to study thermal effects such as the Unruh \cite{Fulling:1972md,Davies:1974th,Unruh1976,DeWitt1979,Birrell1982,Crispino2008, Fulling2014}, Unruh-like~\cite{Letaw:1980yv, Good:2020hav, Biermann2020, Parry:2024jmm}, Hawking \cite{Hawking:1975vcx,Hartle:1976tp} and Gibbons-Hawking \cite{Gibbons1977,Hongwei2011} effects, to study the imprint of early-universe initial conditions on late-time cosmology \cite{Ford1989,Garay2014,Jonsson:2014lja,Toussaint2021,Toussaint2024,Wilkinson2024}, to probe the entanglement structure of the vacuum \cite{Valentini1991,Reznik2003,Reznik2005,Kerstjens2015}, to investigate superpositions of spacetimes \cite{Foo:2021exb, Foo:2022dnz, Goel:2024vtr} and detector worldlines \cite{Barbado:2020snx,Foo:2020xqn}, to understand quantum reference frames \cite{Giacomini:2022hco}, and to develop covariant measurement schemes in quantum field theory \cite{Fewster:2018qbm,Polo-Gomez:2021irs,Polo-Gomez:2025nuy}.

As an example of the operational role of particle detectors, consider the Unruh effect \cite{Fulling:1972md,Davies:1974th,Unruh1976,DeWitt1979,Birrell1982,Crispino2008, Fulling2014}. 
In field-theoretic terms, the restriction of the Minkowski vacuum to a boost-invariant wedge of Minkowski spacetime (known as a Rindler wedge) is a Kubo-Martin-Schwinger (KMS) state of modular temperature $1/(2\pi)$ with respect to the boost flow parameter \cite{Unruh1976,Bisognano:1975ih,Bisognano:1976za}. The integral curves of the boost Killing vector are the uniformly linearly accelerating hyperbolae (Rindler worldlines), 
and the KMS temperature for a Rindler worldline of proper acceleration $a$ is the Unruh temperature $T_U = a/(2\pi)$. 
In operational terms, if a uniformly linearly accelerating observer is modelled as a particle detector localised to a Rindler worldline and the field is prepared initially in the Minkowski vacuum, then the late-time asymptotic state of the \textit{detector\/} is a KMS state at the Unruh temperature with respect to the proper time of the detector \cite{SEWELL198023,Sewell:1982zz,DeBievre2006}.
An ensemble of such detectors therefore serves as a thermometer from which the temperature may be read out in an experiment.

In the detector description of the Unruh effect, and indeed in many other applications, the interaction between the detector and the field is often taken to be so weak that the back-action on the field can be neglected. 
In the present paper, we address the back-action in a perturbative formalism that is covariant and manifestly causal, ensuring that the stress-energy of the field at a spacetime point depends on the interactions only in the causal past of the spacetime point. 
We apply the formalism to inertial and uniformly linearly accelerated detector motion in Minkowski space. We find that the flux of energy from the detector to the field, or from the field to the detector, precisely accounts for the energy in the transitions in the detector. 

\vspace{-0.3cm}
\subsection{Context: a brief history of back-action in the Unruh effect}

To set the present paper in context, we review here selected past work on back-action in the Unruh effect. 

In the perturbative analysis of Ref.~\cite{Unruh1984}, it was found that a transition in the detector, upwards or downwards, is accompanied by the emission of a Minkowski particle. 
When transitions in the detector are in an equilibrium with the surrounding quantum field, a number of detector analyses in $1+1$ dimensions have found a polarisation cloud around the detector but no emissions that propagate to future timelike infinity \cite{Grove1986, Kolbenstvedt1988, Raine1991, Unruh1992, Massar1993}. 
The separation into detector emissions and vacuum polarisation has been argued to be largely semantic~\cite{Unruh1992,Audretsch:1994gg}: the detector-field interaction modifies the field's state, and whether this change is ascribable to detector emissions or vacuum polarisation depends on the observable considered. A~lucid review with further references is given in~\cite{Lin2006}. 
A~similar history exists also for the classical electromagnetic field generated by a charge on a Rindler trajectory: a recent review is in~\cite{Garfinkle:2019vej}. 

A non-perturbative analysis of the back-action in the Unruh effect in $3+1$ dimensions was given in~\cite{Lin2006, Lin2007}, finding both transient regimes and equilibrium regimes, but finding no radiation from the detector at Minkowski null infinity for the system in equilibrium. In this analysis the Unruh effect comes out as transient in nature: the interaction has changed the state of the field and the Rindler observer no longer experiences a thermal bath, as the coherence of the initial Minkowski vacuum has broken down. Related observations about the role of decoherence were made in~\cite{Unruh1984, Unruh1992}. 

Non-perturbative treatments in which the Unruh effect persists in the long time limit include open quantum system analyses using a Markovian approximation \cite{Benatti2004} and nonperturbative spectral techniques~\cite{DeBievre2006}. The relation between detector state and energy flow was addressed in~\cite{Ford:2005sh}. 

The relation between the Unruh thermal bath and acceleration-induced radiation has also been studied in the context of Bremsstrahlung \cite{Higuchi:1992we, Higuchi:1992td, Paithankar:2020akh} and Larmor radiation \cite{Portales-Oliva:2022yqq, Vacalis:2023wfq, Gallock-Yoshimura:2025dxk}. A~recent analysis of the causality of the state update process can be found in~\cite{Garcia-Chung:2021map}.

\vspace{-0.1cm}
\subsection{This paper: covariant perturbative back-action}

In the present paper, we revisit the field back-action from a particle detector in a covariant perturbative formalism that allows us to first localise the interaction between the detector and the field in a compact spacetime region, and to subsequently consider the limit in which the interaction duration becomes arbitrarily long. 
This allows us to consider the energy density and the energy flux of the field, as encoded in the renormalised stress-energy tensor (SET), without an a priori assumption of an equilibrium between the detector and the field. 

We evaluate the renormalised SET by the perturbative techniques of covariant curved spacetime quantum field theory~\cite{Hollands2015}. These techniques do not rely on a global mode decomposition of the field, any nonlocal particle counting, state reduction, measurements of detector transitions, or asymptotic equilibrium conditions. Our core results are for a massless, minimally coupled scalar field in $(3+1)$-dimensional Minkowski spacetime, initially prepared in the Minkowski vacuum, interacting with a pointlike detector on an arbitrary worldline, with explicit applications to detectors that follow inertial or Rindler worldlines. Our formalism is however broad enough to accommodate other initial states for the field, other spacetime dimensions, spacetime curvature, field mass and curvature coupling, and the formalism is generalisable from the SET to other local field observables. 

We emphasise that the formalism is manifestly causal, in the sense that the SET at a spacetime point $\mathsf{x}$ depends on the field-detector interaction only in the causal past of~$\mathsf{x}$. In particular, $\mathsf{x}$ is not restricted to specific asymptotic regions in the spacetime, and we make no assumptions about the detector-field interaction in an asymptotic future. 

\vspace{-0.2cm}

\subsection{Structure and results}

We begin in Sec.\ \ref{sec:field-correlators} by considering the covariant evolution of a real scalar field coupled linearly to a spatially pointlike detector, working to second order in perturbation theory in the coupling strength. We assume that the spacetime is globally hyperbolic, of dimension two or higher, and we assume that the field is initially in a zero-mean Gaussian state satisfying the Hadamard regularity condition~\cite{Decanini2008,Fewster:2013lqa,Hollands2015}. 
The evolution is causal, ensuring that the field at a spacetime point $\mathsf{x}$ is affected only by the interaction in the causal past of~$\mathsf{x}$.  
Using this formalism, we evaluate the expectation value of the field, referred to as the mean field value, and the field's Wightman function, without conditioning these expectation values on measurement of the detector. 
The formalism is general enough to accommodate spacetime curvature, 
mass of the scalar field, and arbitrary internal dynamics of the detector, such as a two-state system or a harmonic oscillator. 

Having established this formalism, we address in Sec.\ \ref{sec:examples-of-systems} the deterministic versus fluctuating character in the internal dynamics of the detector. 
The detector's two-point function decomposes into the sum of a deterministic contribution, which is expressible in terms of the one-point function, and a fluctuating contribution, which is the covariance~\cite{Hsiang2022}.
We first show that the back-action on the field's two-point function has a similar decomposition, with the deterministic and fluctuating contributions arising precisely from those of the detector. 
We then specialise to the two-level Unruh-DeWitt (UDW) and harmonic oscillator (HO) detector models~\cite{Unruh1976, DeWitt1979}, which approximate the dipole interaction between atomic orbitals and the quantum electromagnetic field in processes where angular momentum exchange is negligible~\cite{Martinez2013, Alhambra2014}. 
When prepared in an energy eigenstate, each of two detector models gives a fully fluctuating back-action. By contrast, when prepared in a coherent state, the HO detector's back-action contains both a deterministic contribution that scales with the square of the coherent amplitude and a fluctuating contribution that is independent of the coherent amplitude, and the fluctuating contribution is identical to that of the UDW detector prepared in its ground state. In the rest of the paper we specialise to the UDW detector initialised in an energy eigenstate. 

In Sec.\ \ref{sec:stress-energy} we apply the groundwork laid in Sections \ref{sec:field-correlators} and \ref{sec:examples-of-systems} to the core issue of the paper, the expectation value of the SET of the field due to the interaction with the UDW detector. We specialise to a massless field in $(3+1)$-dimensional Minkowski spacetime, initially prepared in the Minkowski vacuum, and a two-level UDW detector, initially prepared in an energy eigenstate. 
We find the field's Wightman function for a detector trajectory that is assumed real analytic for technical simplicity but is otherwise arbitrary, 
and for an arbitrary 
smooth compactly-supported switching function that determines the time-dependence of the detector's coupling strength to the field. 
We then use this Wightman function to evaluate the field's SET by point-splitting renormalisation~\cite{Christensen1976,Christensen1978}, 
showing that the SET is well defined everywhere in the complement of the detector's worldline. 

In Section \ref{sec:inertial-motion}, 
we consider a detector following an inertial trajectory. We work in the limit in which the interaction has been operating at constant strength since the asymptotic past, and we show that the SET remains well defined in the complement of the detector's worldline in this limit. 
The energy density is strictly positive, regardless of whether the detector was initially prepared in its ground state or in its excited state, and it has the same asymptotic small distance divergence in both cases, proportional to~$1/r^4$, where $r$ is the distance to the detector's trajectory. The large distance falloff of the energy density is faster when the detector was initially prepared in its ground state. The energy flux is nonzero only when the detector was initially prepared in its excited state, and the direction of the flux is then away from the detector. 

We find that the energy flux is closely related to the transitions in the detector. 
A detector initially prepared in its ground state may have a non-vanishing excitation probability, due to switch-on and switch-off effects, but the detector's excitation rate vanishes in the long interaction time limit under rather general conditions, as we describe in Appendix~\ref{app:detector-excitation-probability}\null. This vanishing excitation rate is consistent with the vanishing energy flux in the field. 
For a detector initially prepared in its excited state, by contrast, the de-excitation rate is non-vanishing: we find that the energy flux through a sphere around the detector then equals the product of the detector's energy gap and de-excitation rate. The flux from the detector hence exactly accounts for the energy lost by the detector in its downward transitions.

For the energy density, the relation to the detector's transitions is more subtle. When the detector is initially prepared in its ground state, so that its excitation rate vanishes, we nevertheless find a strictly positive energy density: we hence interpret this nonvanishing energy density as a vacuum polarisation effect~\cite{Schwinger1951}, rather than as a detector emission effect. 
When the detector is initially prepared in its excited state, the energy density is larger, and it may be understood as a combination of vacuum polarisation and detector emissions. 

In Section \ref{sec:uniform-acceleration}, we consider a detector following a Rindler worldline, 
working again in the long interaction duration limit, and showing that the SET remains well defined everywhere in the causal future of the worldline, except on the worldline itself. We consider three distinct notions of energy density and energy flux, adapted to three distinct families of observers in the spacetime: 
Minkowski energy density and flux, adapted to static Minkowski observers; Rindler energy density and flux, adapted to co-accelerated observers on orbits of the boost Killing vector that generates the detector's trajectory; and Milne energy density and flux, adapted to inertial observers on worldlines orthogonal to the boost Killing vector. 
In each timelike plane parallel to the plane of the detector's worldline, the Milne observers emerge from the origin of that plane. 
The Rindler and Milne notions of energy density and flux are adapted to the boost symmetry that the detector-field system inherits in the long interaction duration limit from the boost invariance of the Minkowski vacuum, and the Minkowski energy density and flux provide the perspective of a lab frame. 

For the energy density, we restrict the analysis to the timelike plane that contains the detector's worldline. 
We show that while all three notions of energy density show similarities to the case of the inertial detector, they also display key differences that mirror the Unruh effect experienced by the detector on the Rindler wordline~\cite{Unruh1976}. 
In particular, all three notions of energy density contain Planckian factors in the places where the case of the inertial detector energy density contains step functions, consistently with the Planckian factors that appear in the detector's transition rate~\cite{Unruh1976,Birrell1982}. At large excitation gaps and at large de-excitation gaps, the asymptotic behaviour of the three energy densities is similar to that for the inertial detector. 
Near the detector's trajectory, 
the Minkowski energy density and the Rindler energy density share with the inertial detector case the $1/r^4$ divergence, independent of the detector's energy gap, indicating that the divergence is a geometric effect originating from the point-like nature of the detector. 

A novel feature with the accelerated detector is the appearance of two regions of negative energy density when the detector is initially prepared in its ground state. The first region is near the left-going branch of the Rindler horizon.  This region is present in all three notions of energy density, and the Rindler and Milne observers interpret it respectively as negative energy density near the Rindler horizon and near the initial Milne singularity. This region shrinks rapidly as the excitation gap grows. The second region is present in the Minkowski and Milne energy densities, sufficiently far in the future of the trajectory, and the regions of negative Minkowski and Milne energy densities coincide in the large excitation gap limit. 

For the Rindler energy flux, we give the explicit expression not just in the timelike plane of the detector's worldline but everywhere in the causal future of the detector's worldline. 
We verify that the energy gained (respectively lost) by the detector in its excitations (de-excitations) exactly matches the flux of Rindler energy from the field to the detector (from the detector to the field). 
While the energy responsible for the detector's excitations due to the Unruh effect comes ultimately from the external agent who keeps the detector in acceleration~\cite{Unruh1984}, this match shows that from the Rindler frame perspective the detector's energy gains and losses can be attributed in full to a flow of Rindler energy from and to the field. 

For the Milne and Minkowski energy fluxes, we consider only the timelike plane of the detector's worldline. The direction of the Milne energy flux is away from the detector for detector de-excitations and towards the detector for detector excitations, consistently with the Rindler energy flux. The direction of the Minkowski energy flux shows a more complicated and time-dependent structure, due to the nonstationarity of the detector's trajectory in the Minkowski frame. 

Overall, these results show that our covariant and manifestly causal approach to second-order perturbation theory gives a renormalised SET that faithfully captures the back-action of a pointlike UDW detector on the quantum field when the final state of the detector is not measured. In spite of the singular character of a pointlike interaction beyond first order in perturbation theory~\cite{Fewster2024}, the SET remains well defined away from the detector's worldline even when the interaction operates at constant strength since the infinite past, and it displays the characteristics of the Unruh effect for the uniformly linearly accelerated trajectory. The pointlike character of the detector becomes a limitation only near the detector's worldline, where the energy density diverges too fast for the field's total energy to be defined. However, this limitation does not extend to the energy flux around the detector, which remains well defined, and for the inertial and uniformly linearly accelerated trajectories it precisely accounts for the energy that the detector loses or gains in its transitions. 

Sec.\ \ref{sec:discussion} presents a summary and a brief discussion. Details of technical calculations are deferred to six appendices. 

\subsection{Notation}

We use a mostly positive sign convention such that timelike square distances are negative, and units in which $c = \hbar = 1$. In asymptotic expansions, $f(x) = O(x)$ denotes that $f(x)/x$ is bounded as $x\to 0$, and $o(x)$ denotes that $f(x)/x \to 0$ as $ x \to 0$. 
 
The Heaviside theta function $\Theta(x)$ is defined as
\begin{equation} \label{eq:Theta-def}
    \Theta(x) = 
    \begin{cases}
        1 & \text{ for } x\geq0 \\
        0 & \text{ for } x<0,
    \end{cases}
\end{equation}
and the signum function $\sgn(x)$ is defined as
\begin{align}
    \sgn(x) = 
    \begin{cases}
         1 & \text{ for } x>0 \\
         0 & \text{ for } x=0 \\
        -1 & \text{ for } x<0.
    \end{cases}
    \label{eq:signum-function}
\end{align}
The Fourier transform is defined as
\begin{equation}
    \widehat{f}(\omega) = \int_{-\infty}^\infty \mathrm{d} t \, \mathrm{e}^{-\ii\omega t} f(t).
    \label{eq:fourier-def}
\end{equation} 

\section{Field correlation functions}\label{sec:field-correlators}

In this section, we first describe the field-detector interaction, without yet specifying a particular detector model, and then compute the one- and two-point correlation functions of the field, evaluated in the final state of the field subsystem, given that the field is prepared in a zero-mean Gaussian state. This provides one characterisation of the back-action of the detector on the field, and furthermore, the two-point function will allow us to compute the renormalised SET in Sec. \ref{sec:stress-energy}.

\subsection{Field-detector model}
We consider a real quantum Klein-Gordon field $\phi$ in a globally hyperbolic spacetime $(\mathcal{M},g)$. The field $\phi$ is an operator-valued distribution acting on a Hilbert space $\mathcal{H}_F$ that carries a representation of the covariant canonical commutation relations \cite{Hollands2015}
\begin{equation} \label{eq:CCR}
    \left[\phi(\mathsf{x}),\phi(\mathsf{y})\right]  = iE(\mathsf{x},\mathsf{y}) \mathds{1}_F,
\end{equation}
where $\mathsf{x},\mathsf{y}\in \mathcal{M}$, and $\mathds{1}_F$ is the identity operator on $\mathcal{H}_F$. $E(\mathsf{x},\mathsf{y})$ is the retarded-minus-advanced Green's function of the Klein-Gordon equation i.e.
\begin{equation} \label{eq:E+-E-}
    E(\mathsf{x},\mathsf{y}) = E^+(\mathsf{x},\mathsf{y}) - E^-(\mathsf{x},\mathsf{y}),
\end{equation}
where $E^+$ and $E^-$ denote the retarded and advanced solutions, respectively. $E^+(\mathsf{x},\mathsf{y})$ is non-zero only when a future-directed causal curve runs from $\mathsf{x}$ to $\mathsf{y}$, and $E^-(\mathsf{x},\mathsf{y})$ is non-zero only when a past-directed causal curve runs from $\mathsf{x}$ to $\mathsf{y}$. Note that the global hyperbolicity of the spacetime ensures the existence and uniqueness of $E^\pm$~\cite{Bar2008}.

We take the detector to be a spatially pointlike quantum system with Hilbert space $\mathcal{H}_D$ and identity~$\mathds{1}_D$, moving through $\mathcal{M}$ on a real-analytic timelike trajectory~$\mathsf{Z}(\tau)$, parametrised by the proper time $\tau$. The real-analyticity will be used in Sec. \ref{sec:stress-energy} to analytically continue the proper time parameter into the complex plane. At this point, we do not specify the dimension of $\mathcal{H}_D$; the formalism described in this section applies equally to, for example, a harmonic oscillator detector or a two-level detector.

The total Hilbert space of the combined field-detector system is $\mathcal{H}_D\otimes \mathcal{H}_F$, and we work throughout in the interaction picture with respect to the free dynamics on both factors. The field and the detector interact via the interaction Hamiltonian

\begin{align}
    H_I(\tau) := \lambda \chi(\tau) \mu(\tau) \otimes  \phi_\mathsf{Z}(\tau), \label{eq:interaction-hamiltonian}
\end{align}
where $\lambda$ is a real-valued coupling constant,
and $\phi_\mathsf{Z}(\tau) := \phi\!\left(\mathsf{Z}(\tau)\right)$ is the field operator pulled back to the detector's trajectory. 
The Hermitian operator $\mu(\tau)$ is the detector's monopole moment operator. 
$\chi(\tau)$~is a real-valued switching function that specifies how the interaction is turned on and off. We assume $\chi$ to be smooth. To begin with, we take $\chi$ to have compact support, and later we consider the limit in which the coupling has been active since the asymptotic past.

Suppose that in the asymptotic past, before the interaction is switched on, the field-detector system is prepared in a state described by the density matrix $\rho_i$ acting on the total Hilbert space $\mathcal{H}_D\otimes \mathcal{H}_F$. Then, in the interaction picture, the state of the combined system at proper time $\tau$ is given by
\begin{equation}
    \rho(\tau) = \mathcal{U}(\tau) \rho_i \, \mathcal{U}^\dagger(\tau),
\end{equation}
where $\mathcal{U}(\tau)$ is the unitary time-evolution operator given by
\begin{equation}
    \mathcal{U}(\tau) = \mathcal{T}\exp\!\left( -i \int_{-\infty}^\tau \mathrm{d} \tau' H_I(\tau') \right),
\end{equation}
where $\mathcal{T} \exp$ denotes the time-ordered exponential.

We now denote the final state at time $\tau_f$ as $\rho_f = \rho(\tau_f)$. Assuming that the coupling constant $\lambda$ is small, we may expand the final state $\rho_f$ in a perturbative Dyson series in $\lambda$ as
\begin{subequations}
\label{eq:dyson-series}
    \begin{align}
        \rho_f &= \rho_i + \rho^{(1)} + \rho^{(2)} + O(\lambda^3),\\
        \rho^{(1)} &= -i\int_{-\infty}^{\tau_f}\mathrm{d}\tau\; [H_I(\tau), \rho_i] ,\\
        \rho^{(2)} &= -\int_{-\infty}^{\tau_f}\mathrm{d}\tau \int_{-\infty}^{\tau}\mathrm{d}\tau' \Bigl[H_I(\tau), \bigl[H_I(\tau'), \rho_i\bigr]\Bigr],
    \end{align}
\end{subequations}
where $\rho^{(1)}$ is of order $\lambda$ and 
$\rho^{(2)}$ is of order~$\lambda^2$. 
When $\rho_f$ is used for calculating the one- and two-point correlation functions of the field, the choice of $\tau_f$ relative to the spacetime points is constrained by causality, in the sense that the detector cannot influence correlators at points that are not in the causal future of the interaction region. We shall address this causality below in Secs.\ \ref{subsec:mean-field-value}
and~\ref{subsec:two-point-function}\null. We also verify that covariance is maintained in the system in the sense that $\tau_f$ drops out of field observable expectation values.

From this point on, we assume that the initial state of the combined system factorises as $\rho_i=  \rho_{i,D} \otimes \rho_{i,F}$, where $\rho_{i,D}$ and $\rho_{i,F}$ are the initial states of the detector and field subsystems, respectively. That is, the detector and the field are initially completely uncorrelated. We further assume that the field's initial state $\rho_{i,F}$ is a zero-mean Gaussian state. Thus all odd $n$-point correlation functions of the field in the state $\rho_{i,F}$ vanish, and all even $n$-point correlation functions decompose, by Wick's theorem, into sums of products of Wightman functions in the initial state $\rho_{i,F}$~\cite{Hollands2015},  
\begin{align}
W(\mathsf{x}, \mathsf{x}') = \tr_F(\phi(\mathsf{x})\phi(\mathsf{x}')\rho_{i,F}), 
\label{eq:wightman-initial}
\end{align}
where $\tr_F$ denotes the trace over $\mathcal{H}_F$.
For example, the four-point function may be written as
\begin{align}
\label{eq:Wick_4point}        
    \tr_F(\phi(\mathsf{x_1})\phi(\mathsf{x_2})\phi(\mathsf{x_3})\phi(\mathsf{x_4})\rho_{i,F}) &= W(\mathsf{x}_1,\mathsf{x}_2)W(\mathsf{x}_3,\mathsf{x}_4) \notag \\
&\hspace{0.3cm}+ W(\mathsf{x}_1,\mathsf{x}_3)W(\mathsf{x}_2,\mathsf{x}_4)\notag \\
&\hspace{0.3cm}+ W(\mathsf{x}_1,\mathsf{x}_4)W(\mathsf{x}_2,\mathsf{x}_3).
\end{align}

\vspace{-0.3cm}
\subsection{Mean field value}\label{subsec:mean-field-value}

We begin by calculating the one-point correlation function of the field evaluated in the state $\rho_f$ to leading non-trivial order in perturbation theory, with the assumption that no measurement is made on the detector. This one-point function is given by
\begin{align}
    \braket{\phi(\mathsf{x})} := \tr\Bigl[\bigl(\mathds{1}_D\otimes\phi(\mathsf{x})\bigr)\rho_f\Bigr].
\end{align}
In the notation, we suppress the state in the expectation value for brevity; $\braket{\cdot}$ denotes the expectation value in the final state of the field subsystem.

Since the initial state of the combined system is by assumption $\rho_i = \rho_{i,D}\otimes \rho_{i,F}$, and the initial state of the field $\rho_{i,F}$ is zero-mean, it follows from \eqref{eq:interaction-hamiltonian} and \eqref{eq:dyson-series} that 
\begin{subequations}
\label{eq:phi-mean-and-sup-1}
    \begin{align}
        \braket{\phi(\mathsf{x})} &= \lambda\braket{\phi(\mathsf{x})}^{(1)} + O(\lambda^2), \label{eq:phi_mean}\\
        \braket{\phi(\mathsf{x})}^{(1)} &= -\int_{-\infty}^{\tau_f}\mathrm{d}\tau\; \chi(\tau)\braket{\mu(\tau)}_i E\!\left(\mathsf{Z}(\tau),\mathsf{x}\right), \label{eq:phi-sup-1}
    \end{align}
\end{subequations}
where $\braket{\mu(\tau)}_i := \tr_D(\mu(\tau)\rho_{i,D})$ with $\tr_D$ denoting the trace acting on~$\mathcal{H}_D$, and in \eqref{eq:phi-sup-1} we have used~\eqref{eq:CCR}.

Two remarks are in order. 

First, the order $\lambda$ contribution $\braket{\phi(\mathsf{x})}^{(1)}$ in \eqref{eq:phi_mean} can be written as
\begin{equation}
    \lambda\braket{\phi(\mathsf{x})}^{(1)} = \tr\!\left[ \big(\mathds{1}_D\otimes \phi(\mathsf{x})\big) \rho^{(1)}\right],
\end{equation}
where $\rho^{(1)}$ is the order $\lambda$ term in the Dyson expansion \eqref{eq:dyson-series}. Since $\tr[\rho_f] = \tr[\rho_i]=1 $, it follows that ${\tr[\rho^{(n)}]=0}$ for all $n\geq 1$, so $\rho^{(n)}$ is not a density matrix for ${n\ge1}$. Accordingly, $\braket{\phi(\mathsf{x})}^{(1)}$ should be understood as the coefficient of $\lambda$ in the perturbative expansion of~$\braket{\phi(\mathsf{x})}$, rather than as an expectation value in a state by itself.

Second, if the field's initial state is not zero-mean, equation \eqref{eq:phi_mean} also contains an order $\lambda^0$ contribution, namely the free one-point function in the field's initial state. The order $\lambda$ term is still as in \eqref{eq:phi-mean-and-sup-1} and gives the leading order effect of the interaction on~$\braket{\phi(\mathsf{x})}$.

\begin{figure}
    \centering
    \includegraphics[width=0.9\linewidth]{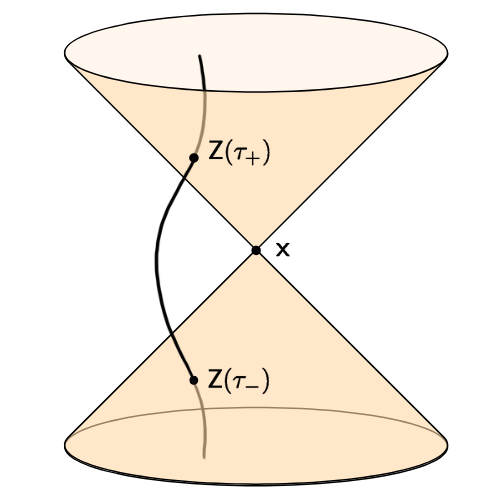}
    \caption{A spacetime diagram showing the detector's trajectory $\mathsf{Z}(\tau)$ alongside a spacetime point $\mathsf{x}$ that is not on the trajectory. The point $\mathsf{Z}(\tau_-)$ is the latest point on the trajectory that intersects the past null cone of~$\mathsf{x}$; any earlier intersections, as may exist for example in spatially compact spacetimes, are beyond the spacetime region shown. Similarly, the point $\mathsf{Z}(\tau_+)$ is the earliest point on the trajectory that intersects the future null cone of~$\mathsf{x}$; any later intersections are beyond the spacetime region shown. For some trajectories and some~$\mathsf{x}$, $\mathsf{Z}(\tau_-)$ and/or $\mathsf{Z}(\tau_+)$ do not exist; in this case we understand $\tau_-$ as $-\infty$ and $\tau_+$ as~$\infty$. 
    Note that $\mathsf{Z}(\tau)$ is spacelike separated from $\mathsf{x}$ if and only if $\tau_- < \tau < \tau_+$.}
    \label{fig:spacetime-diagram}
\end{figure}

We now turn to the implementation of causality in our system. 

We assume that the point $\mathsf{x}$ is not on the trajectory $\mathsf{Z}(\tau)$ of the detector. 
To establish notation, we denote by $\tau_-$ the latest value of $\tau$ where the trajectory intersects the past null cone of~$\mathsf{x}$, and we denote by $\tau_+$ the earliest value of $\tau$ where the trajectory intersects the future null cone of~$\mathsf{x}$. The situation is depicted in Fig.~\ref{fig:spacetime-diagram}. 
For some trajectories and some $\mathsf{x}$, $\tau_-$ and/or $\tau_+$ do not exist; in this case we understand $\tau_-$ as $-\infty$ and $\tau_+$ as~$\infty$. Note that $\mathsf{Z}(\tau)$ is spacelike separated from $\mathsf{x}$ if and only if $\tau_- < \tau < \tau_+$. 
Writing $E^\pm_\mathsf{Z}(\tau; \mathsf{x}) := E^\pm(\mathsf{Z}(\tau), \mathsf{x})$, the support properties of $E^\pm$ then imply 
\begin{subequations}
\label{eq:support-of-E}
    \begin{align}
        E^+_\mathsf{Z}(\tau; \mathsf{x}) &= 0 \;\forall\tau > \tau_- \label{eq:support-of-E:plus} , \\
        E^-_\mathsf{Z}(\tau; \mathsf{x}) &= 0 \;\forall\tau < \tau_+ . \label{eq:support-of-E:minus}
    \end{align} 
\end{subequations}

Recall now that the expectation value in \eqref{eq:phi-sup-1} refers to the perturbative expansion in the final state in the Cauchy evolution of the combined system, and this final state depends on the choice of the final Cauchy hypersurface. We have however not yet specified how the final hypersurface relates to the spacetime point~$\mathsf{x}$, and \eqref{eq:phi-sup-1} still involves~$\tau_f$, which is the detector's proper time on the final hypersurface.

We implement causality in \eqref{eq:phi-sup-1} by evolving the combined system 
to a final Cauchy hypersurface that intersects the detector's trajectory at a point that is spacelike separated from the point~$\mathsf{x}$. The choice of this hypersurface thus determines~$\tau_f$, 
and $\tau_- < \tau_f < \tau_+$. 
Using~\eqref{eq:E+-E-} and \eqref{eq:support-of-E:minus}, 
\eqref{eq:phi-sup-1} hence reduces to 
\begin{align}
    \braket{\phi(\mathsf{x})}^{(1)} = -\int_{-\infty}^{\tau_f}\mathrm{d}\tau \, \chi(\tau)\braket{\mu(\tau)}_iE^+_\mathsf{Z}(\tau; \mathsf{x}).
    \label{eq:phi-1-interm}
\end{align}
As $\tau_f > \tau_-$, \eqref{eq:support-of-E:plus} shows that we may replace the upper limit of the integral in \eqref{eq:phi-1-interm} by infinity, finding 
\begin{align}
\braket{\phi(\mathsf{x})}^{(1)} = -\int_{-\infty}^{\infty}\mathrm{d}\tau \, \chi(\tau)\braket{\mu(\tau)}_iE^+_\mathsf{Z}(\tau; \mathsf{x}). 
    \label{eq:mean-field-value}
\end{align}

We emphasise that the result \eqref{eq:mean-field-value} for $\braket{\phi(\mathsf{x})}^{(1)}$ depends on the spacetime point $\mathsf{x}$ but it contains no other reference to the final Cauchy hypersurface, and $\braket{\phi(\mathsf{x})}^{(1)}$ depends on the interaction only in the causal past of~$\mathsf{x}$. 
In particular, $\tau_f$ has dropped out of~\eqref{eq:mean-field-value}, 
and causality has been implemented in a covariant manner. 
In words, we may think of 
$\braket{\phi(\mathsf{x})}^{(1)}$ \eqref{eq:mean-field-value} 
in terms of 
a perturbative expansion of the mean field value 
in the equivalence class of final states on Cauchy hypersurfaces whose intersection with the trajectory is spacelike-separated from~$\mathsf{x}$. Note that this equivalence class is distinct for each~$\mathsf{x}$.

Finally, note that since $E^+$ is state-independent, 
the result \eqref{eq:mean-field-value} for $\braket{\phi(\mathsf{x})}^{(1)}$ does not depend on the field's initial state, holding even when the initial state is not zero-mean or Gaussian. 

\vspace{-0.2cm}
\subsection{Two-point function}\label{subsec:two-point-function} 

We next turn to the two-point correlation function of the field evaluated in the state $\rho_f$ to order $\lambda^2$ in perturbation theory, with the assumption that no measurement is made on the detector. This two-point function is given by 
\begin{align}
    \braket{\phi(\mathsf{x})\phi(\mathsf{x}')} := \tr\Bigl[\bigl(\mathds{1}_D\otimes\phi(\mathsf{x})\phi(\mathsf{x}')\bigr)\rho_f\Bigr] . 
\label{eq:twopoint-raw}
\end{align}
Again, we suppress the state in the expectation value for brevity; $\braket{\cdot}$ denotes the expectation value in the final state of the field subsystem.

We assume $\mathsf{x}$ and $\mathsf{x}'$ to be distinct and not on the detector's trajectory. 
Substituting \eqref{eq:interaction-hamiltonian} and \eqref{eq:dyson-series} into \eqref{eq:twopoint-raw} gives 
\begin{align}
    \braket{\phi(\mathsf{x})\phi(\mathsf{x}')} &= \braket{\phi(\mathsf{x})\phi(\mathsf{x}')}^{(0)}
    +\lambda\braket{\phi(\mathsf{x})\phi(\mathsf{x}')}^{(1)} \notag\\
    &\quad+ \lambda^2\braket{\phi(\mathsf{x})\phi(\mathsf{x}')}^{(2)} 
    + O(\lambda^3) . 
\label{eq:phi-phi-lambdasq}
\end{align}
As discussed below \eqref{eq:phi-mean-and-sup-1}, in the Dyson series \eqref{eq:dyson-series}, each perturbative term $\rho^{(n)}$ with $n \geq 1$ has vanishing trace and is thus not a density matrix by itself. 
Accordingly, $\braket{\phi(\mathsf{x})\phi(\mathsf{x'})}^{(n)}$ for $n\ge1$ should be understood as the coefficient of $\lambda^n$ in the perturbative expansion of $\braket{\phi(\mathsf{x})\phi(\mathsf{x'})}$ \eqref{eq:phi-phi-lambdasq}, rather than as an expectation value in a state by itself.

Using the assumption that the initial state of the combined system is $\rho_i = \rho_{i,D}\otimes \rho_{i,F}$, we find
\begin{widetext}
\begin{subequations}
\begin{align}
    \braket{\phi(\mathsf{x})\phi(\mathsf{x}')}^{(1)} &= i\int_{-\infty}^{\tau_f}\mathrm{d}\tau \, \chi(\tau)\braket{\mu(\tau)}_i 
    \tr\Bigl(\bigl[\phi_\mathsf{Z}(\tau),\phi(\mathsf{x})\phi(\mathsf{x}')\bigr]\rho_{i,F} \Bigr) , 
\label{eq:phiphi-one}
\\
    \braket{\phi(\mathsf{x})\phi(\mathsf{x}')}^{(2)} &= \int_{-\infty}^{\tau_f}\mathrm{d}\tau\int_{-\infty}^{\tau}\mathrm{d}\tau' \, \chi(\tau)\chi(\tau')\braket{\mu(\tau)\mu(\tau')}_i\tr\Bigl(\bigl[\phi(\mathsf{x})\phi(\mathsf{x}'),\phi_\mathsf{Z}(\tau)\bigr]\phi_\mathsf{Z}(\tau')\rho_{i,F}\Bigr) + (\mathsf{x} \leftrightarrow \mathsf{x}')^* ,  \label{eq:two-point-correlator-unsimplified}
\end{align}
\end{subequations}
\end{widetext}
where $(\mathsf{x} \leftrightarrow \mathsf{x}')$ represents the interchange of $\mathsf{x}$ and $\mathsf{x}'$ and the asterisk denotes complex conjugation.

By assumption, the initial field state $\rho_{i,F}$ is a Gaussian state. Therefore, since \eqref{eq:phiphi-one} contains three-point functions of the field in the state $\rho_{i,F}$, we find that $\braket{\phi(\mathsf{x})\phi(\mathsf{x}')}^{(1)} = 0$. In \eqref{eq:two-point-correlator-unsimplified}, we may use Wick's theorem \eqref{eq:Wick_4point} to write
\begin{align}
    \tr\Bigl(\bigl[\phi(\mathsf{x})\phi(\mathsf{x}'), \phi_\mathsf{Z}(\tau)\bigr]\phi_\mathsf{Z}(\tau')\rho_{i,F}\Bigr)&\notag\\
    &\hspace{-3cm}=\bigl[\phi(\mathsf{x}), \phi_\mathsf{Z}(\tau)\bigr]W(\mathsf{x}', \mathsf{Z}(\tau')) + (\mathsf{x} \leftrightarrow \mathsf{x}').
\label{eq:4point-interm-2}
\end{align}
Substitution of \eqref{eq:CCR}, \eqref{eq:E+-E-} and \eqref{eq:4point-interm-2} into
\eqref{eq:two-point-correlator-unsimplified} then yields 
\begin{align}
    \braket{\phi(\mathsf{x})\phi(\mathsf{x}')}^{(2)} &= 2\int_{-\infty}^{\infty}\mathrm{d}\tau \, \chi(\tau)E^+_\mathsf{Z}(\tau; \mathsf{x})A(\tau; \mathsf{x}')  \notag\\
    &\qquad+ (\mathsf{x} \leftrightarrow \mathsf{x}') , 
    \label{eq:two-point-emissions}
\end{align}
where
\begin{align}
    A(\tau; \mathsf{x}') &= \im \! \left[\int_{-\infty}^\tau\mathrm{d}\tau' \, \chi(\tau')\braket{\mu(\tau')\mu(\tau)}_iW(\mathsf{Z}(\tau'), \mathsf{x}')\right],   \label{eq:A}
\end{align}
where we have implemented causality by evolving the combined system to a final Cauchy hypersurface that intersects the detector's trajectory at a point that is spacelike-separated from both $\mathsf{x}$ and~$\mathsf{x}'$. Such hypersurfaces exist when $\mathsf{x}$ and $\mathsf{x}'$ are sufficiently close to each other, which we assume here and from now on. Note that the assumption of spacelike separation ensures that $\tau_f$ in \eqref{eq:two-point-correlator-unsimplified} can be replaced by infinity in \eqref{eq:two-point-emissions}, so that \eqref{eq:two-point-emissions} and \eqref{eq:A} no longer involve~$\tau_f$, and causality has been implemented in a covariant manner. 
We may think of 
$\braket{\phi(\mathsf{x})\phi(\mathsf{x}')}^{(2)}$ \eqref{eq:two-point-emissions}
in terms of the equivalence class of final states on Cauchy hypersurfaces whose intersection with the trajectory is spacelike-separated from both $\mathsf{x}$ and $\mathsf{x}'$, and this equivalence class is distinct for each pair $(\mathsf{x},\mathsf{x}')$.  

\vspace{-0.7cm}

The remaining subtlety in \eqref{eq:two-point-emissions} and \eqref{eq:A} is that these formulae may become ambiguous if there are values of $\tau$ for which $\mathsf{Z}(\tau)$ is on the past lightcones of both $\mathsf{x}$ and $\mathsf{x}'$, because of the product $E^+_\mathsf{Z}(\tau; \mathsf{x})A(\tau; \mathsf{x}')$ in \eqref{eq:two-point-emissions}. A point-splitting analysis of local observables will then need to take the coincidence limit in a way that avoids this ambiguity. We shall see in Sec.\ \ref{sec:stress-energy} that this issue arises in $(3+1)$-dimensional Minkowski spacetime, and we show there how the ambiguity can be avoided.

\section{Deterministic and fluctuating contributions to back-action}\label{sec:examples-of-systems}

In this section, we consider the deterministic and fluctuating contributions to a two-point function, where the deterministic contribution is expressible in terms of the one-point function, and the fluctuating contribution is the covariance~\cite{Hsiang2022}.
We first show that the deterministic and fluctuating contributions to the field back-action originate entirely from the deterministic and fluctuating contributions to the two-point function of the detector's monopole moment operator, respectively. 

We then introduce the two-level Unruh-DeWitt (UDW) and harmonic oscillator (HO) detector models~\cite{Unruh1976, DeWitt1979}. These detectors approximate the dipole interaction between atomic orbitals and the quantum electromagnetic field in processes where angular momentum exchange is negligible~\cite{Martinez2013, Alhambra2014}, and they have been extensively used in acceleration effect analyses in quantum field theory~\cite{Birrell1982,Hu:2012jr}. 
We show that when prepared in an energy eigenstate, both the HO detector and the UDW detector give a fully fluctuating back-action. By contrast, the HO detector prepared in a coherent state gives a deterministic back-action that scales with the coherent amplitude, and a fluctuating back-action that is independent of the coherent amplitude and identical to that of a UDW detector prepared in its ground state.

\subsection{Decomposition of the back-action}\label{subsec:classical-LQS}

In the back-action contribution to the field's two-point function $\braket{\phi(\mathsf{x})\phi(\mathsf{x}')}^{(2)}$, given by \eqref{eq:two-point-emissions} and~\eqref{eq:A}, 
the detector enters only through $\braket{\mu(\tau)\mu(\tau')}_i$, 
the two-point function of the monopole moment operator evaluated in the initial state of the detector. We may split this two-point function into the sum of a deterministic part and a fluctuating part as follows. 

Recall that the covariance of operators $A$ and $B$ in a state $\rho$ is defined by 
\begin{align}
    \cov_\rho\!\left( A,B \right) &= \braket{AB}_\rho - \braket{A}_\rho \!\braket{B}_\rho . \label{eq:covariance-definition}
\end{align}
The two-point function $\braket{\mu(\tau)\mu(\tau')}_\rho$ hence decomposes as 
\begin{equation} \label{eq:<mumu>_split}
    \braket{\mu(\tau)\mu(\tau')}_\rho = \braket{\mu(\tau)}_\rho \braket{\mu(\tau')}_\rho + \cov_\rho\!\left( \mu(\tau),\mu(\tau') \right) .
\end{equation}
If the detector, prepared in the state $\rho$, behaves as a deterministic source with c-number monopole, so that ${\mu(\tau) = \braket{\mu(\tau)}_\rho \mathds{1}_D}$, then the covariance vanishes. We therefore interpret the first term on the right-hand side of \eqref{eq:<mumu>_split} as the deterministic part of $\braket{\mu(\tau)\mu(\tau')}_\rho$ and the second term as the non-deterministic, or fluctuating, part~\cite{Hsiang2022}.

Using \eqref{eq:<mumu>_split} with $\rho$ being the initial state of the detector, we decompose the two-point function \eqref{eq:two-point-emissions} as 
\begin{align}
    \braket{\phi(\mathsf{x})\phi(\mathsf{x}')}^{(2)} &= 2\biggl[\int_{-\infty}^\infty\mathrm{d}\tau \, \chi(\tau)\braket{\mu(\tau)}_i E^+_\mathsf{Z}(\tau; \mathsf{x})\notag\\
    &\quad\times\int_{-\infty}^\tau\mathrm{d}\tau'\chi(\tau')\braket{\mu(\tau')}_i \im \! \left[W(Z(\tau'), \mathsf{x}')\right] \notag\\
    &\quad+(\mathsf{x} \leftrightarrow \mathsf{x}')\biggr] + \braket{\phi(\mathsf{x})\phi(\mathsf{x}')}_{\rm Cov}^{(2)}, 
    \label{eq:phi-phi-classical-int1}
\end{align}
where we have used that $\braket{\mu(\tau)}_i$ is real-valued, and 
$\braket{\phi(\mathsf{x})\phi(\mathsf{x}')}_{\rm Cov}^{(2)}$ denotes the contribution from the covariance term in~\eqref{eq:<mumu>_split}. The explicit expression for $\braket{\phi(\mathsf{x})\phi(\mathsf{x}')}_{\rm Cov}^{(2)}$ will not be needed in what follows. 

In the first term in the square brackets in~\eqref{eq:phi-phi-classical-int1}, 
we use \eqref{eq:CCR} and \eqref{eq:E+-E-} to write 
\begin{align}
2 \im \! \left[W(Z(\tau'), \mathsf{x}')\right] = 
E(\mathsf{Z}(\tau'), \mathsf{x}')
= E^+_\mathsf{Z}(\tau'; \mathsf{x}'),
\end{align}
where the last equality follows because $E^-_\mathsf{Z}(\tau'; \mathsf{x}') = 0$ everywhere in the integration range by~\eqref{eq:support-of-E:minus}.
Proceeding similarly in the $(\mathsf{x} \leftrightarrow \mathsf{x}')$ term in~\eqref{eq:phi-phi-classical-int1},
and combining the two terms via a relabelling of the integration variables, we can write \eqref{eq:phi-phi-classical-int1} as 
\begin{align}
    \braket{\phi(\mathsf{x})\phi(\mathsf{x}')}^{(2)} &= \int_{-\infty}^\infty\mathrm{d}\tau \, \chi(\tau)\braket{\mu(\tau)}_i E^+_\mathsf{Z}(\tau; \mathsf{x})\notag\\
    &\quad\times\int_{-\infty}^\infty\mathrm{d}\tau' \, \chi(\tau')\braket{\mu(\tau')}_i E^+_\mathsf{Z}(\tau'; \mathsf{x}') \notag\\
    &\quad + \braket{\phi(\mathsf{x})\phi(\mathsf{x}')}_{\rm Cov}^{(2)} \notag\\
    &= \braket{\phi(\mathsf{x})}^{(1)}\braket{\phi(\mathsf{x}')}^{(1)} + \braket{\phi(\mathsf{x})\phi(\mathsf{x}')}_{\rm Cov}^{(2)} ,  \label{eq:classical-emissions}
\end{align}
where $\braket{\phi(\mathsf{x})}^{(1)}$ \eqref{eq:mean-field-value} is the leading-order back-action contribution to the mean field value.

From \eqref{eq:phi_mean}, \eqref{eq:phi-phi-lambdasq} and \eqref{eq:classical-emissions} we see that 
\begin{equation}
    \cov\!\left(\phi(\mathsf{x}),\phi(\mathsf{x}')\right) = W(\mathsf{x}, \mathsf{x}') + \lambda^2 \braket{\phi(\mathsf{x})\phi(\mathsf{x}')}_{\rm Cov}^{(2)} + O(\lambda^3) , 
\label{eq:cov-x-xprime-expansion}
\end{equation}
noting that the Wightman function $W(\mathsf{x}, \mathsf{x}')$ \eqref{eq:wightman-initial} is equal to the covariance of $\phi(\mathsf{x})$ in the field's initial state. 
The covariance on the left-hand side of \eqref{eq:cov-x-xprime-expansion} is evaluated in the equivalence class of final states determined by $\mathsf{x}$ and $\mathsf{x}'$ as discussed below~\eqref{eq:A}.  
Hence, the leading-order effect of the interaction on the covariance of the field is given exactly by the contribution from the covariance of the detector. The deterministic contribution to the detector's monopole moment operator creates only deterministic back-action, and similarly for the fluctuating contribution.

\subsection{Unruh-DeWitt detector}\label{subsec:udw-detector}

We now consider the deterministic and fluctuating contributions to the back-action when the detector is a two-level UDW detector~\cite{Unruh1976, DeWitt1979}. 

The UDW detector's Hilbert space is $\mathcal{H}_D \cong \mathbb{C}^2$, spanned by the orthonormal basis $\left\{ \ket{0}_D, \ket{1}_D \right\}$. The detector's free Hamiltonian is $H_D = E \sigma^+\sigma^-$, where $E\in \mathbb{R}$ is the energy gap, and $\sigma^-$ and $\sigma^+$ are respectively the lowering and raising ladder operators, satisfying $\{\sigma^-,\sigma^+\} = \mathds{1}_D$. The ladder operators act on $\mathcal{H}_D$ as 
\begin{align}
    \sigma^+ \ket{0}_D &= \ket{1}_D, \quad\quad \sigma^+ \ket{1}_D = 0 \notag \\
    \sigma^- \ket{0}_D &= 0, \hspace{0.55cm}\quad\quad \sigma^- \ket{1}_D = \ket{0}_D.
\end{align}
Therefore, $H_D\ket{0}_D = 0$ and $H_D \ket{1}_D = E \ket{1}_D$. 
For $E>0$, the states $\ket{0}_D$ and $\ket{1}_D$ are respectively the ground and excited states of the detector; for $E<0$, the roles are reversed; for $E=0$, both states have energy zero and the system is degenerate. 

We take the monopole moment operator to be
\begin{equation}
    \mu(\tau) = \mathrm{e}^{-i E \tau}\sigma^- +\mathrm{e}^{i E \tau}\sigma^+.
    \label{eq:two-level}
\end{equation}
We verify in Appendix \ref{app:udw-nonclassicality} 
that when the detector is nondegenerate, $E\ne0$, the detector has no state $\rho$ in which the fluctuating part in $\braket{\mu(\tau)\mu(\tau')}_\rho$ vanishes. 

We now assume that the detector's initial state is~$\ket{0}_D$. 
We then have 
$\braket{\mu(\tau)}_i = 0$ and
\begin{align}
\cov_{i}(\mu(\tau),\mu(\tau')) = \braket{\mu(\tau)\mu(\tau')}_i 
    = \mathrm{e}^{-iE(\tau - \tau')} , 
\label{eq:mu-mu-twolevel}
\end{align}
so that 
$\braket{\mu(\tau)\mu(\tau')}_i$ consists of only the fluctuating part. In this sense, the UDW detector initialised in an energy eigenstate is entirely fluctuating, and the observations of Sec.~\ref{subsec:classical-LQS} show that its back-action on the quantum field is entirely fluctuating. 

Finally, we reiterate that since the sign of $E$ is not fixed,  initialising the detector in the state~$\ket{0}_D$ allows us to consider both excitations and de-excitations, and the distinction is encoded in the sign of~$E$. 
The case $E>0$ is when the detector starts in the ground state and gains energy when it transitions, whereas the case $E<0$ is when the detector starts in the excited state and loses energy when it transitions. 

\subsection{Harmonic oscillator detector}\label{subsec:SHO}

For comparison, we now consider the deterministic and fluctuating contributions to the back-action when the detector is a harmonic oscillator~(HO). 

The HO detector's Hilbert space $\mathcal{H}_D \cong \ell^2(\mathbb{N}_0)$ is spanned by the orthonormal basis $\left\{ \ket{n}_D \right\}_{n\in \mathbb{N}_0}$. The detector's free Hamiltonian is $H_D = \Omega a^\dagger a$, where $\Omega>0$ is the energy-level spacing, and $a$ and $a^\dagger$ are respectively the lowering and raising ladder operators, satisfying $[a,a^\dagger]=\mathds{1}_D$. The ladder operators act on $\mathcal{H}_D$ as 
\begin{align}
a\ket{n}_D &= \sqrt{n}\ket{n - 1}_D, 
    & a^\dagger\ket{n-1}_D &= \sqrt{n}\ket{n}_D, 
\end{align}
for $n\in \mathbb{Z}_{\geq 1}$, and $a\ket{0}_D = 0$. Therefore, $H_D \ket{0}_D = 0$ and $H_D \ket{n}_D = n \Omega \ket{n}_D$. 

We take the monopole moment operator to be 
\begin{align}
    \mu(\tau) = \mathrm{e}^{-i\Omega\tau}a + \mathrm{e}^{i\Omega\tau}a^\dagger . \label{eq:sho}
\end{align}
We verify in Appendix \ref{app:ho-nonclassicality} 
that the detector has no state $\rho$ in which the fluctuating part in $\braket{\mu(\tau)\mu(\tau')}_\rho$ vanishes. 

Suppose first that the detector's initial state is the energy eigenstate~$\ket{n}_D$, with energy eigenvalue~$n\Omega$. We then have 
$\braket{\mu(\tau)}_i = 0$ and
\begin{align}
\cov_{i}(\mu(\tau),\mu(\tau')) &= \braket{\mu(\tau)\mu(\tau')}_i 
\notag\\
&= 2 n \cos(\Omega(\tau-\tau')) 
+ \mathrm{e}^{-i\Omega(\tau - \tau')} , 
\label{eq:mu-mu-HO-eigen}
\end{align}
so that 
$\braket{\mu(\tau)\mu(\tau')}_i$ consists of only the fluctuating part. The back-action on the quantum field is entirely fluctuating. 

Suppose then that the detector's initial state is the coherent state~$\ket{\alpha}_D$, 
defined as a normalised eigenstate of the lowering operator, 
\begin{align}
    a\ket{\alpha}_D = \alpha\ket{\alpha}_D, 
\end{align}
with the eigenvalue $\alpha\in\mathbb{C}$, known as the coherent amplitude. 
This is a state in which the HO most closely resembles a classical oscillator~\cite{Klauder1985}. 
We then have 
$\braket{\mu(\tau)}_i = 2 \re \! \left(\alpha\mathrm{e}^{-i\Omega\tau}\right)$ and 
\begin{align}
    \braket{\mu(\tau)\mu(\tau')}_i &= 4 \re\Bigl(\alpha\mathrm{e}^{-i\Omega\tau}\Bigr) \re\Bigl(\alpha\mathrm{e}^{-i\Omega\tau'}\Bigr) + \mathrm{e}^{-i\Omega(\tau - \tau')} , \label{eq:mumu-sho}
\end{align}
so that 
\begin{align}
    \cov_{i}(\mu(\tau),\mu(\tau')) &= \mathrm{e}^{-i\Omega(\tau - \tau')} . \label{eq:mumu-cov}
\end{align}
For $\alpha\ne0$, $\braket{\mu(\tau)\mu(\tau')}_i$ hence has both a deterministic part and a fluctuating part. The deterministic part comes from the coherent amplitude~$\alpha$; this is in line with the description of the coherent states as displaced vacua~\cite{Klauder1985}. The fluctuating part comes from the oscillator's ground state. 
The deterministic part dominates when $|\alpha|\gg1$, whereas the fluctuating part dominates when $|\alpha|\ll1$. 

In this paper we focus on the fluctuating part of the back-action and will therefore not consider the HO coherent states with $\alpha\ne0$ further. 

When prepared in an energy eigenstate, both the HO detector and the UDW detector give a fully fluctuating back-action. 
When the detectors are prepared in their respective ground states, their back-action is identical, as seen from \eqref{eq:mu-mu-twolevel} and~\eqref{eq:mu-mu-HO-eigen}. By contrast, when the UDW detector is prepared in its excited state, its back-action is not identical to that of the HO detector prepared in any of its excited states, not even the lowest one, as seen from \eqref{eq:mu-mu-twolevel} and~\eqref{eq:mu-mu-HO-eigen}: the physical reason for this that a HO detector prepared in an excited state can undergo transitions both upwards and downwards, and both of these transitions contribute to our second-order perturbation theory analysis. 

For the rest of this paper we consider the UDW detector, leaving further consideration of the HO detector to future work.

\section{Stress-energy tensor in \texorpdfstring{$(3+1)$}{3+1} Minkowski space}\label{sec:stress-energy}

In this section, we specialise to a massless minimally coupled scalar field in $(3+1)$-dimensional Minkowski spacetime, initially prepared in the Minkowski vacuum. We find an integral expression for the field's renormalised SET for an arbitrary detector trajectory, working under the technical assumption that the trajectory is real analytic over the interval of proper time in which the switching function is non-vanishing. 

\vspace{-0.5cm}
\subsection{Point-splitting renormalisation}

We renormalise the SET via the point-splitting prescription~\cite{Christensen1976, Christensen1978, Ford1989}, which in Minkowski spacetime is equivalent to normal-ordering~\cite{Wald:1977up}. In the point-splitting prescription in Minkowski spacetime, the renormalised SET in a state $\rho$ is defined with respect to a reference state $\rho_0$ as
\begin{equation}
    \braket{T_{\mu\nu}(\mathsf{x})}_{\text{ren},\rho} =\! \lim_{\mathsf{x'}\to \mathsf{x}}\left(\mathbb{T}_{\mu\nu}\!\left[ \braket{\phi(\mathsf{x})\phi(\mathsf{x'})}_\rho - \braket{\phi(\mathsf{x})\phi(\mathsf{x'})}_{\rho_0}  \right]\right),
\end{equation}
where 
\begin{align}
    \mathbb{T}_{\mu\nu} &= \frac{1}{2}\partial_\mu\partial_{\nu'} + \frac{1}{2}\partial_{\nu}\partial_{\mu'} - \frac{1}{2}g_{\mu\nu}\partial_\rho\partial^{\rho'}
    \label{eq:tmunu-bidistribution-operator}
\end{align}
is the point-split stress-energy differential operator, with unprimed partial derivatives acting on $\mathsf{x}$ and primed partial derivatives acting on~$\mathsf{x}'$.
This prescription yields a well-defined renormalised SET if both $\rho$ and $\rho_0$ are Hadamard states; that is, states for which the two-point function is locally given by the Minkowski vacuum two-point function, which contains all the short-distance singularities \cite{Decanini2008}. Such states are thus considered to be physically relevant.

We assume that the initial state of the free field is a Hadamard state. 
For the free scalar field, the Hadamard property is known to be stable under Cauchy evolution: if the two-point function has Hadamard short-distance structure in a neighbourhood of one Cauchy surface, then it has Hadamard short-distance structure everywhere~\cite{Fulling:1978ht, Sahlmann:2000zr}. In perturbatively interacting quantum field theory with local interactions given by Wick polynomials smeared with smooth, compactly supported spacetime test functions, the perturbative algebraic framework likewise shows that the physically relevant interacting states are precisely those whose two-point function retains the Hadamard singularity structure of the underlying free theory~\cite{Hollands:2001qe,Fewster2015,Rejzner:2016hdj}. 

However, our detector coupling is supported along a timelike worldline, not smeared spatially, and the above stability results do not apply. It is therefore not clear \textit{a priori\/} whether the final state of the field is Hadamard. We shall nevertheless use point-splitting renormalisation of the SET in the final state of the field. The resulting renormalised SET is well-defined to second order in the coupling constant at all spacetime points that are not on the detector's worldline; at this order, any potential additional singularities generated by the pointlike interaction are confined to the worldline itself and do not affect our analysis away from it.

With the reference Hadamard field state taken to be the Minkowski vacuum, the renormalised SET in the final state of the field to second order in $\lambda$ is
\begin{subequations}
    \begin{align}
        \braket{T_{\mu\nu}(\mathsf{x})}_{\rm ren} &= \lambda^2  \braket{T_{\mu\nu}(\mathsf{x})}^{(2)} + O(\lambda^3), 
        \label{eq:T2-expansion}\\
        \braket{T_{\mu\nu}(\mathsf{x})}^{(2)} &= \lim_{\mathsf{x}'\to\mathsf{x}}\mathbb{T}_{\mu\nu}\left[\braket{\phi(\mathsf{x})\phi(\mathsf{x}')}^{(2)}\right],\label{eq:T2-pointsplit}
    \end{align}
\end{subequations}
where we have suppressed the final field-state in the notation for brevity. As discussed in Sec.~\ref{subsec:two-point-function}, the order $\lambda$ contribution to the two-point function vanishes for zero-mean Gaussian initial field states, and hence $\braket{T_{\mu\nu}}^{(1)}=0$. 
As with $\braket{\phi(\mathsf{x})}^{(n)}$
and $\braket{\phi(\mathsf{x})\phi(\mathsf{x'})}^{(n)}$, we note that $\braket{T_{\mu\nu}(\mathsf{x})}^{(2)}$ is the coefficient of $\lambda^2$ in the perturbative expansion~\eqref{eq:T2-expansion}, not an expectation value in a state by itself.

\subsection{\texorpdfstring{$\braket{T_{\mu\nu}(\mathsf{x})}^{(2)}$}{Evaluating the leading order contribution}}

We take the detector to be a two-level UDW detector, in the notation of Sec.~\ref{subsec:udw-detector}\null. From \eqref{eq:two-point-emissions} and \eqref{eq:mu-mu-twolevel} we find 
\begin{align}
    \braket{\phi(\mathsf{x})\phi(\mathsf{x}')}^{(2)} &= 2\im\Biggl[\int_{-\infty}^\infty\mathrm{d}\tau \, \chi(\tau)\mathrm{e}^{-iE\tau}E^+_\mathsf{Z}(\tau; \mathsf{x}) \notag\\
    &\qquad\times\int_{-\infty}^\tau\mathrm{d}\tau' \, \chi(\tau')\mathrm{e}^{iE\tau'}W(\mathsf{Z}(\tau'), \mathsf{x}')\Biggr] \notag\\
    &\quad + (\mathsf{x} \leftrightarrow \mathsf{x}'). \label{eq:two-point-UDW-general}
\end{align}
For a massless scalar field, initialised in the Minkowski vacuum, we have \cite{Birrell1982,Kay:1988mu}
\begin{align}
    W(\mathsf{Z}(\tau), \mathsf{x}) &= \frac{1}{4\pi^2}\frac{1}{(\mathsf{Z}(\tau - i\epsilon) -  \mathsf{x})^2} , 
\label{eq:Mink-Wightman}
\end{align}
understood in the distributional sense $\epsilon\to0_+$. 
We have used the real analyticity of the trajectory to group 
$-i\epsilon$ with~$\tau$: 
this will simplify the technical steps that follow. 

It is at this point that we introduce the function
\begin{align}
    f(\tau; \mathsf{x}) := (\mathsf{Z}(\tau) - \mathsf{x})^2 , 
\label{eq:f-func-def}
\end{align}
and denote its partial derivatives by 
\begin{subequations}
\label{eq:f-derivatives-def}
\begin{align}
f'(\tau; \mathsf{x}) &:= \partial_\tau f(\tau; \mathsf{x}) = 2 \dot{\mathsf{Z}}^\mu(\tau) (\mathsf{Z}(\tau) - \mathsf{x})_\mu \, , 
\\
f_\mu(\tau; \mathsf{x}) &:= \partial_\mu f(\tau; \mathsf{x})
= - 2 (\mathsf{Z}(\tau) - \mathsf{x})_\mu \, . 
\end{align}
\end{subequations}
We denote by $\tau_-$ the function of $\mathsf{x}$ 
defined in Sec.\ \ref{subsec:mean-field-value} and illustrated in Fig.~\ref{fig:spacetime-diagram}, and we denote by $\tau_-'$ this function evaluated at~$\mathsf{x}'$.  
It follows that 
$f(\tau_-; \mathsf{x}) =0$ and $f(\tau_-'; \mathsf{x}') =0$, and further that $f'(\tau_-; \mathsf{x}) >0$ and $f'(\tau_-'; \mathsf{x}') >0$. 

In this notation, the retarded propagator $E^+_\mathsf{Z}(\tau; \mathsf{x})$ is given by 
\begin{align}
    E^+_\mathsf{Z}(\tau; \mathsf{x}) &= 2\Theta(t- \mathsf{Z}^0(\tau))\im \Bigl[W(\mathsf{Z}(\tau), \mathsf{x})\Bigr] \notag\\
    &=\frac{1}{2\pi}\frac{\delta(\tau_- - \tau)}{f'(\tau_-; \mathsf{x})} , 
\label{eq:EplusZ-Mink}
\end{align}
using \eqref{eq:Mink-Wightman} and \eqref{eq:f-func-def}, and isolating the distributional contribution in $W(\mathsf{Z}(\tau), \mathsf{x})$ by the Sokhotski-Plemelj identity, 
\begin{align}
\lim_{\epsilon \to 0_+} 
\frac{1}{\tau \pm i \epsilon}
= 
\mp i\pi \delta(\tau) + {\rm P.V.} \left(\frac{1}{\tau}\right) , 
\label{eq:sokhotski}
\end{align}
where P.V.\ stands for the Cauchy principal value. 

Recall that, by assumption, $\mathsf{x}$ and $\mathsf{x}'$ are distinct and not on the detector's trajectory. We now assume that $\mathsf{x}$ and $\mathsf{x}'$ are timelike separated. It follows that $\tau_-$ and $\tau_-'$ do not coincide. This implies that \eqref{eq:two-point-UDW-general} is well defined, because the value of $\tau$ at which the inner integral ends on a singularity of $W(\mathsf{Z}(\tau'), \mathsf{x}')$ does not coincide with the singularity of $E^+_\mathsf{Z}(\tau; \mathsf{x})$. Using~\eqref{eq:EplusZ-Mink}, \eqref{eq:two-point-UDW-general} then reduces to 
\begin{align}
    \braket{\phi(\mathsf{x})\phi(\mathsf{x}')}^{(2)} &= \im\biggl[\frac{\chi(\tau_-)\mathrm{e}^{iE\tau_-}}{4\pi^3 f'(\tau_-; \mathsf{x})}\int_{-\infty}^{\tau_-}\mathrm{d}\tau \, \frac{\chi(\tau)\mathrm{e}^{-iE\tau}}{f(\tau - i\epsilon; \mathsf{x}')}\biggr]\notag\\
    &\qquad + (\mathsf{x} \leftrightarrow \mathsf{x}') . \label{eq:mink-phi-E-sub}
\end{align}

When $\tau_-' < \tau_-$, the integrand in the term written out in \eqref{eq:mink-phi-E-sub} contains a distributional singularity, and the $\epsilon\to0_+$ limit can be taken by the Sokhotski-Plemelj identity \eqref{eq:sokhotski}. 
In the $(\mathsf{x} \leftrightarrow \mathsf{x}')$ term, the corresponding integrand is nonsingular, 
and the $\epsilon\to0_+$ limit can be taken directly under the integral. 
When $\tau_- < \tau_-'$, the roles of the two terms are reversed: 
the term written out in \eqref{eq:mink-phi-E-sub} has a nonsingular integrand and the integrand in the $(\mathsf{x} \leftrightarrow \mathsf{x}')$ term has a distributional singularity. 
We write the outcome of the $\epsilon\to0_+$ limit as 
\begin{align}
    \braket{\phi(\mathsf{x})\phi(\mathsf{x}')}^{(2)}\! &= \im\biggl[\frac{\chi(\tau_-)\mathrm{e}^{iE\tau_-}}{4\pi^3 f'(\tau_-; \mathsf{x})}\biggl(\dashint_{-\infty}^{\tau_-}\mathrm{d}\tau \, \frac{\chi(\tau)\mathrm{e}^{-iE\tau}}{f(\tau; \mathsf{x}')}\notag\\
    &+i\pi\frac{\chi(\tau_-')\mathrm{e}^{-iE\tau_-'}}{f'(\tau_-'; \mathsf{x}')}\Theta(\tau_- - \tau_-')\biggr)\biggr] \!+\! (\mathsf{x} \leftrightarrow \mathsf{x}') , 
    \label{eq:two-point-function-mink}
\end{align}
which covers both $\tau_-' < \tau_-$ and $\tau_- < \tau_-'$, 
with the understanding that
$\dashint$ denotes the Cauchy principal value integral in the term in which the denominator has a linear zero, and a normal integral in the term in which the denominator does not have a zero.  

What remains is to substitute \eqref{eq:two-point-function-mink} into \eqref{eq:T2-pointsplit} and~\eqref{eq:tmunu-bidistribution-operator}, evaluate the derivatives, and take the coincidence limit. 
Defining 
\begin{align}
    I_{\mu\nu}(\mathsf{x}) &:= \lim_{\mathsf{x}'\to\mathsf{x}}\partial_\mu\partial_{\nu'}\braket{\phi(\mathsf{x})\phi(\mathsf{x}')}^{(2)} , 
\label{eq:Imunu-def}
\end{align}
we show in Appendix \ref{app:derivatives-of-correlator} that 
\begin{align}
    I_{\mu\nu}\!
    & = \frac{1}{4\pi^3f'(\tau_{-})} \Biggl[  - E\partial_{\tau}\left.\left( \frac{\alpha_{\mu}(\tau)\alpha_{\nu}(\tau)}{f'(\tau)} \right) \right|_{\tau = \tau_-}   \notag\\
    &\quad + \pi\frac{E^{2}\alpha_{\mu}(\tau_{-})\alpha_{\nu}(\tau_{-}) + \alpha'_{\mu}(\tau_{-})\alpha'_{\nu}(\tau_{-})}{f'(\tau_{-})}\notag\\
    &\quad - 2\int _{-\infty}^{\tau_{-}} \mathrm{d}\tau \biggl(E^{2}   \frac{\sin(E(\tau - \tau_{-}))}{f(\tau)}\alpha_{(\mu}(\tau_{-})\alpha_{\nu)}(\tau) \notag\\
    &\quad +\! E\cos(E(\tau - \tau_{-})) \frac{\alpha_{(\mu}'(\tau_{-})\alpha_{\nu)}(\tau) - \alpha_{(\mu}'(\tau)\alpha_{\nu)}(\tau_{-})}{f(\tau)}\notag\\
    &\quad  + \frac{\sin(E(\tau - \tau_{-}))}{f(\tau)} \alpha_{(\mu}'(\tau_{-})\alpha_{\nu)}'(\tau) \biggr)  \Biggr] , \label{eq:masterequation}
\end{align}
with
\begin{align}
    \alpha_\mu(\tau) := \chi(\tau)\frac{f_\mu(\tau; \mathsf{x})}{f'(\tau; \mathsf{x})} , 
    \label{eq:alpha-mu-def}
\end{align}
recalling that $f$ and its derivatives were defined in \eqref{eq:f-func-def} and~\eqref{eq:f-derivatives-def},  recalling that $\tau_-$ is the function of $\mathsf{x}$ defined in Sec.\ \ref{subsec:mean-field-value} and illustrated in Fig.~\ref{fig:spacetime-diagram}, 
and suppressing in the notation the dependence on~$\mathsf{x}$. 
The indices in parentheses denote symmetrisation, in the convention $A_{(\mu\nu)} := \frac{1}{2}\left(A_{\mu\nu} + A_{\nu\mu}\right)$.
From \eqref{eq:T2-pointsplit}, \eqref{eq:tmunu-bidistribution-operator} and \eqref{eq:Imunu-def} we then find 
\begin{align}
    \braket{T_{\mu\nu}(\mathsf{x})}^{(2)} = I_{\mu\nu}(\mathsf{x}) - \frac{1}{2}g_{\mu\nu} g^{\sigma\rho}I_{\rho\sigma}(\mathsf{x}) .
    \label{eq:Tmunu-from-masterequation}
\end{align}

Note that $\braket{T_{\mu\nu}(\mathsf{x})}^{(2)}$ \eqref{eq:Tmunu-from-masterequation} only depends on the interaction in the causal past of the spacetime point~$\mathsf{x}$, thus respecting causality. This property is inherited from the causality of the two-point function~\eqref{eq:two-point-emissions}. 

Note also that formula \eqref{eq:masterequation} for $I_{\mu \nu} (\mathsf{x})$ is free of singularities. While the factor $1/f(\tau)$ in the integrand has an inverse linear singularity
at the integration endpoint, $\tau=\tau_-$, this singularity is cancelled by the linear zero in the sine factor in two of the terms and by the combination $\alpha_{(\mu}'(\tau_{-})\alpha_{\nu)}(\tau) - \alpha_{(\mu}'(\tau)\alpha_{\nu)}(\tau_{-})$ in the remaining term. The distributional aspects of the two-point functions have hence been fully resolved, and $\braket{T_{\mu\nu}(\mathsf{x})}^{(2)}$ 
\eqref{eq:Tmunu-from-masterequation} is manifestly well defined.

\section{Inertial detector}\label{sec:inertial-motion}

In this section we consider the SET $\braket{T_{\mu\nu}(\mathsf{x})}^{(2)}$ \eqref{eq:Tmunu-from-masterequation} for an inertial UDW detector. 

\subsection{Evaluation of the stress-energy tensor}

As a simplifying assumption, we take the limit $\chi(\tau) = 1$ such that that the detector has been operating at constant coupling strength since the asymptotic past: while formula \eqref{eq:Tmunu-from-masterequation} was obtained under the technical assumption that $\chi$ has compact support, we shall see that the formula remains well defined in this limit. 
As we take this infinite interaction duration limit after the perturbative expansion in the coupling constant, the limit is not Markovian~\cite{Benatti2004, DeBievre2006}, but should be understood to occupy an asymptotic regime where the smallness of the coupling constant overrides any nonperturbative effects that may grow with the interaction duration. 
We leave effects due to a strictly finite interaction duration subject to future work.

We work in spherical polar coordinates $(t, r, \theta, \varphi)$ in which our detector is static at $r=0$, and we set $\chi=1$. 
From \eqref{eq:f-func-def}, 
\eqref{eq:f-derivatives-def}
and \eqref{eq:alpha-mu-def} we find 
\begin{subequations}
\label{eq:inertial-f-and-alpha}
    \begin{align}
        f(\tau; \mathsf{x}) &= r^2 - (\tau - t)^2 \\
        \tau_- &= t - r \\
        f'(\tau; \mathsf{x}) &= -2(\tau - t) \\
        \alpha_t(\tau; \mathsf{x}) &= -1 \\
        \alpha_r(\tau; \mathsf{x}) &= \frac{r}{t - \tau} \\
        \alpha_\theta(\tau; \mathsf{x}) &= 
        \alpha_\varphi(\tau; \mathsf{x}) = 0 , 
    \end{align}
\end{subequations}
in which the spherical symmetry of the system is apparent. Substituting \eqref{eq:inertial-f-and-alpha} into \eqref{eq:masterequation} gives 
\begin{subequations}
    \label{eq:inertial-i-munu}
    \begin{align}
         I_{tt} &= \frac{1}{8\pi^3}\biggl[\frac{E^{2}}{r^{2}}\int_{0}^{\infty} \mathrm{d}\tau  \frac{\sin(E\tau)}{\tau + 2r} + \frac{\pi E^{2}}{r^{2}}\Theta(-E) - \frac{E}{2r^{3}}\biggr] \\
         I_{rr} & = \frac{1}{8\pi^3}\biggl[\left(\frac{1}{r^{4}} - \frac{E^2}{r^2}\right)\int_{0}^{\infty}\mathrm{d}\tau  \frac{\sin(E\tau)}{\tau + 2r} + \frac{\pi E^{2}}{r^{2}}\Theta(-E) \notag\\
         &  + \frac{2E}{r^{3}}\int_{0}^{\infty} \mathrm{d}\tau  \frac{\cos(E\tau)}{\tau + 2r} + \frac{E}{2r^{3}} + \frac{\pi}{r^{4}}\Theta(-E)\biggr] \\
         I_{tr} &= \frac{E^{2}}{8\pi^2 r^{2}}\Theta(-E) \\
         I_{\theta\mu} &= I_{\varphi\mu} = 0 , 
    \end{align}
\end{subequations}
where $\Theta(x)$ is the Heaviside step function~\eqref{eq:Theta-def}. 
The integrals in \eqref{eq:inertial-i-munu} exist as improper Riemann integrals for $E\ne0$, and they can be expressed in terms of the sine and cosine integrals~\cite{DLMF}: the limit $\chi=1$ has hence produced a well-defined outcome. 
$\braket{T_{\mu\nu}(\mathsf{x})}^{(2)}$ is given by \eqref{eq:Tmunu-from-masterequation} with~\eqref{eq:inertial-i-munu}.

\subsection{Energy density}\label{subsec:inertial-energy-density}

\begin{figure}
    \centering
    \includegraphics[width=\linewidth]{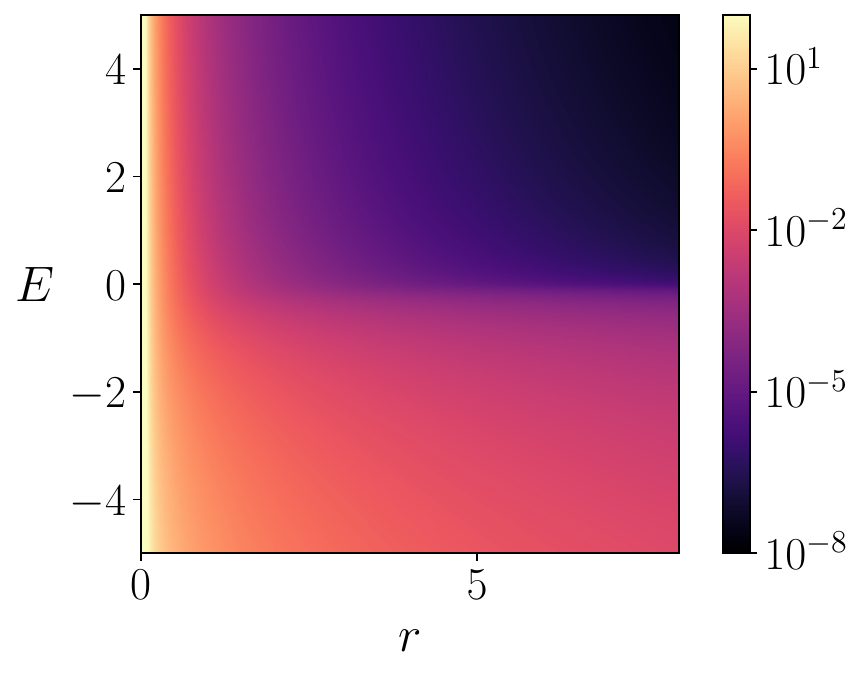}
    \caption{The energy density $\braket{T_{tt}}^{(2)}$ of the field for an inertial detector in the long-time limit~\eqref{eq:inertial-T_(tt)}, as a function of the radial distance $r$ and the energy gap~$E$, in arbitrary units. Clearly seen are the different fall-off rates in $r$ for a detector initially in its ground state ($E > 0$) and a detector initially in its excited state ($E < 0$). Also seen is the expected property that larger de-excitation gaps emit more energy into the field. An $r^{-4}$ singularity, which is independent of the energy gap, is seen close to $r=0$.}
    \label{fig:static-energy-density}
\end{figure}

For the energy density, $T_{tt}$, we find 
\begin{align}
    \braket{T_{tt}}^{(2)} & = \frac{1}{2}I_{tt} + \frac{1}{2}I_{rr} \notag\\
        &= \frac{1}{16\pi^2}\left(\frac{2 E^2}{r^2}
        + \frac{1}{r^4}\right)
        \Theta(-E) 
        \notag\\
        &\hspace{3ex} + 
        \frac{1}{16\pi^3}\Biggl(\frac{2E}{r^3}\int_{0}^{\infty}\mathrm{d}\tau \, \frac{\cos(E\tau)}{\tau + 2r} 
        \notag\\
        &\hspace{13ex} 
        + \frac{1}{r^4}\int_{0}^{\infty}\mathrm{d}\tau \, \frac{\sin(E\tau)}{\tau + 2r} \Biggr) . \label{eq:inertial-T_(tt)}
\end{align}
A plot of $\braket{T_{tt}}^{(2)}$ \eqref{eq:inertial-T_(tt)} is shown in Fig.~\ref{fig:static-energy-density}. 

Three observations are in order. 

First, $\braket{T_{tt}}^{(2)}$ is positive, continuous, and strictly decreasing in~$E$, as we show in Appendix~\ref{app:T00pos}\null. 
It is  asymptotic to ${(16 \pi^3 E r^5)}^{-1}$ as $E\to\infty$ and to 
${(8 \pi^2)}^{-1}E^2/r^2$ as $E\to-\infty$, as follows from integration by parts in~\eqref{eq:inertial-T_(tt)}~\cite{Wong2001}. 

That $\braket{T_{tt}}^{(2)}$ is strictly positive for $E>0$ may be surprising in view of the fact that a detector initially prepared in its ground state would have a vanishing excitation rate in the long interaction time limit, were the detector's final state measured~\cite{Birrell1982}. 
There is however no contradiction, for two reasons. First, if the detector's final state is measured, the detector's excitation \emph{probability\/}, rather than the excitation rate, is sensitive to the way in which the long interaction time limit is taken, and there exist long time limit implementations in which the excitation probability remains finite in the long time limit, due to residual switching effects, as we show in Appendix~\ref{app:detector-excitation-probability}\null. 
Second, in our situation the detector's final state is not measured. 
These properties suggest that $\braket{T_{tt}}^{(2)}$ for $E>0$ is not directly related to transitions of the detector, but rather it is a leading order correction to the ground state of the coupled system. This is an explicit instance of vacuum polarisation~\cite{Schwinger1951}, with the binding energy of the system coming from the interaction Hamiltonian and being dispersed into the field.

Second, $\braket{T_{tt}}^{(2)}$ does differ for $E>0$ and $E<0$, and the distinction shows up strongly in the far-field fall-off. 
When the UDW detector starts in its excited state ($E < 0$), $\braket{T_{tt}}^{(2)}$ falls off as a multiple of $r^{-2}$ and scales in $E$ proportionally to~$E^2$;  
when the detector starts in its ground state ($E > 0$), the fall-off is stronger, a multiple of~$r^{-5}$, and the $E$-dependence is suppression proportional to $E^{-1}$ at large~$E$. 

Third, as $r\to0$, we have 
\begin{align}
\braket{T_{tt}}^{(2)} \sim \frac{1}{32 \pi^2 r^4} ,
\label{eq:Ttt-small-r-as}
\end{align}
which shows that 
$\braket{T_{tt}}^{(2)}$ diverges to positive infinity close to the detector's worldline, proportionally to~$r^{-4}$, 
and the leading term in this divergence is independent of~$E$.  
The singularity is so strong that the total energy in a sphere surrounding the detector is infinite. 
This infinity is a consequence of the spatially pointlike interaction between the detector and the field.

\begin{figure}
    \centering
    \includegraphics[width=\linewidth]{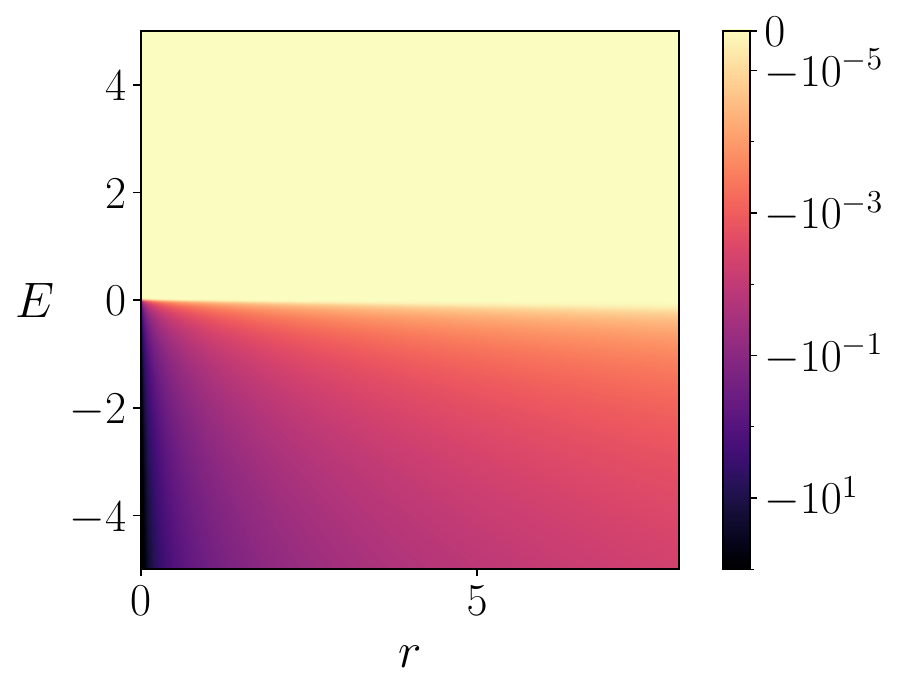}
    \caption{The energy flux $\braket{T_{tr}}^{(2)}$ of the field for an inertial detector in the long-time limit~\eqref{eq:Ttr-inertial-final}. $\braket{T_{tr}}^{(2)}$~is nonvanishing only for a detector initially in its excited state ($E < 0$), and it is continuous across $E=0$. Its magnitude is a multiple $E^2/r^2$, and its negative sign indicates an energy flux away from the detector.}
    \label{fig:static-energy-flux}
\end{figure}

\subsection{Energy flux}\label{subsec:inertial-energy-flux}

For the energy flux, $T_{tr}$, we find 
\begin{align}
    \braket{T_{tr}}^{(2)} &= -\frac{E^2}{8\pi^2r^2}\Theta(-E) = -\frac{1}{4\pi r^2}\;\frac{|E|}{2\pi}\Theta(-E)\; |E| . 
\label{eq:Ttr-inertial-final}
\end{align}
A plot of $\braket{T_{tr}}^{(2)}$ \eqref{eq:Ttr-inertial-final} is shown in Fig.~\ref{fig:static-energy-flux}.

The flux is nonzero only when $E<0$, that is, when the detector was initially in the excited state, and the sign shows that the energy is then flowing from the detector towards the infinity: the energy flux through a spherical shell with outward unit normal $n^\mu = {(\partial_r)}^\mu$, as seen by a static observer with four-velocity $u^\mu = {(\partial_t)}^\mu$, is equal to ${-\braket{T_{\mu \nu}}^{(2)}u^\mu n^\nu = -\braket{T_{tr}}^{(2)}}$~\cite{Misner:1973prb}. 
In the last expression in~\eqref{eq:Ttr-inertial-final}, we have written $\braket{T_{tr}}^{(2)}$ as the product of the detector's transition rate, $|E|(2\pi)^{-1}\Theta(-E)$~\cite{Birrell1982}, the energy gap absolute value~$|E|$, and the inverse area ${(4\pi r^2)}^{-1}$ of a sphere of radius~$r$. This suggests that the flux can be attributed to emissions from the detector as the detector decays from its excited state to its ground state, even though the detector's final state is not measured in our situation.

\section{Uniformly linearly accelerated detector}\label{sec:uniform-acceleration}

In this section, we examine the SET $\braket{T_{\mu\nu}(\mathsf{x})}^{(2)}$ \eqref{eq:Tmunu-from-masterequation} for a uniformly linearly accelerated UDW detector.

\subsection{Rindler trajectory and the long interaction time limit}

\begin{figure}
    \centering
    \resizebox{0.45\textwidth}{!}{%
        \begin{tikzpicture}
        \def\a{0.5}
            \draw[->] (-3,0) -- (3,0) node[below] {$z$};
            \draw[->] (0,-3) -- (0,3) node[left] {$t$};
            \draw[->] (3,-3) -- (-3,3) node[below left] {$u$};
            \draw[->] (-3,-3) -- (3,3) node[below right] {$v$};
            \draw[->,scale=0.5,domain=-3.3:3.3,smooth,variable=\t] plot ({cosh(\t * \a) / \a},{sinh(\t * \a) / \a}) node[below=2.5ex] {$\mathsf{Z}(\tau)$};
            \draw node[above] at (1.5,0) {$R$};
            \draw node[above] at (-1.5,0) {$L$};
            \draw node[left] at (0,1.5) {$F$};
            \draw node[left] at (0,-1.5) {$P$};
        \end{tikzpicture}
    }
    \caption{A spacetime diagram of the Rindler trajectory~\eqref{eq:linear-acc-trajectory}, labelled here~$\mathsf{Z}(\tau)$, in the $(t, z)$ plane. The lightlike coordinates are $u = t - z$ and $v = t + z$. The left and right Rindler wedges are labelled respectively as $L$ and $R$, and the future and past wedges are labelled respectively as $F$ and~$P$. The causal future of the trajectory is at $v>0$. The future branch of the Rindler horizon is at $u=0$, $v>0$.}
    \label{fig:rindler-trajectory}
\end{figure}
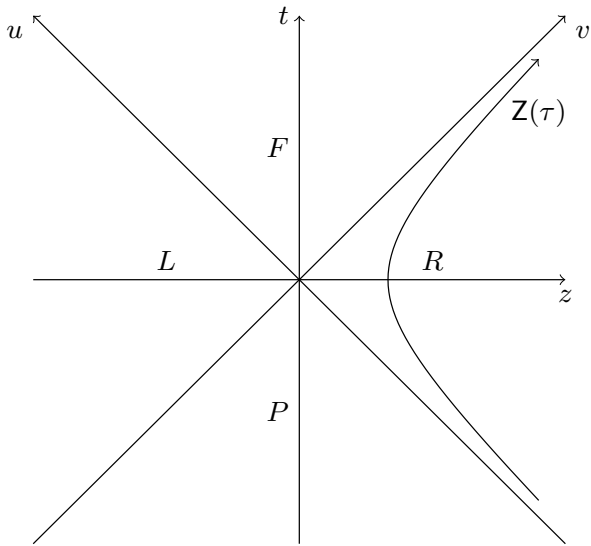

In the Minkowski coordinates $(t, x, y, z)$, the uniformly linearly accelerated trajectory reads 
\begin{align}
    Z(\tau) = \left( \frac{1}{a}\sinh(a\tau), 0, 0, \frac{1}{a}\cosh(a\tau) \right) , 
\label{eq:linear-acc-trajectory}
\end{align}
where the positive constant $a$ is the proper acceleration. 
The trajectory is in the $(t,z)$ plane, and therein in the quadrant $|t| < z$. A spacetime diagram is shown in Fig.~\ref{fig:rindler-trajectory}. The trajectory is known as the Rindler trajectory, and the quadrant $|t| < z$ as the right Rindler wedge~\cite{Rindler:1960zz}. 

In the perturbative setting of our $\braket{T_{\mu\nu}(\mathsf{x})}^{(2)}$~\eqref{eq:Tmunu-from-masterequation}, we assume that the detector has been operating at constant coupling strength since the asymptotic past, and we implement this long interaction time limit by taking $\chi(\tau) = 1$. As was the case for the inertial trajectory, this limit occupies an asymptotic regime where the smallness of the coupling constant overrides any nonperturbative long interaction time effects. While the impact of a finite duration for the interaction is well understood from the detector's perspective~\cite{Svaiter1992,Higuchi1993,Sriramkumar1994,Fewster2016,Parry:2025wub}, we leave finite time effects from the field's perspective subject to future work. Since the Unruh effect can be found in the long-time limit of a perturbative interaction, we choose $\chi(\tau) = 1$ such that the results of this paper could be considered a reasonable candidate for its emissive counterpart.

\vspace{0.3cm}

\subsection{Stress-energy}\label{subsec:SET-formulas}
\vspace{-0.2cm}

In the Minkowski coordinates $(t, x, y, z)$, the Rindler trajectory is given in \eqref{eq:linear-acc-trajectory}. 
In the null coordinates $(u,v,x,y)$, where $u = t-z$ and $v = t+z$, 
the Rindler trajectory is at the hyperbola $uv = -a^2$, on the branch where $u<0$ and $v>0$, as shown in the spacetime diagram of Fig.~\ref{fig:rindler-trajectory}.
We now find the SET in the causal future of the trajectory, $v>0$, noting that the causality of our model guarantees the SET to vanish for $v < 0$. For simplicity, we consider the SET only in the timelike plane of the detector's worldline, $x=y=0$, and only the components in this plane. We express the SET both in the Minkowski coordinates and in two sets of coordinates adapted to the boost isometries that leave the Rindler trajectory invariant. The three coordinate systems will be utilised for analysing the SET in Sections \ref{subsec:rindler-observers}--\ref{subsec:minkowski-observers}.

We evaluate $\braket{T_{tt}(\mathsf{x})}^{(2)}$ and $\braket{T_{tz}(\mathsf{x})}^{(2)}$ in Appendix~\ref{app:accelerated-emissions}\null. We find
\begin{widetext}
\begin{subequations}
\label{eq:rindler-stress-energy-mink}
\begin{align}
    \braket{T_{tt}(\mathsf{x})}^{(2)} & = \frac{v^{2} + u^{2}}{4\pi^3a^{2}{(uv + a^{-2})}^{4}}\Biggl[ \frac{2\pi}{1 - \mathrm{e}^{ 2\pi E/a }} + \frac{a}{E} 
- 2a(uv)^{S} \int _{0}^{\infty} \mathrm{d}\tau \frac{ \sin(E\tau)}{a^{-2S}\mathrm{e}^{ a\tau } + (uv)^{S} }  \Biggr]  
\notag\\
        & \quad
+\frac{1}{2\pi^2a^4\zeta^2{(uv + a^{-2})}^{2}} \frac{E^{2}}{1 - \mathrm{e}^{ 2\pi E/a }}
+\frac{E(v^{2} + u^{2})(uv)^{S - 1}}{2\pi^3a^2{|uv + a^{-2}|}^{3}}\int_{0}^{\infty} \mathrm{d}\tau  \frac{\cos(E\tau)}{a^{-2S}\mathrm{e}^{ a\tau } + (uv)^{S}} ,\label{eq:rindler-T_(tt)}
\\[5mm]
    \braket{T_{tz}(\mathsf{x})}^{(2)} &= \frac{u^2 - v^2}{4\pi^3a^2{(uv + a^{-2})}^4}\Biggl[ \frac{2\pi}{1 - \mathrm{e}^{ 2\pi E/a }} + \frac{a}{E} 
- 2a(uv)^{S} \int _{0}^{\infty} \mathrm{d}\tau \frac{ \sin( E\tau )}{a^{-2S}\mathrm{e}^{ a\tau } + (uv)^{S} }  \Biggr] \notag\\
    &\quad 
+ \frac{S}{2\pi^2a^{4}{\zeta^2(uv + a^{-2})}^2}\frac{E^2}{1 - \mathrm{e}^{2\pi E/a}} 
+ \frac{E(u^2 - v^2)(uv)^{S - 1}}{2\pi^3a^2{|uv + a^{-2}|}^3} \int_{0}^{\infty}\mathrm{d}\tau  \frac{\cos(E\tau)}{a^{-2S}\mathrm{e}^{a\tau} + (uv)^S} , 
\label{eq:rindler-T_(tz)}
\end{align}
\end{subequations}
\end{widetext}
where 
\begin{align}
    \zeta = \begin{cases}
        v & S = 1 \\
        u & S = -1
    \end{cases}
\label{eq:zeta-def}
\end{align}
and $S=1$ to the left of the trajectory, 
$u> - a^2/v$, and $S=-1$ to the right of the trajectory, $u< - a^2/v$. 
We show in Appendix \ref{app:accelerated-emissions} also that $\braket{T_{zz}}^{(2)} = \braket{T_{tt}}^{(2)}$. 
Note that $\braket{T_{tt}}^{(2)}$ and $\braket{T_{tz}}^{(2)}$ are well defined for $v>0$, everywhere except on trajectory, $uv = -a^2$. In particular, $\braket{T_{tt}}^{(2)}$ and $\braket{T_{tz}}^{(2)}$ are well defined on the future Rindler horizon, $u=0$, where $S=1$ in~\eqref{eq:rindler-stress-energy-mink}. 

The symmetries of the SET become more transparent in coordinates adapted to the invariance of the Rindler trajectory under the boosts in $(t,z)$. 
In the right Rindler wedge, $|t| < z$,  convenient coordinates are a version of the Rindler coordinates that we denote by $(\eta,\xi,x,y)$, related to the Minkowski coordinates by 
\begin{align}
t & = \xi \sinh\eta, & z &= \xi \cosh\eta, 
\label{eq:rindler-coords-def}
\end{align}
where $\xi>0$ and $-\infty<\eta<\infty$. The metric reads 
\begin{align}
ds^2 = - \xi^2 \mathrm{d}\eta^2 + \mathrm{d}\xi^2 + \mathrm{d}x^2 + \mathrm{d}y^2 . 
\label{eq:rindler-metric}
\end{align}
Noting that on the $t=0$ hypersurface we have $\mathrm{d}t = \xi \mathrm{d}\eta$ and $\mathrm{d}z = \mathrm{d}\xi$, we may read off $\braket{T_{\eta\eta}}^{(2)}$ and $\braket{T_{\eta\xi}}^{(2)}$ by matching with \eqref{eq:rindler-stress-energy-mink} at $t=0=\eta$ and then noting that the expressions hold for all $\eta$ by the invariance of the Rindler trajectory and the switching under the boost Killing vector $\partial_\eta = z\partial_t + t\partial_z$. We find 
\begin{widetext}
\begin{subequations}
\label{eq:rindler-coordinates-stress-energy}
\begin{align}
     \braket{T_{\eta\eta}(\mathsf{x})}^{(2)} & = \frac{a^6\xi^4}{2\pi^3{(a^2\xi^2 - 1)}^{4}}\Biggl[ \frac{2\pi}{1 - \mathrm{e}^{ 2\pi E/a }} + \frac{a}{E} 
+ 2a \int _{0}^{\infty} \mathrm{d}\tau \frac{ \sin(E\tau)}{{(a^2\xi^2)}^{-S}\mathrm{e}^{ a\tau } - 1 }  \Biggr]  \notag\\
        & \quad
+\frac{1}{2\pi^2{(a^2\xi^2 - 1)}^{2}} \frac{E^2}{1 - \mathrm{e}^{ 2\pi E/a }} 
+\frac{E a^4\xi^2}{\pi^3{|a^2\xi^2 - 1|}^{3}}\int_{0}^{\infty} \mathrm{d}\tau  \frac{\cos(E\tau)}{{(a^2\xi^2)}^{-S} \mathrm{e}^{ a\tau } - 1} , 
        \label{eq:rindler-T_(eta-eta)}
\\
\braket{T_{\eta\xi}(\mathsf{x})}^{(2)} & = \frac{S}{2\pi^2\xi{(a^2\xi^2 - 1)}^2}\frac{E^2}{1 - \mathrm{e}^{2\pi E/a}} . \label{eq:rindler-T_(eta-xi)}
\end{align}
\end{subequations}
\end{widetext}

Similarly, in the future wedge, $|z| < t$, convenient coordinates are a version of the Milne coordinates~\cite{Milne1932nature,Ellis1988}, which we denote by $(\sigma,\rho,x,y)$, related to the Minkowski coordinates by 
\begin{align}
t = \sigma \cosh\rho, 
\hspace{2ex}
z = \sigma \sinh\rho, 
\label{eq:Milnecoords-def}
\end{align}
where $\sigma>0$ and $-\infty<\rho<\infty$. The metric reads 
\begin{align}
ds^2 = - \mathrm{d}\sigma^2 + \sigma^2 \mathrm{d}\rho^2 + \mathrm{d}x^2 + \mathrm{d}y^2 . 
\end{align}
We may read off $\braket{T_{\sigma\sigma}}^{(2)}$ and $\braket{T_{\sigma\rho}}^{(2)}$ by matching with \eqref{eq:rindler-stress-energy-mink} at $z=0=\rho$, where $\mathrm{d}t = \mathrm{d}\sigma$ and  $\mathrm{d}z = \sigma\mathrm{d}\rho$,  
and noting that the expressions hold for all $\rho$ by invariance under the boost Killing vector $z\partial_t + t\partial_z = \partial_\rho$. We find 
\begin{widetext}
\begin{subequations}
\label{eq:milne-coordinates-stress-energy}
\begin{align}
    \braket{T_{\sigma\sigma}(\mathsf{x})}^{(2)} & = \frac{a^6 \sigma^2} {2\pi^3{(a^2\sigma^2 + 1)}^{4}}\Biggl[ \frac{2\pi}{1 - \mathrm{e}^{ 2\pi E/a }} + \frac{a}{E} 
- 2a  \int _{0}^{\infty} \mathrm{d}\tau \frac{ \sin( E\tau )}{{(a^2 \sigma^2)}^{-1} \mathrm{e}^{ a\tau } + 1 }  \Biggr]  \notag\\
        & \quad
+\frac{1}{2\pi^2\sigma^2{(a^2\sigma^2 + 1)}^{2}} \frac{E^{2}}{1 - \mathrm{e}^{ 2\pi E/a }} 
+\frac{E a^4 }{\pi^3 {(a^2 \sigma^2 + 1)}^{3}}\int_{0}^{\infty} \mathrm{d}\tau  \frac{\cos(E\tau)}{{(a^2 \sigma^2)}^{-1} \mathrm{e}^{ a\tau } + 1} ,  
\label{eq:rindler-T_(sigma-sigma)}
\\
\braket{T_{\sigma\rho}(\mathsf{x})}^{(2)} & = \frac{1}{2\pi^2\sigma{(a^2\sigma^2 + 1)}^2}\frac{E^{2}}{1 - \mathrm{e}^{2\pi E/a}} . \label{eq:rindler-T_(sigma-rho)}
\end{align}
\end{subequations}
\end{widetext}

\subsection{Rindler frame in the Rindler wedge}\label{subsec:rindler-observers}

We analyse the SET first in the Rindler wedge, in the Rindler coordinates $(\eta, \xi, x, y)$. 
We consider initially the plane of the detector's worldline, $x=y=0$, where $\braket{T_{\eta\eta}}^{(2)}$ and $\braket{T_{\eta\xi}}^{(2)}$ are given by~\eqref{eq:rindler-coordinates-stress-energy}, and $\braket{T_{\xi\xi}}^{(2)} = \xi^{-2}\braket{T_{\eta\eta}}^{(2)}$. For the energy flux, in Section~\ref{sec:rindlerframe-flux}, we consider also a generalisation off the plane of the detector's worldline.  

We recall that worldlines of constant $\xi$ are orbits of the boost Killing vector~$\partial_\eta$, and the detector's worldline has $\xi = 1/a$. As the Rindler coordinates are Fermi normal coordinates about the detector's trajectory~\cite{Fermi1922,Manasse1963}, observers on the worldlines of constant $\xi$ can be considered to share the frame of the detector. Note that the SET \eqref{eq:rindler-coordinates-stress-energy} seen by the observers of constant $\xi$ is independent of their coordinate time $\eta$, as a consequence of the boost invariance of the system.

\subsubsection{Energy density}

\begin{figure}
    \centering
    \includegraphics[width=\linewidth]{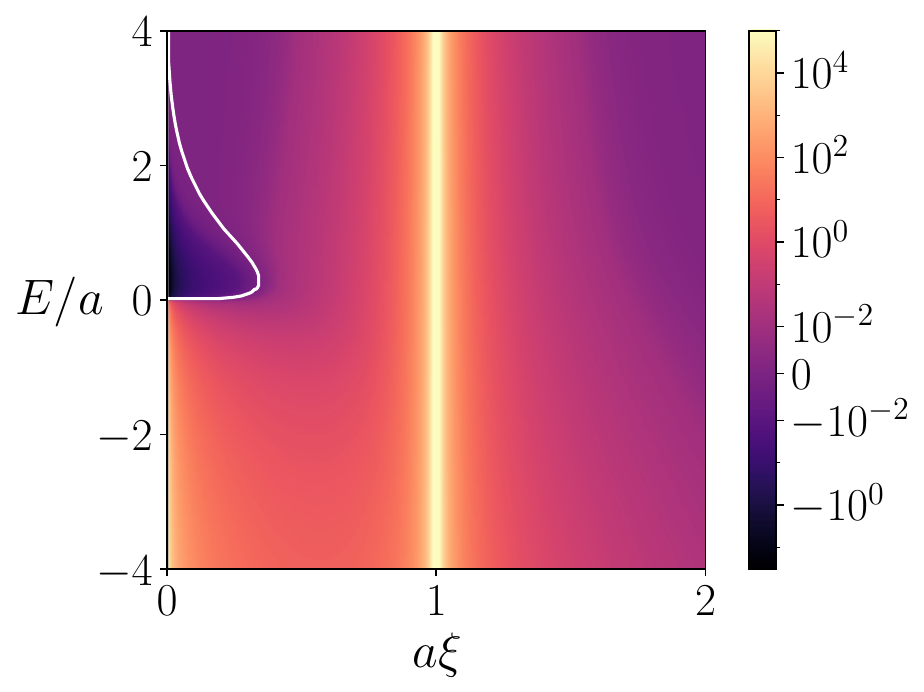}
    \caption{$a^{-4}\xi^{-2}\braket{T_{\eta\eta}}^{(2)}$ as a function of $a\xi$ and~$E/a$, obtained from~\eqref{eq:rindler-T_(eta-eta)}. At the detector's trajectory, $a\xi \to 1$, there is divergence to~$\infty$, with the asymptotics~\eqref{eq:Tetaeta-norm-trajectory-as}. At $\xi\to0$, there is divergence to $-\sgn(E)\infty$, with the asymptotics~\eqref{eq:Tetaeta-norm-horizon-as}. The curve where $\braket{T_{\eta\eta}}^{(2)}$ passes through zero is shown in white. At $a\xi\to\infty$, there is a $\xi^{-6}$ falloff~\eqref{eq:Tetaeta-norm-infty-as}, with a positive coefficient.}
    \label{fig:rindler-t_eta-eta}
\end{figure}

Figure \ref{fig:rindler-t_eta-eta} shows a plot of $\xi^{-2}\braket{T_{\eta\eta}}^{(2)}$ as a function of $a\xi$ and~$E/a$, evaluated from~\eqref{eq:rindler-T_(eta-eta)}. This is the energy density seen by the observers on the uniformly accelerated worldlines of constant~$\xi$. 

We note first that the formula for $\xi^{-2}\braket{T_{\eta\eta}}^{(2)}$, obtained from~\eqref{eq:rindler-T_(eta-eta)}, 
has significant structural similarity with the formula for the inertial detector trajectory $\braket{T_{tt}}^{(2)}$~\eqref{eq:inertial-T_(tt)}. 
In particular, the gap-dependent step functions in $\braket{T_{tt}}^{(2)}$ are replaced by Plankian factors in $\xi^{-2}\braket{T_{\eta\eta}}^{(2)}$. This is similar to what happens with the response functions of an UDW detector on the two trajectories~\cite{Birrell1982}. 
Also, the integral terms in $\braket{T_{tt}}^{(2)}$ and $\xi^{-2}\braket{T_{\eta\eta}}^{(2)}$ are highly similar, giving a contribution that is odd in $E$ with a $1/E$ falloff at large~$|E|$. It is these integral terms that most closely mirror the vacuum polarisation effects in quantum electrodynamics~\cite{Uehling1935,Schwinger1951}. 

Second, on approaching the Rindler detector's trajectory, $a\xi\to1$, $\xi^{-2}\braket{T_{\eta\eta}}^{(2)}$ diverges to positive infinity as 
\begin{align}
    \xi^{-2}\braket{T_{\eta\eta}}^{(2)} \sim \frac{1}{32\pi^2{(\xi - a^{-1})}^{4}} ,
    \label{eq:Tetaeta-norm-trajectory-as}
\end{align}
This is identical to the near-trajectory divergence for the inertial trajectory, given in~\eqref{eq:Ttt-small-r-as}, 
recalling that $|\xi-a^{-1}|$ is the proper distance to the detector's trajectory on a hypersurface of constant~$\eta$. The divergence is clearly visible both in Figure \ref{fig:rindler-t_eta-eta}
and in Figure~\ref{fig:static-energy-density}.

\begin{figure}
    \centering
    \includegraphics[width=\linewidth]{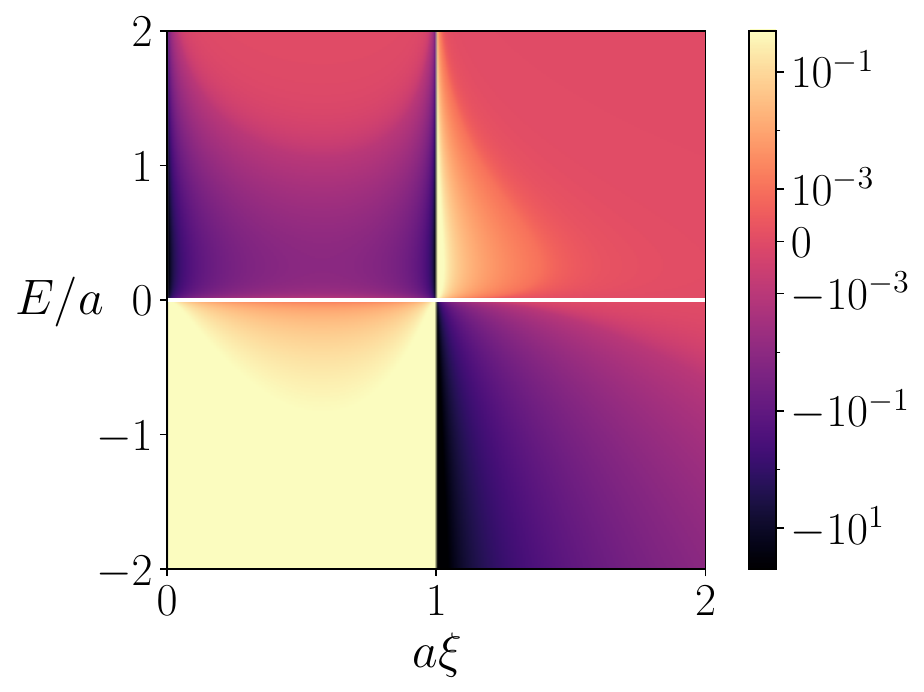}
    \caption{$a^{-4}\xi^{-1}\braket{T_{\eta\xi}}^{(2)}$ as a function of $a\xi$ and~$E/a$, obtained from~\eqref{eq:rindler-T_(eta-xi)}. There is an energy flux away from the detector for $E<0$ and towards the detector for $E>0$. The flux vanishes for $E=0$, shown in white. At the detector's trajectory, $a\xi \to 1$, there is a divergence to $\sgn(E) \sgn(a\xi-1)\infty$. At $\xi\to0$ there is a divergence proportional to $-\xi^{-2} \sgn(E)$, and at $\xi\to\infty$ there is a falloff proportional to $\xi^{-6}\sgn(E)$.}
    \label{fig:rindler-t_eta-xi}
\end{figure}

Third, at the Rindler infinity, $a\xi\to\infty$, the asymptotic behaviour of $\xi^{-2}\braket{T_{\eta\eta}}^{(2)}$ is 
\begin{align}
    \xi^{-2}\braket{T_{\eta\eta}}^{(2)} \sim \frac{1}{2\pi^2 a^2\xi^6} \left( \frac{2 + E^2/a^2}{1 - \mathrm{e}^{ 2\pi E/a }}+ \frac{a}{\pi E} \right) , 
    \label{eq:Tetaeta-norm-infty-as}
\end{align}
which has a $\xi^{-6}$ falloff whose coefficient is a positive strictly decreasing function of~$E$. 
This falloff is stronger than that in $\braket{T_{tt}}^{(2)}$, and the acceleration has softened the distinction between excitations and deexcitations that occurs in $\braket{T_{tt}}^{(2)}$, as is visible on comparing the large $r$ region in Figure \ref{fig:rindler-t_eta-eta}
and the large $\xi$ region in Figure~\ref{fig:static-energy-density}. 

Fourth, as $\xi\to0$, we have 
\begin{align}
    \xi^{-2}\braket{T_{\eta\eta}}^{(2)} \sim \frac{1}{2\pi^2 \xi^2}\frac{E^{2}}{1 - \mathrm{e}^{ 2\pi E/a }} , 
    \label{eq:Tetaeta-norm-horizon-as}
\end{align}
which diverges to positive infinity for $E<0$ and negative infinity for $E>0$. When $E>0$, 
$\xi^{-2}\braket{T_{\eta\eta}}^{(2)}$
is hence negative for sufficiently small~$\xi$.
This region of negative Rindler energy density is shown in Figure~\ref{fig:rindler-t_eta-eta}: 
it shrinks as $E$ increases to large positive values, as one may expect from the exponential suppression in the coefficient in \eqref{eq:Tetaeta-norm-horizon-as} as $E\to\infty$. We shall comment further on this region when we investigate the perspectives of Milne and Minkowski observers below.

Fifth, consider large positive and negative values of the gap. As $E\to\infty$, repeated integration by parts in \eqref{eq:rindler-T_(eta-eta)} gives~\cite{Wong2001}
\begin{align}
    \xi^{-2}\braket{T_{\eta\eta}}^{(2)} \sim \frac{a^5}{2\pi^3E}\frac{a^2\xi^2(3 + a^2\xi^2)}{{|a^2\xi^2 - 1|}^5} ,  
    \label{eq:Tetaeta-norm-Etoinfty}
\end{align}
showing a $1/E$ suppression with a positive coefficient.
As $E\to-\infty$, we have
\begin{align}
    \xi^{-2}\braket{T_{\eta\eta}}^{(2)} \sim \frac{E^{2}}{2\pi^2 \xi^2{(a^2\xi^2 - 1)}^{2}} , 
    \label{eq:Tetaeta-norm-Etominusinfty}
\end{align}
showing an $E^2$ increase. 

Sixth, note that the $\xi\to\infty$ asymptotic form of the $\xi$-dependent coefficients in 
\eqref{eq:Tetaeta-norm-Etoinfty}
and 
\eqref{eq:Tetaeta-norm-Etominusinfty} gives for $\xi^{-2}\braket{T_{\eta\eta}}^{(2)}$ a result identical to that from the  $E\to\pm\infty$ asymptotic form of the $E$-dependent coefficient in~\eqref{eq:Tetaeta-norm-infty-as}, showing that the $\xi\to\infty$ and $E\to\pm\infty$ asymptotics commute. 

Seventh, on approaching the detector's trajectory, $a\xi\to1$, the expressions in 
\eqref{eq:Tetaeta-norm-Etoinfty}
and 
\eqref{eq:Tetaeta-norm-Etominusinfty} have the asymptotic form ${(16 \pi^3 E )}^{-1}\bigl|\xi - a^{-1}\bigr|^{-5}$ and
${{(8 \pi^2)}^{-1}E^2\bigl(\xi - a^{-1}\bigr)^{-2}}$, respectively. This is identical to the $E\to\pm\infty$ results found for 
$\braket{T_{tt}}^{(2)}$ in Section \ref{subsec:inertial-energy-density} on identifying $|\xi - a^{-1}|$ with~$r$.

\subsubsection{Energy flux} 
\label{sec:rindlerframe-flux}

Figure \ref{fig:rindler-t_eta-xi} shows a plot of $\xi^{-1}\braket{T_{\eta\xi}}^{(2)}$, 
in the plane of the detector's wordline, evaluated from~\eqref{eq:rindler-T_(eta-xi)}. 
The energy flux in the direction of increasing $\xi$ seen by an observer at constant $\xi$ is equal to $-\xi^{-1}\braket{T_{\eta\xi}}^{(2)}$
(cf.~the discussion in Section \ref{subsec:inertial-energy-flux} for the inertial detector trajectory). 

For de-excitations, $E<0$, the direction of the flux is away from the detector, as for the inertial trajectory. For excitations, $E>0$, there is a nonvanishing flux, and the flux is towards the detector: this is a key difference from the inertial trajectory, and it is seen clearly in Figure~\ref{fig:rindler-t_eta-xi}.  
We observe from \eqref{eq:rindler-T_(eta-xi)}
that  $\xi^{-1}\braket{T_{\eta\xi}}^{(2)}$ falls off as $\xi^{-6}$ when $\xi\to\infty$ and diverges as $\xi^{-2}$ when $\xi\to0$. 

The near-trajectory asymptotic form of $\xi^{-1}\braket{T_{\eta\xi}}^{(2)}$ is 
\begin{align}
\xi^{-1} \braket{T_{\eta\xi}}^{(2)} &\sim - \frac{\sgn(a\xi-1)}{8\pi^2{|\xi - a^{-1}|}^2}\frac{E^2}{1 - \mathrm{e}^{2\pi E/a}} 
\notag\\
&= 
- \frac{\sgn(a\xi-1)}{4\pi{|\xi - a^{-1}|}^2} \times 
\frac{E}{2\pi \bigl( 1 - \mathrm{e}^{2\pi E/a} \bigr)} 
\times E . 
\label{eq:rindlercoords-energyflux-neartraj}
\end{align}
We show in Appendix 
\ref{appsec:rindlerflux-transverse} that in a neighbourhood of the detector's trajectory in the spacetime, with $x$ and $y$ not necessarily vanishing, \eqref{eq:rindlercoords-energyflux-neartraj} generalises to the near-trajectory asymptotic form 
\begin{align}
\xi^{-1} \braket{T_{\eta b}}^{(2)} 
&\sim  
- \frac{X_b}{4\pi X^3} \times 
\frac{E}{2\pi \bigl( 1 - \mathrm{e}^{2\pi E/a} \bigr)} 
\times E , 
\label{eq:rindlercoords-energyflux-neighbourhood}
\end{align}
where $b \in \{\xi,x,y \}$ and 
\begin{subequations}
\begin{align}
X_\xi &= \xi-a^{-1} 
, 
\ \
X_x = x , 
\ \
X_y = y, 
\\
X &= \sqrt{X_\xi^2 + X_x^2 + X_y^2} . 
\end{align}
\end{subequations}
We recognise $X$ as the proper distance to the detector on a hypersurface of constant~$\eta$, 
and we note that the three-vector 
$(4\pi)^{-1} X_b/X^3$ that appears in \eqref{eq:rindlercoords-energyflux-neighbourhood} is spatially isotropic and has unit flux through a two-sphere around the detector on a hypersurface of constant~$\eta$. 
The spatial isotropy might have been expected by the isotropy of the response of a direction-sensitive detector~\cite{Grove1985}. 
We also recognise the remaining factors in \eqref{eq:rindlercoords-energyflux-neighbourhood} as the detector's transition rate $(2\pi)^{-1} E / \bigl( 1 - \mathrm{e}^{2\pi E/a} \bigr)$ and the detector's energy gap~$E$. 

These observations show that the energy gained (respectively lost) by the detector in its excitations (de-excitations) exactly matches the flux of energy from the field to the detector (from the detector to the field), just as we saw in Section \ref{subsec:inertial-energy-flux}
to happen with the detector's de-excitations for the inertial trajectory. 
While the conventional wisdom is that the energy responsible for the detector's excitations due to the Unruh effect comes ultimately from the external agent who keeps the detector in acceleration~\cite{Unruh1984}, the match shows that from the Rindler frame perspective the detector's energy gains and losses can be attributed in full to a flow of Rindler energy from and to the field. 

\begin{figure}
    \centering
    \includegraphics[width=\linewidth]{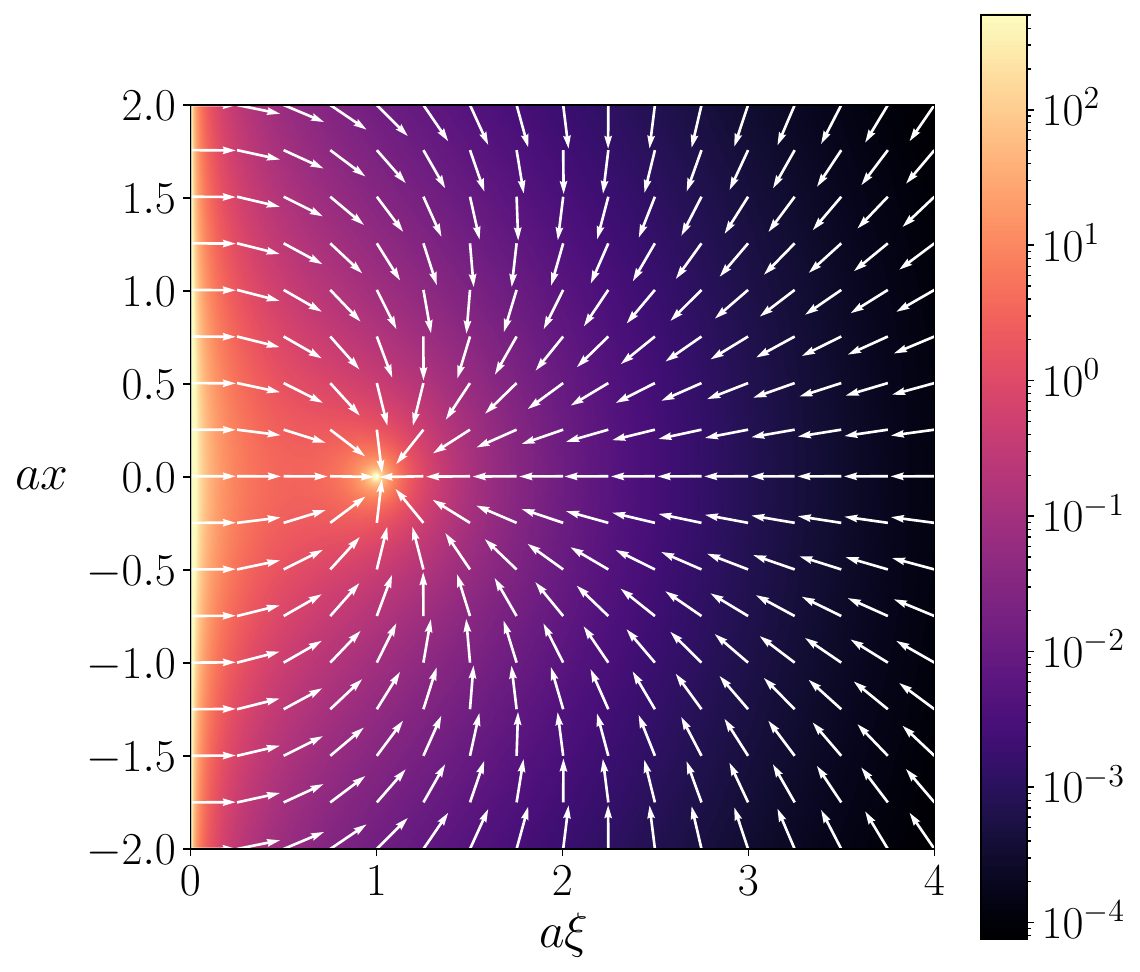}
    \caption{$a^{-4}\xi^{-1} \braket{T_{\eta b}}^{(2)}$ in the plane $y=0$ as a function of $a\xi$ and~$ax$, 
    for $E>0$, 
    evaluated from formula \eqref{eq:app-fluxvector} in Appendix~\ref{app:accelerated-emissions}. 
    The near-trajectory asymptotic form is shown in~\eqref{eq:rindlercoords-energyflux-neighbourhood}. The colour encodes the magnitude, in units of ${(2\pi)}^{-1} (E/a)^2 {\bigl| 1 - \mathrm{e}^{2\pi E/a} \bigr|}^{-1}$, which is $E/a^2$ times the detector's transition rate. The equal-length arrows indicate the direction of the energy flux, $-\xi^{-1} \braket{T_{\eta b}}^{(2)}$. For $E<0$, the plot is the same except that the arrows are reversed. The detector is the bright dot at $a\xi = 1$, $ax = 0$. 
    Note the flow of energy from the Rindler horizon ($a\xi=0$) to the detector for $E>0$, and from the detector to the Rindler horizon for $E<0$.}
    \label{fig:rindler-flux-direction}
\end{figure}

Fig.~\ref{fig:rindler-flux-direction} shows a plot of $\xi^{-1} \braket{T_{\eta b}}^{(2)}$ in the plane $y=0$ as a function of $a\xi$ and~$ax$, 
evaluated from formula \eqref{eq:app-fluxvector} in Appendix~\ref{app:accelerated-emissions}, which generalises the near-trajectory formula \eqref{eq:rindlercoords-energyflux-neighbourhood} to general $(X_\xi, X_x, X_y)$. By the rotational symmetry in the transverse directions $(x,y)$, the choice of the plane $y=0$ has no loss of generality. As seen from~\eqref{eq:app-fluxvector}, the $E$-dependence of $\xi^{-1} \braket{T_{\eta b}}^{(2)}$ is again in the multiplicative factor $(2\pi)^{-1} E^2 / \bigl( 1 - \mathrm{e}^{2\pi E/a} \bigr)$, 
which is the product of the detector's transition rate $(2\pi)^{-1} E / \bigl( 1 - \mathrm{e}^{2\pi E/a} \bigr)$ and the detector's energy gap~$E$, and hence has different signs for $E>0$ and $E<0$, and vanishes for a gapless detector, $E=0$. 
The case $E>0$, shown in Fig.~\ref{fig:rindler-flux-direction}, has energy flowing from the Rindler horizon to the detector; conversely, the case $E<0$ has energy flowing from the detector to the Rindler horizon. In this sense, when viewed in the Rindler frame, the energy gained by the detector in its excitations originates from the Rindler horizon, and the energy lost by the detector in its de-excitations is absorbed by the Rindler horizon.

\subsection{Milne frame in the future wedge}\label{subsec:milne-observers}

We next consider the SET in the future wedge, in the Milne coordinates $(\sigma, \rho, x, y)$. We consider only the plane 
$x=y=0$, where $\braket{T_{\sigma\sigma}}^{(2)}$ and $\braket{T_{\sigma\rho}}^{(2)}$ are given by~\eqref{eq:milne-coordinates-stress-energy}, and $\braket{T_{\rho\rho}}^{(2)} = \sigma^2\braket{T_{\sigma\sigma}}^{(2)}$. 

The boost Killing vector $\partial_\rho$ is now spacelike, 
and the boost invariance of the system now has the consequence that $\braket{T_{\sigma\sigma}}^{(2)}$ and $\braket{T_{\sigma\rho}}^{(2)}$ are independent of~$\rho$. 
Observers on the inertial worldlines of constant $\rho$ are known as Milne observers~\cite{Milne1932nature,Ellis1988}, and they all originate from the initial coordinate singularity at $t=z=0$, where $\sigma=0$. 
The energy density seen by the Milne observers is $\braket{T_{\sigma\sigma}}^{(2)}$, 
and the energy flux towards increasing $\rho$ seen by them is $- \sigma^{-1}\braket{T_{\sigma\rho}}^{(2)}$.

By contrast to the Rindler observers, for whom the boost symmetry makes the SET invariant under time evolution, 
the boost symmetry in the future wedge maps the Milne observers onto each other. All the Milne observers hence see the same SET, but they see the SET evolve in their proper time~$\sigma$. 

\begin{figure}
    \centering
    \includegraphics[width=\linewidth]{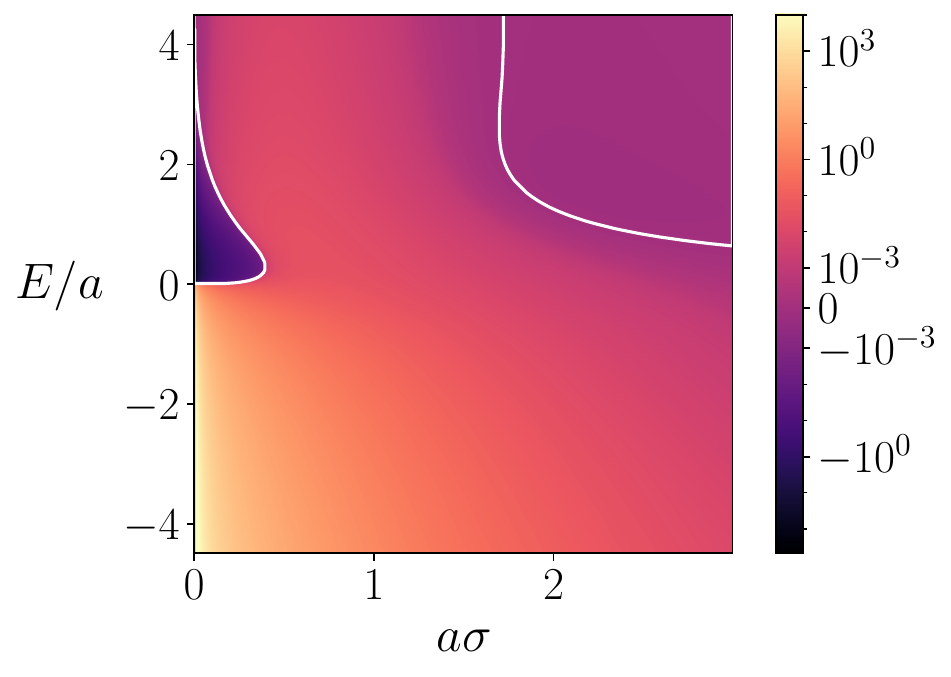}
    \caption{$a^{-4}\braket{T_{\sigma\sigma}}^{(2)}$ \eqref{eq:rindler-T_(sigma-sigma)}, as a function of $a\sigma$ and~$E/a$. Note that in the spacetime $\sigma$ increases to the future, whereas in the plot $\sigma$ increases to the right. At $\sigma\to0$, there is divergence to $-\sgn(E)\infty$, with the asymptotics shown in~\eqref{eq:Tsigmasigma-sigmasmall-as}. The large positive and negative $E$ asymptotics are as shown in \eqref{eq:Tsigmasigma-Elargepos-as} and~\eqref{eq:Tsigmasigma-Elargeneg-as}. The curves where $\braket{T_{\sigma\sigma}}^{(2)}$ passes through zero are shown in white. }
    \label{fig:rindler-t_sigma-sigma}
\end{figure}

\begin{figure}
    \centering
    \includegraphics[width=\linewidth]{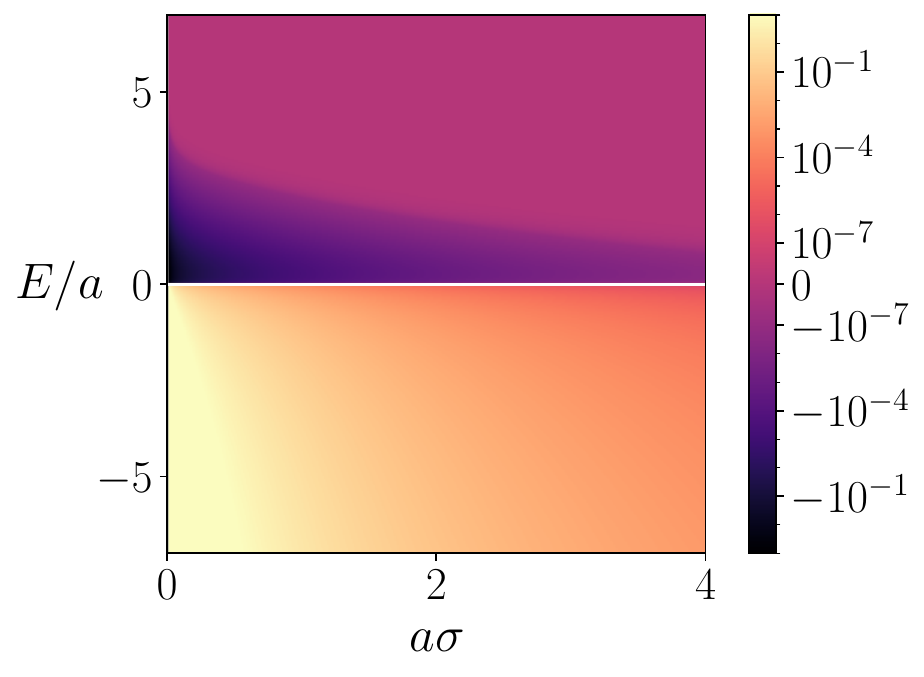}
    \caption{$a^{-4}\sigma^{-1}\braket{T_{\sigma\rho}}^{(2)}$ as a function of $a\sigma$ and~$E/a$, obtained from~\eqref{eq:rindler-T_(sigma-rho)}. There is a leftward flux, away from the detector, for $E<0$, and a rightward flux, towards the detector, for $E>0$. The flux vanishes for $E=0$, shown in white. At $\sigma\to0$ there is a divergence proportional to $-\sigma^{-2} \sgn(E)$, and at $\sigma\to\infty$ there is a falloff proportional to $-\sigma^{-6}\sgn(E)$.}
    \label{fig:rindler-t_sigma-rho}
\end{figure}

\subsubsection{Energy density}

Figure \ref{fig:rindler-t_sigma-sigma} shows a plot of 
$\braket{T_{\sigma\sigma}}^{(2)}$ \eqref{eq:rindler-T_(sigma-sigma)} as a function of $a\sigma$ and $E/a$. This is the energy density seen by the Milne observers. 

At the Milne infinity $a\sigma\to\infty$, in the far future, the leading term in the falloff of $\braket{T_{\sigma\sigma}}^{(2)}$ consists of terms proportional to $\sigma^{-6}$ and to the real and imaginary parts of $\sigma^{-6} {(a\sigma)}^{2iE/a}$, with $E$-dependent coefficients, as can be verified by writing the integrals in \eqref{eq:rindler-T_(sigma-sigma)} in terms of the Hurwitz-Lerch zeta function $\Phi(-a^2\sigma^2,1,1 - iE/a)$ and using Theorem 3 in~\cite{Ferreira2004}. 

At the Milne initial coordinate singularity $\sigma\to0$, we have 
\begin{align}
    \braket{T_{\sigma\sigma}}^{(2)}\sim\frac{1}{2\pi^2\sigma^2} \frac{E^{2}}{1 - \mathrm{e}^{ 2\pi E/a }} , 
    \label{eq:Tsigmasigma-sigmasmall-as}
\end{align}
which diverges to positive infinity for $E<0$ and negative infinity for $E>0$. When $E>0$, there is hence a negative energy density for sufficiently small~$\sigma$, similarly to what we saw in the Rindler wedge at small~$\xi$. This region of negative Milne energy density is visible in Figure~\ref{fig:rindler-t_sigma-sigma}. 

As $E\to\infty$, repeated integration by parts in \eqref{eq:rindler-T_(sigma-sigma)} gives~\cite{Wong2001}
\begin{align}
    \braket{T_{\sigma\sigma}}^{(2)}\sim\frac{a^5}{2\pi^3E}\frac{a^2\sigma^2(3 - a^2\sigma^2)}{{(a^2\sigma^2 + 1)}^5} ,
    \label{eq:Tsigmasigma-Elargepos-as}
\end{align}
mirroring the $1/E$ suppression in~\eqref{eq:Tetaeta-norm-Etoinfty}, but with a coefficient that is negative for $a\sigma > \sqrt{3}$. This region of negative energy density for large $\sigma$ and $E$ is visible in Figure~\ref{fig:rindler-t_sigma-sigma}, including the asymptotic behaviour where the zero-crossing approaches $a\sigma = \sqrt{3}$ for large positive~$E$. While this region of negative energy density is not compact, the $\sigma^{-6}$ suppression in the coefficient of $1/E$ in \eqref{eq:Tsigmasigma-Elargepos-as} indicates that the energy density does not take large negative values. Numerically, we find that the value of $\braket{T_{\sigma\sigma}}^{(2)}$ in this region is not below $\sim -10^{-5}\,a^{4}$ in the parameter ranges that we have investigated. 
This suggests that the negative energy density region may not violate quantum energy inequalities~\cite{Fewster2012}. We will comment further on this region in Section \ref{subsec:minkowski-observers} where we investigate Minkowski observers.

As $E\to-\infty$, 
\begin{align}
    \braket{T_{\sigma\sigma}}^{(2)}\sim\frac{E^{2}}{2\pi^2\sigma^2{(a^2\sigma^2 + 1)}^{2}} ,
    \label{eq:Tsigmasigma-Elargeneg-as}
\end{align}
which increases proportionally to~$E^2$, 
like the Rindler frame energy density~\eqref{eq:Tetaeta-norm-Etominusinfty}. Note that the coefficient in \eqref{eq:Tsigmasigma-Elargeneg-as} is positive and falls off at a rate of $\sigma^{-6}$ for large~$\sigma$. These properties are visible in  Figure~\ref{fig:rindler-t_sigma-sigma}.

\subsubsection{Energy flux}


Figure \ref{fig:rindler-t_sigma-rho}  shows a plot of $\sigma^{-1}\braket{T_{\sigma\rho}}^{(2)}$, evaluated from~\eqref{eq:rindler-T_(sigma-rho)}. 
The energy flux in the direction of increasing $\rho$ seen by an observer at constant $\rho$ is equal to $-\sigma^{-1}\braket{T_{\sigma\rho}}^{(2)}$. 

For de-excitation gaps the flux is left-moving, away from the detector, and for excitation gaps the flux is right-moving, towards the detector. This is consistent with the direction of the flux seen by the Rindler observers, found in Section~\ref{subsec:rindler-observers}\null. 
We further recognise from \eqref{eq:rindler-T_(sigma-rho)} that the $E$-dependence of the flux is the product of the detector's transition rate $(2\pi)^{-1} E / \bigl( 1 - \mathrm{e}^{2\pi E/a} \bigr)$ and the detector's energy gap~$E$, as for the Rindler observers. 

The $\sigma^{-6}$ falloff in the far future agrees with the falloff of the energy density, but it does not involve the oscillatory factors that appear in the energy density. At the Milne initial singularity, $\sigma\to0$, there is a divergence proportional to~$\sigma^{-2}$.

\subsection{Minkowski frame}\label{subsec:minkowski-observers}

We now consider the SET in all of the causal future of the detectors worldline, $t+z>0$, in the Minkowski coordinates $(t,x,y,z)$. We again specialise to the plane $x=y=0$, where $\braket{T_{tt}}^{(2)}$ and $\braket{T_{tz}}^{(2)}$ are given by~\eqref{eq:rindler-stress-energy-mink}, and $\braket{T_{zz}}^{(2)} = \braket{T_{tt}}^{(2)}$. 

$\braket{T_{tt}}^{(2)}$ and $\braket{T_{tz}}^{(2)}$ are respectively the energy density and the negative of the energy flux seen by static Minkowski observers, whose worldlines are integral curves of~$\partial_t$.
These observers do not follow a trajectory adapted to the boost symmetry of our system, as the Rindler and Milne observers do, and the energy density and flux seen by them have hence a more complicated dependence on the position in the spacetime. This family of observers provides the perspective of an inertial frame in which typical experimental apparatus measuring the field's stress-energy might be operating. Note that this family of observers exists in the whole region $t+z>0$, and the family hence ties together the Rindler and Milne analyses of Sections \ref{subsec:rindler-observers}
and~\ref{subsec:milne-observers}\null. We shall in particular see that $\braket{T_{tt}}^{(2)}$ and $\braket{T_{tz}}^{(2)}$ are regular across the future Rindler horizon, $t=z>0$, where neither the Rindler nor the Milne coordinates are well defined.

\subsubsection{Energy density}

\begin{figure}
    \centering
    \includegraphics[width=\linewidth]{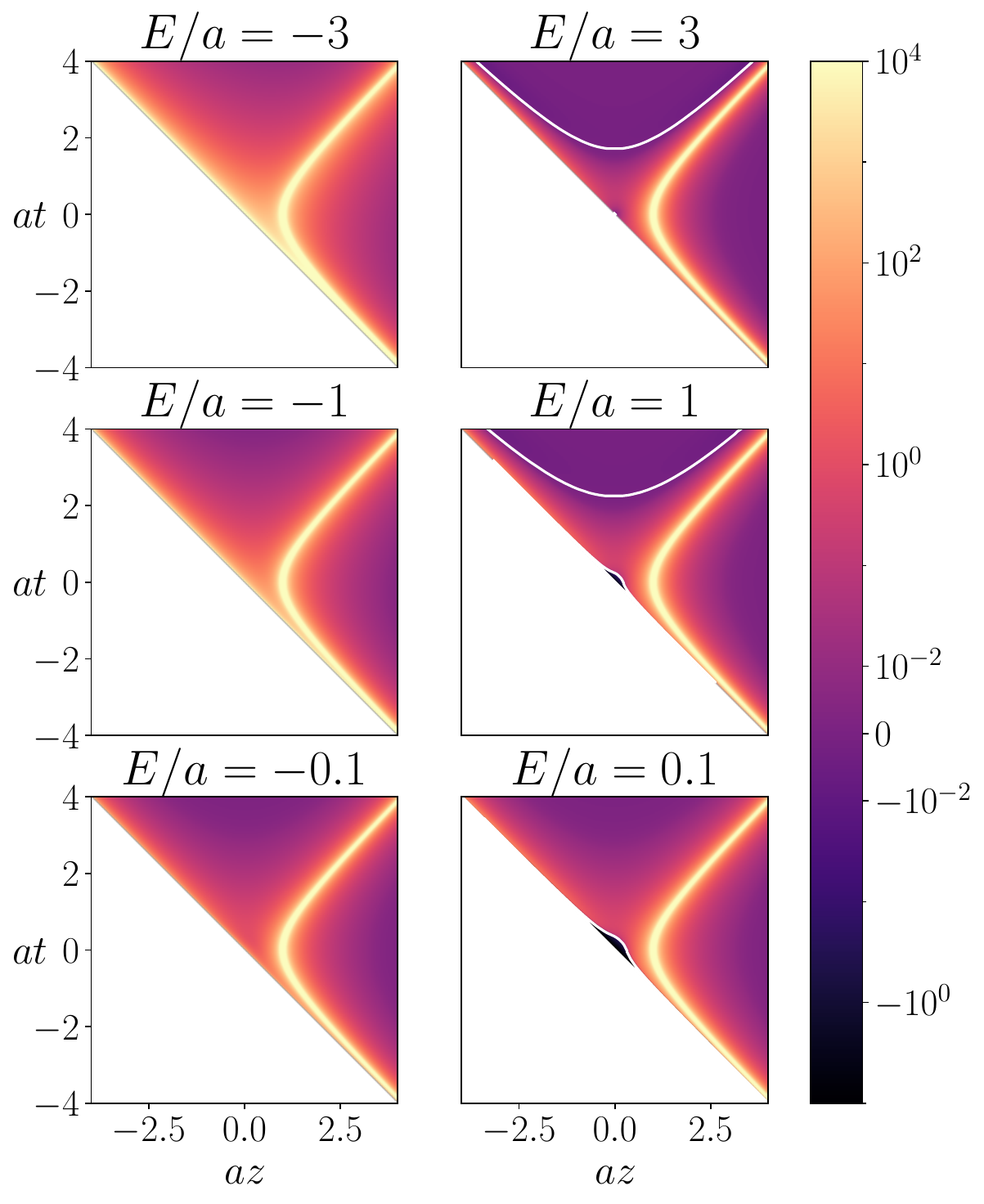}
    \caption{Spacetime plots of $a^{-4} \braket{T_{tt}}^{(2)}$ \eqref{eq:rindler-T_(tt)}, for selected values of~$E/a$. There is a divergence to $\infty$ at the detector's trajectory, with the asymptotics shown in~\eqref{eq:Ttt-neartrajectory-asymp}. The divergence to $-\sgn(E)\infty$ as $t+z\to0_+$, with the asymptotics shown in~\eqref{eq:Ttt-smallv-asymp}, is borderline visible within the scale of the plots. 
    Curves where $\braket{T_{tt}}^{(2)}$ passes through zero are shown in white.}
    \label{fig:rindler-t_tt}
\end{figure}

Fig.~\ref{fig:rindler-t_tt} shows spacetime plots of the Minkowski energy density $\braket{T_{tt}}^{(2)}$ \eqref{eq:rindler-T_(tt)}, for selected values of~$E/a$. 

Close to the detector's worldline, 
$\braket{T_{tt}}^{(2)}$ is asymptotic to 
\begin{align}
\braket{T_{tt}}^{(2)}
& \sim 
\frac{a^2 \bigl(v^{2} + u^{2}\bigr)}{64\pi^2{\left(\sqrt{-uv} - a^{-1}\right)}^{4}} , 
    \label{eq:Ttt-neartrajectory-asymp}
\end{align}
where we recall that the detector's worldline is at $uv = -a^{-2}$ with $u<0$ and $v>0$, 
we have used 3.911.2 in \cite{Gradshteyn2007} in the first integral in~\eqref{eq:rindler-T_(tt)}, and we have noted that the second integral in \eqref{eq:rindler-T_(tt)} diverges logarithmically and hence gives only a subleading contribution. 
$\braket{T_{tt}}^{(2)}$ hence diverges at the detector's wordline, as is clearly visible in the plots of Figure~\ref{fig:rindler-t_tt}. 
On approaching the wordline in any spacelike hyperplane, \eqref{eq:Ttt-neartrajectory-asymp} shows that the divergence is proportional to the power $-4$ of the proper distance to the worldline, in agreement with the divergence seen in \eqref{eq:Tetaeta-norm-trajectory-as}
in the Rindler frame. In particular, in the hyperplane $\eta=0$ we have $-u = v = z$, $\braket{T_{tt}}^{(2)} = \xi^{-2}\braket{T_{\eta\eta}}^{(2)}$, and \eqref{eq:Ttt-neartrajectory-asymp} reduces exactly to~\eqref{eq:Tetaeta-norm-trajectory-as}. 
Note that the divergence is again independent of the energy gap, reinforcing the evidence that the divergence is geometric and originates from the point-like nature of the detector.

\begin{figure}
    \centering
    \includegraphics[width=\linewidth]{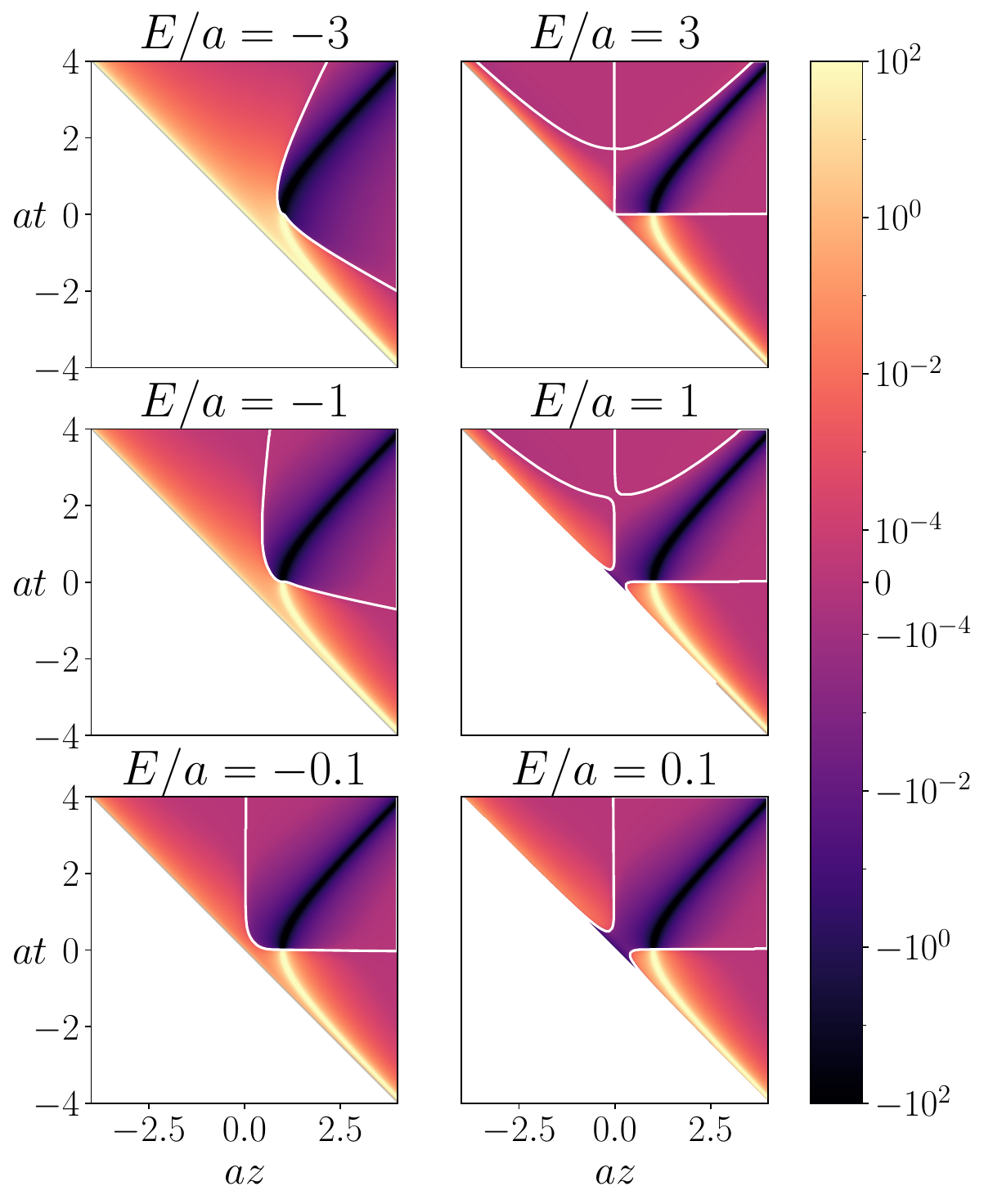}
    \caption{Spacetime plots of $a^{-4}\braket{T_{tz}}^{(2)}$ \eqref{eq:rindler-T_(tz)}, for selected values of~$E/a$. There is a divergence to $-\sgn(t)\infty$ at the detector's trajectory, with the asymptotics shown in~\eqref{eq:Ttz-neartrajectory-asymp}. The divergence to $-\sgn(E)\infty$ as $t+z\to0_+$, with the asymptotics shown in~\eqref{eq:Ttz-smallv-asymp}, is not clearly visible within the scale of the plots. 
    Curves where $\braket{T_{tz}}^{(2)}$ passes through zero are shown in white. Positive (respectively negative) values of $\braket{T_{tz}}^{(2)}$ correspond to a leftward (rightward) energy flux.}
    \label{fig:rindler-t_tz}
\end{figure}

$\braket{T_{tt}}^{(2)}$ also diverges as $v\to0$, at the boundary of the causal future of the detectors worldline. Taking $v\to0$ with constant~$u$, the asymptotic behaviour is 
\begin{align}
    \braket{T_{tt}}^{(2)}\sim\frac{1}{2\pi^2 v^2} \frac{E^{2}}{1 - \mathrm{e}^{ 2\pi E/a }} , 
    \label{eq:Ttt-smallv-asymp}
\end{align}
which diverges to $-\sgn(E)\infty$. 
The divergence is borderline visible in Figure~\ref{fig:rindler-t_tt}, within the scale of the plots. 
This divergence is expected, 
since the detector has been accelerating and interacting since the asymptotic past. Similar divergences exist for an accelerating point charge that interacts with the electromagnetic field~\cite{Boulware1980, Garfinkle:2019vej}.

Note that the small $v$ asymptotics of 
$\braket{T_{tt}}^{(2)}$ \eqref{eq:Ttt-smallv-asymp} is closely similar to the small $\xi$ asymptotics of 
$\xi^{-2}\braket{T_{\eta\eta}}^{(2)}$ \eqref{eq:Tetaeta-norm-horizon-as} 
and the small 
$\sigma$ asymptotics of 
$\braket{T_{\sigma\sigma}}^{(2)}$~\eqref{eq:Tsigmasigma-sigmasmall-as}. In particular, for $E>0$, all three are negative, as is visible 
for 
$\xi^{-2}\braket{T_{\eta\eta}}^{(2)}$ and $\braket{T_{\sigma\sigma}}^{(2)}$
in Figures \ref{fig:rindler-t_eta-eta} and~\ref{fig:rindler-t_sigma-sigma}, 
and for $\braket{T_{tt}}^{(2)}$
in Figure~\ref{fig:rindler-t_tt}, 
right-hand column, most clearly in the $E/a = 0.1$ panel. 
In each case, the region of negative energy density becomes smaller as $E$ increases. 

At large positive and negative~$E$, $\braket{T_{tt}}^{(2)}$ has the asymptotics 
\begin{subequations}
\begin{align}
    \braket{T_{tt}}^{(2)} &\sim \frac{u^2 + v^2}{4\pi^3aE}\frac{3a^{-2}-uv}{{|uv + a^{-2}|}^5}  \ \ \text{as}\ E\to\infty, 
    \label{eq:rindler-T_tt-largeposE-as}
\\
    \braket{T_{tt}}^{(2)} &\sim \frac{E^2}{2\pi^2a^4\zeta^2{(uv + a^{-2})}^2} \ \ \text{as}\ E\to-\infty, 
    \label{eq:rindler-T_tt-largenegE-as}
\end{align}
\end{subequations}
as can be verified from~\eqref{eq:rindler-T_(tt)}, by repeated integration by parts~\cite{Wong2001}. 
The $E^2$ growth for $E<0$ and the $1/E$ suppression for $E>0$ are consistent with what we saw in the Rindler and Milne frames. 

As $E\to\infty$, \eqref{eq:rindler-T_tt-largeposE-as} shows that $\braket{T_{tt}}^{(2)}$ is negative to the future of the hyperbola $uv = 3 a^{-2}$: for finite positive~$E$, the right-hand-side panels in Figure \ref{fig:rindler-t_tt} show a sign change already when $E/a = 3$ and $E/a=1$. This sign change is similar to that in $\braket{T_{\sigma\sigma}}^{(2)}$ at large positive~$E$, seen in Figure~\ref{fig:rindler-t_sigma-sigma}. 
As $E\to\infty$, the sign change in $\braket{T_{\sigma\sigma}}^{(2)}$ and 
$\braket{T_{tt}}^{(2)}$ occurs on the same hyperbola, $uv = 3 a^{-2}$, as seen from \eqref{eq:Tsigmasigma-Elargepos-as} and~\eqref{eq:Milnecoords-def}. 

As in the Milne case, we find that $\braket{T_{tt}}^{(2)}$ does not take large negative values in the numerical ranges investigated, and we note that the coefficient of $1/E$ in \eqref{eq:rindler-T_tt-largeposE-as} has $t^{-6}$ suppression as $t\to\infty$ with fixed~$z$. This suggests that the region of negative $\braket{T_{tt}}^{(2)}$ may not violate quantum energy inequalities~\cite{Fewster2012}.

\subsubsection{Energy flux}

Figure \ref{fig:rindler-t_tz} shows spacetime plots of $\braket{T_{tz}}^{(2)}$~\eqref{eq:rindler-T_(tz)}, for selected values of $E/a$. 
The energy flux in the direction of increasing $z$ as seen by a static Minkowski observer is $-\braket{T_{tz}}^{(2)}$. 

A striking feature in Figure \ref{fig:rindler-t_tz} is the qualitative difference between the $t<0$ half and the $t>0$ half of the Rindler quadrant, especially near the detector's trajectory, and the qualitative difference between the $z<0$ half and the $z>0$ half in the future quadrant. The reason for these differences is that the
Lorentz boost that relates the Minkowski flux to the Rindler and Milne frame SETs depends on the spacetime position. 
In the Rindler quadrant, the Rindler energy density contributes to $\braket{T_{tz}}^{(2)}$ with a coefficient that is positive for $t<0$ and negative for $t>0$, and this contribution dominates near the detector's trajectory, as seen from \eqref{eq:Tetaeta-norm-trajectory-as}
and~\eqref{eq:rindlercoords-energyflux-neartraj}. This leads to the large positive (respectively negative) near-trajectory values of $\braket{T_{tz}}^{(2)}$ seen for $t<0$ ($t>0$) in Figure~\ref{fig:rindler-t_tz}. 
Similarly, in the future quadrant, the Milne energy density contributes to $\braket{T_{tz}}^{(2)}$ with a coefficient that has different signs for $z<0$ and $z>0$. 

Close to the detector's worldline, the asymptotic form of 
$\braket{T_{tz}}^{(2)}$ is
\begin{align}
\braket{T_{tz}}^{(2)}\sim\frac{a^2 \bigl(u^2 - v^2\bigr)}{64\pi^2 {\left(\sqrt{-uv} - a^{-1}\right)}^{4}}, 
    \label{eq:Ttz-neartrajectory-asymp}
\end{align}
showing the divergence to $\infty$ for $u^2-v^2>0$ ($t<0$) and to $-\infty$ for $u^2-v^2<0$ ($t>0$). 
As discussed above, this asymptotic form comes from the Rindler frame energy density \eqref{eq:Tetaeta-norm-trajectory-as} by a Lorentz boost. When a static observer sees the detector move past, the observer hence sees a large energy flux in the direction of the detector's motion. 

At the boundary of the causal future of the detector's wordline, taking $v\to0$ with fixed~$u$, we have 
\begin{align}
    \braket{T_{tz}}^{(2)}\sim\frac{1}{2\pi^2 v^2}\frac{E^2}{1 - \mathrm{e}^{2\pi E/a}} , 
    \label{eq:Ttz-smallv-asymp}
\end{align}
which diverges to $-\sgn(E)\infty$. This divergence is not clearly visible in Figure \ref{fig:rindler-t_tz} within the scale of the plots. Note the similarity between the asymptotics in~\eqref{eq:Ttz-smallv-asymp}, the $\xi\to0$ asymptotics of $\xi^{-1}\braket{T_{\eta\xi}}^{(2)}$ found from~\eqref{eq:rindler-T_(eta-xi)}, and the 
$\sigma\to0$ asymptotics of $\sigma^{-1}\braket{T_{\sigma\rho}}^{(2)}$ found from~\eqref{eq:rindler-T_(sigma-rho)}. 

For large de-excitation gaps, $E\to-\infty$, we see from \eqref{eq:rindler-T_(tz)} that the asymptotic form of 
$\braket{T_{tz}}^{(2)}$ is
\begin{align}
    \braket{T_{tz}}^{(2)}\sim\
    \frac{S E^2}{2\pi^2a^{4}{\zeta^2(uv + a^{-2})}^2}, 
    \label{eq:Ttz-E-to-minusinfty-asymp}
\end{align}
where we recall that $S=1$ to the left of the detector's worldline and $S=-1$ to the right of the worldline, and $\zeta$ is defined in~\eqref{eq:zeta-def}. 
The flux is hence leftward to the left of the worldline and rightward to the right of the worldline, similarly to what we found for the Rindler and Milne frame fluxes. 
In Figure~\ref{fig:rindler-t_tz}, this behaviour is visible in the top left panel, $E/a = -3$, far from the detector's worldline, whilst near the trajectory the asymptotic behaviour \eqref{eq:Ttz-neartrajectory-asymp} dominates. The white curve in the panel shows where the two effects cancel and make the flux vanish. 

For large excitation gaps, 
repeated integration by parts in \eqref{eq:rindler-T_(tz)} shows that the asymptotic form of 
$\braket{T_{tz}}^{(2)}$ is
\begin{align}
    \braket{T_{tz}}^{(2)} \sim \frac{2(u^2 - v^2)}{a\pi^3E}\frac{3a^{-2}-uv}{{|uv + a^{-2}|}^5} .
    \label{eq:rindler-T_tz-largeposE-as}
\end{align}
This expression crosses zero on the lines $t = 0$ and $z = 0$, and on the hyperbola $t = \sqrt{z^2 + 3a^{-2}}$. These crossings are clearly visible Figure~\ref{fig:rindler-t_tz}, in the top right panel, $E/a = 3$, and the $E/a=1$ and $E/a=0.1$ panels show the approach to this behaviour. 

This contrast between large excitation gaps and large de-excitation gaps can be attributed to different effects dominating the two cases, recalling how the Minkowski energy flux gets contributions from both the energy density and the energy flux in the Rindler and Milne frames. 
For large de-excitation gaps, the dominant terms are radiative, in that obey an inverse square law in distance; for large excitation gaps, by contrast, the dominant terms have a faster spatial falloff and can be interpreted as vacuum polarisation. The sign changes across $t =0 $ and $z = 0$ as $E\to\infty$ are a direct consequence of the
spacetime dependence of the boost that relates the Minkowski frame to the Rindler frame and the Milne frame.

\section{Discussion}\label{sec:discussion}

\subsection{Summary of results}
In this paper, we have investigated the back-action from a spatially pointlike particle detector on a real quantum scalar field, as characterised by  expectation values of field observables, without conditioning on a measurement of the final state of the detector. 
We assumed the field to be initially in a zero-mean Gaussian Hadamard state in a globally hyperbolic spacetime and evaluated the field's one-point and two-point functions in second-order perturbation theory using the manifestly causal techniques of covariant curved spacetime quantum field
theory~\cite{Hollands2015}. These techniques allow a full control of the spacetime localisation of the interaction, and they do not rely on mode decompositions, non-local particle countings, or asymptotic equilibrium assumptions. 

As an intermediate result, we showed that the back-action in the field's two-point function decomposes into a deterministic part and a fluctuating part, and these parts come from the deterministic and fluctuating parts of the detector's monopole moment operator, respectively. 
For the UDW detector and the HO detector initialised in an energy eigenstate, the deterministic part vanishes. In comparison, the HO detector initialised in a coherent state produces back-action whose deterministic part scales with the square of the coherent amplitude, while the fluctuating contribution is identical to that of the UDW detector initialised in its ground state. In the rest of the paper, we specialised to the UDW detector initialised in an energy eigenstate. 

Using the two-point function, we computed the renormalised SET for a minimally coupled massless scalar field in $(3+1)$-dimensional Minkowski spacetime for a general real-analytic detector trajectory, with the field initially prepared in the Minkowski vacuum. The restriction to real-analytic trajectories was made for technical convenience, and we expect the results to extend to $C^n$ trajectories with sufficiently high~$n$. 

We then presented detailed analytic and numerical results for the SET for an inertial detector and a uniformly linearly accelerated detector, each switched on in the asymptotic past. For the inertial detector, we considered the energy density and the energy flux in the Minkowski frame of the detector. For the accelerated detector, we considered both the Minkowski frame, giving the perspective of inertial observers, and two frames adapted to the boost symmetry of the detector's trajectory: a Rindler frame, giving the perspective of co-accelerarating observers, and a Milne frame, giving the perspective of inertial observers on worldlines orthogonal to the boost Killing vector that generates the detector's trajectory. 

For the accelerated detector, the Rindler frame SET analysis revealed that the energy flux into and out of the detector accounts exactly for the energy gained and lost by the detector in its transitions due to the Unruh effect. 
For the inertial detector, the Minkowski frame analysis showed that the same accounting holds for the outward flux associated with the detector's de-excitations when the detector is initially prepared in its excited state; when the detector is initially prepared in the ground state, the same accounting holds trivially in the sense that both the flux and the excitation rate vanish. 

For the accelerated detector, the Milne frame energy flux shares the gap dependence of the Rindler frame energy flux, and is likewise directed towards the detector for excitation gaps and away from the detector for de-excitation gaps. The Minkowski frame energy flux is more intricate since the system is not invariant under Minkowski time or space translations, 
and near the detector the sign of the Minkowski energy flux is dominated by the contribution from the boosted Rindler frame energy density, which changes sign depending on the direction of the detector's motion. 

In the energy density, a key difference between the inertial trajectory and all three notions of energy density for the accelerated trajectory is that the accelerated case formulae feature Planckian factors where the inertial case formula features step functions, 
consistently with the similar factors that appear in the detector's transition rate. Another difference is that the inertial case energy density is strictly positive, whereas the accelerated case has regions of negative energy density, in all three notions of energy density, when the detector is initially prepared in its ground state. One of these regions occurs near the Rindler horizon that bounds the causal future of the trajectory, and is present in all three notions of energy density; the other region is present in the Minkowski and Milne energy densities, and occurs in the far future of the trajectory. 

A similarity between the energy densities for the two detector trajectories is an $r^{-4}$ divergence near the detector. This divergence is independent of the energy gap and it reflects the singular character of the pointlike interaction between the detector and the field~\cite{Fewster2024}. Another similarity is in the large positive gap and large negative gap asymptotics, growing as the square of the gap for large de-excitation gaps and falling off as the inverse of the gap for large excitation gaps. 

\subsection{Outlook and future work}

Our setup builds on the treatment of the back-action of an accelerating detector on a quantum field in Ref.~\cite{Unruh1984}, where it was found that detector transitions are accompanied by the emission and absorption of Minkowski particles. Here the same physics is recast directly at the level of the renormalised stress-energy tensor, without invoking particle notions or conditioning on detector measurements, and we similarly find that detector transitions are correlated with fluxes of energy into and out of the detector. A significant advantage of our treatment is that the renormalised SET is obtained as a genuinely local quantity, defined at each spacetime point rather than only in asymptotic regions, and this allows us to identify divergences and regions of negative energy density that occur for the uniformly accelerated detector. 

From a conceptual point of view, one might ask to what extent the divergences and negative energy density regions for the uniformly accelerated detector are artifacts of the idealised detector model, in which the coupling is spatially pointlike and effectively active from the asymptotic past. The near-worldline divergence is in fact generic and appears already on the level of the two-point function \eqref{eq:two-point-function-mink} for a general worldline; this divergence is due to the singular nature of the pointlike coupling~\cite{Fewster2024}. By contrast, the divergence of the energy density at the Rindler horizon is specific to the accelerating case and is due to the infinite duration of the field-detector interaction. Both of these divergences can be regularised by implementing a compactly supported \textit{spacetime\/} smearing of the interaction. In that setting, the stability results for Hadamard states under Cauchy evolution apply~\cite{Fulling:1978ht,Sahlmann:2000zr,Hollands:2001qe,Fewster2015,Rejzner:2016hdj}, and one expects the renormalised SET to remain finite everywhere, with the short-distance behaviour near the detector controlled by its spatial size. In particular, the finite interaction duration would remove the accumulation of energy density at the Rindler horizon~\cite{Garfinkle:2019vej}.

The negative energy density regions we find for the uniformly accelerated detector prepared initially in its ground state are more subtle. If these regions persist with a compactly supported spacetime smearing, then it would be reasonable to interpret them as a genuine, but constrained effect in a manner similar to the Casimir effect \cite{casimir-1948,Ford2010-ijmpa}. This is because for free fields in Hadamard states, quantum energy inequalities (QEIs) provide state-independent lower bounds on time-averaged energy densities along timelike worldlines, so that any negative energy density must be compensated by positive energy density elsewhere and cannot be made arbitrarily large or long-lived \cite{Fewster2012}. Whether the same is true in the strictly pointlike and infinite-duration limits that we have analysed in this paper is less clear as Hadamard stability is not guaranteed and thus our results lie outside the existing QEI theorems. Our results do show, however, that the negative energy densities in the regions we have identified are strongly suppressed, and thus it is possible that even in this case, any relevant QEIs might not be violated.

Introducing a compact spatial smearing however introduces a certain degree of non-locality to the field-detector interaction, since the coupling at a given proper time samples the field over a spatial hypersurface near the trajectory. This non-locality introduces ambiguities in the time-ordering operation, thus breaking covariance~\cite{Martin-Martinez:2020lul}. Nevertheless, if the detector is prepared initially in a state that is diagonal in the energy eigenbasis, then the effect of this broken covariance does not appear until third or fourth order in perturbation theory \cite{Martin-Martinez:2020lul} and therefore would not affect the covariance of the renormalised SET in second-order perturbation theory. 
The generalisation of our implementation of causality and covariance in Sec.~\ref{sec:field-correlators} would presumably then proceed by replacing the worldline with a worldtube; for a given spacetime point, or pair of spacetime points, one would choose a final Cauchy surface whose entire intersection with the support of the coupling region is spacelike separated from these points, so that the contributions from later interaction regions drop out exactly as in the pointlike case. Such an extension to our analysis would place the model closer to the standard perturbative algebraic framework \cite{Fewster2015} and to spatially smeared detector models in relativistic quantum information \cite{Martin-Martinez:2020pss}.

We have assumed that the final state of the detector is not measured. A possible generalisation would be to analyse how a measurement of the detector affects the field's SET, and how the associated state update fits within recent covariant measurement schemes in quantum field theory \cite{Fewster:2018qbm,Polo-Gomez:2021irs,Polo-Gomez:2025nuy}, providing a local generalisation of the results in~\cite{Garcia-Chung:2021map}. 

Our core results are for inertial and uniformly linearly accelerated detector trajectories in $(3+1)$-dimensional Minkowski spacetime, with the field prepared in the Minkowski vacuum. 
The formalism of Sec.\ \ref{sec:field-correlators} however applies to any timelike detector trajectory in a globally hyperbolic spacetime of arbitrary dimension with the field prepared in a zero-mean Gaussian Hadamard state, and this generalisation may provide an arena for future work. For example, it would be of interest to investigate how the back-action distinguishes thermality due to the Unruh effect and thermality due to an ambient heat bath, that is, to compare the back-action from an accelerated detector where the field is prepared in the Minkowski vacuum, as computed in Sec.~\ref{sec:uniform-acceleration}, with that of a static detector where the field is prepared in a thermal state. The formalism could further be applied to field-detector back-action in other spacetimes of interest, such as black hole spacetimes and cosmological spacetimes, to investigate effects due to curvature, event horizons, cosmological horizons, and cosmological expansion. 
For the stress-energy tensor, or indeed for any field observable that requires renormalisation, these more general settings may introduce renormalisation ambiguities~\cite{Wald:1977up,Wald:1978pj,Wald1994}, whereas the transition probability of a pointlike UDW detector has no ambiguity even in these more general settings as the transition probability needs no renormalisation to quadratic order in perturbation theory.
We leave these questions to future work.

\section*{Acknowledgments}

We thank Rick Perche, Kinjalk Lochan, Chris Fewster, Benito Ju\'arez-Aubry, Gerardo Adesso and the members of the Nottingham Gravity Lab for numerous insightful interactions. 
The work of JL was supported by United Kingdom Research and Innovation Science and Technology Facilities Council [grant numbers ST/S002227/1, ST/T006900/1 and ST/Y004523/1], 
and has benefited from the activities
of COST Action CA23115: Relativistic Quantum Information, funded by COST (European Cooperation
in Science and Technology).
WGU thanks the Natural Sciences and Engineering Research Council of
Canada (NSERC) (Grant No. 5-80441), 
the Hagler Institute, and the Chancellor's
Fund and IQSE at Texas A\&M University, and the Perimeter Institute, for support.
For the purpose of open access, the authors have applied a CC BY public copyright licence to any Author Accepted Manuscript version arising.

\appendix

\section{Transition probability of an inertial UDW detector in the long time limit}\label{app:detector-excitation-probability}

In this appendix, we consider the transition probability and the transition rate of an inertial UDW detector interacting with a massless scalar field in $(3+1)$-dimensional Minkowski spacetime, with the field initially prepared in the Minkowski vacuum. We first assume the detector-field interaction occurs over a finite duration, controlled by a smooth, compactly supported switching function $\chi(\tau)\in C_0^\infty(\mathbb{R})$, and we subsequently consider the limit in which the interaction occurs for a long time. 

With the interaction Hamiltonian given by 
\eqref{eq:interaction-hamiltonian} with~\eqref{eq:two-level}, and working to leading order in perturbation theory, the detector's transition probability on a stationary trajectory is a multiple of the response function~${\mathcal F}_\chi$, given by \cite{Fewster2016}
\begin{align}
\label{eq: stationary RF}
    {\mathcal F}_\chi(E) = \frac{1}{2\pi} \int_{-\infty}^\infty \mathrm{d}\omega\, 
    |{\widehat \chi}(\omega)|^2 \, \widehat {\mathcal W}(\omega+E) , 
\end{align}
where the hat denotes the Fourier transform, with the convention given in~\eqref{eq:fourier-def}, and ${\mathcal W}$ is the pull-back of the field's Wightman function to the detector's stationary trajectory. For an inertial detector in the $(3+1)$-dimensional Minkowski vacuum, we have 
\begin{align}
{\mathcal W}(\tau) = - \frac{1}{4\pi^2} 
\frac{1}{{(\tau - i\epsilon)}^2} , 
\end{align}
where $\epsilon\to0_+$, from which 
\begin{align}
\widehat {\mathcal W}(\omega) = - \frac{\omega}{2\pi} \Theta(-\omega) , 
\label{eq:What-Mink}
\end{align}
where $\Theta$ is the Heaviside function, defined in~\eqref{eq:Theta-def}. From \eqref{eq: stationary RF} and \eqref{eq:What-Mink} we have 
\begin{align}
\label{eq: stationary RF Minkvac}
    {\mathcal F}_\chi(E) = \frac{1}{4\pi^2} \int_E^\infty \mathrm{d} \omega\, (\omega - E) \, 
    |{\widehat \chi}(\omega)|^2  , 
\end{align} 
using the evenness of $|{\widehat \chi}(\omega)|$, which follows because $\chi$ is real-valued. 

Now consider the limit in which the interaction occurs for a long time. Since $\chi(\tau)$ is assumed to have compact support, the long time limit may be formalised \cite{Fewster2016,Parry:2025wub} by considering a family of switching functions $\chi_\eta(\tau)$, parametrised by $\eta>0$, with support that grows proportionally to $\eta$ as $\eta \to \infty$, and further, each $\chi_\eta(\tau)$ remains approximately constant, in an averaged sense, over its support. We shall now consider two examples of such switching families and analyse the long interaction time limit in both cases.

The first example is the adiabatic switching family defined by
\begin{equation}
    \chi_\eta(\tau) = \zeta(\tau/\eta),
\end{equation}
where $\zeta\in C_0^\infty(\mathbb{R})$, 
and we assume that $\zeta$ is not identically vanishing. 
We then have $\widehat{\chi}_\eta(\omega) = \eta \widehat{\zeta}(\eta\omega)$, and \eqref{eq: stationary RF Minkvac} can be written as 
\begin{align}
\label{eq: adiab-F-Minkvac}
    {\mathcal F}_{\chi_\eta}(E) = \frac{1}{4\pi^2} \int_{\eta E}^\infty \mathrm{d} u\, (u-\eta E) \, 
    |{\widehat {\zeta}}(u)|^2 ,  
\end{align}
by the substitution $\omega = u/\eta$. 
Because 
$\zeta \in C^\infty_0(\mathbb{R})$, $\widehat{\zeta}$ is smooth and decays at infinity faster than any inverse power of its argument. \eqref{eq: adiab-F-Minkvac} is hence well defined.

Consider now the limit $\eta\to\infty$ in~\eqref{eq: adiab-F-Minkvac}. 
When $E>0$, the large argument falloff of $\widehat{\zeta}$ implies that ${\mathcal F}_{\chi_\eta}(E) \to 0$ as $\eta\to\infty$. 
This means that the detector's excitation probability vanishes in the long time limit. When $E<0$, we have
\begin{align}
    \frac{{\mathcal F}_{\chi_\eta}(E)}{\eta} &= \frac{|E|}{4\pi^2} \int_{-\eta |E|}^\infty \mathrm{d} u \, 
    |{\widehat {\zeta}}(u)|^2 
    \notag 
    \\
    &\hspace{3ex} 
    + \frac{1}{4\pi^2\eta} 
    \int_{-\eta |E|}^\infty \mathrm{d} u \, u \, 
    |{\widehat {\zeta}}(u)|^2 
    \notag
    \\
    &\xrightarrow[\eta\to\infty]{} 
    \frac{|E|}{4\pi^2} 
    \int_{-\infty}^\infty \mathrm{d} w \, 
    |{\widehat {\zeta}}(w)|^2 , 
    \label{eq: adiab-F-Minkvac-negE}
\end{align}
again using the large argument falloff of~$\widehat{\zeta}$. 
\eqref{eq: adiab-F-Minkvac-negE}~means that the detector's de-excitation probability diverges linearly in~$\eta$ as $\eta\to\infty$, but the de-excitation rate, 
the de-excitation probability per unit time, approaches $|E|$ times a positive constant.  

The second example is the plateau switching family defined by~\cite{Fewster2016,Parry:2025wub}
\begin{equation}
    \chi_\eta(\tau) = \int_{-\infty}^\tau \mathrm{d} \tau' \, \big( \psi(\tau')-\psi(\tau'-\tau_s-\eta \tau_p) \big),
\end{equation}
where $\tau_s$ and $\tau_p$ are positive parameters, and ${\psi \in C^\infty_0(\mathbb{R})}$ is non-negative with $\supp(\psi) =[0,\tau_s]$. 
$\chi_\eta$~is then also smooth and of compact support with ${\supp(\chi_\eta) = [0,2\tau_s+\eta \tau_p]}$. We then have ${\widehat{\chi}_\eta(\omega) = \frac{1}{i \omega}\left(1 - \mathrm{e}^{-i \omega (\tau_s+\eta \tau_p)} \right)\widehat{\psi}(\omega)}$, and \eqref{eq: stationary RF Minkvac} can be written as
\begin{equation} \label{eq: plateau-F-Minkvac}
    \mathcal{F}_{\chi_\eta}(E) = \frac{1}{2\pi^2
    }\int_{E}^\infty \mathrm{d} \omega \, (\omega -E)\frac{1-\cos(\kappa(\eta)\omega)}{\omega^2 } |{\widehat \psi}(\omega)|^2,
\end{equation}
where we have defined $\kappa(\eta) = \tau_s + \eta \tau_p$. Since $\psi\in C_0^\infty(\mathbb{R})$, its Fourier transform $\widehat{\psi}$ is smooth and decays faster than any inverse power of its argument at infinity, and thus \eqref{eq: plateau-F-Minkvac} is well-defined. 

Consider now the $\eta \to \infty$ limit of \eqref{eq: plateau-F-Minkvac}. For $E>0$, we may write \eqref{eq: plateau-F-Minkvac} as
\begin{align}
     \mathcal{F}_{\chi_\eta}(E) &= \frac{1}{2\pi^2}\int_E^\infty \mathrm{d} \omega \, \frac{\omega - E}{\omega^2} |{\widehat \psi}(\omega)|^2 \notag \\
     &\hspace{0.4cm} - \frac{1}{2\pi^2}\int_E^\infty \mathrm{d} \omega \, \cos(\kappa(\eta)\omega) \frac{\omega - E}{\omega^2}|{\widehat \psi}(\omega)|^2,
\label{eq:plat-exc-prob-total}
\end{align}
which again is well-defined due to the strong decay of $\widehat{\psi}$ at infinity. 
The first term in \eqref{eq:plat-exc-prob-total} is independent of~$\eta$.
Repeated integration by parts \cite{Wong2001} shows that the second term is $O(\eta^{-2})$ as $\eta \to \infty$.
Therefore
\begin{equation} \label{eq:plat-exc-prob}
    \mathcal{F}_{\chi_\eta}(E)\xrightarrow[\eta\to\infty]{} \frac{1}{2\pi^2}\int_E^\infty \mathrm{d} \omega \, \frac{\omega - E}{\omega^2} |{\widehat \psi}(\omega)|^2.
\end{equation}
For $E<0$, we may write
\begin{align}
    \mathcal{F}_{\chi_\eta}(E) &= \frac{1}{2\pi^2}\int_{|E|}^\infty \mathrm{d} \omega \, \frac{\omega + |E|}{\omega^2} |{\widehat \psi}(\omega)|^2 \notag \\
     &\hspace{0.2cm} - \frac{1}{2\pi^2}\int_{|E|}^\infty \mathrm{d} \omega \, \cos(\kappa(\eta)\omega) \frac{\omega + |E|}{\omega^2}|{\widehat \psi}(\omega)|^2 \notag \\
     &\hspace{0.4cm}+\frac{|E|}{\pi^2}\int_{0}^{|E|} \mathrm{d} \omega \, \frac{1-\cos(\kappa(\eta)\omega)}{\omega^2 } |{\widehat \psi}(\omega)|^2 ,
\label{eq:plat-decay-prob-total}
\end{align}
where we have used the evenness of $|{\widehat \psi}(\omega)|^2$, and the split is justified by the strong decay of $\widehat{\psi}$ at infinity. 
The first term in \eqref{eq:plat-decay-prob-total} is independent of~$\eta$.
Integration by parts \cite{Wong2001} shows that the second term is $O(\eta^{-1})$ as $\eta \to \infty$. 
The change of variables $v=\kappa(\eta)\omega$  puts the third term in the form 
\begin{align}
    &\frac{1}{\pi^2}\kappa(\eta)|E|\int_0^{\kappa(\eta)|E|} \mathrm{d} v \, \frac{1-\cos v}{v^2}\left|{\widehat \psi}\!\left(\frac{v}{\kappa(\eta)}\right)\right|^2 \notag \\
    &\hspace{0.3cm}=  \frac{1}{\pi^2}\kappa(\eta)|E|\int_0^{\infty} \mathrm{d} v \, \frac{1-\cos v}{v^2}\left|{\widehat \psi}\!\left(\frac{v}{\kappa(\eta)}\right)\right|^2 \notag \\
    &\hspace{0.6cm}- \frac{1}{\pi^2}\kappa(\eta)|E|\int_{\kappa(\eta)|E|}^\infty \mathrm{d} v \, \frac{1-\cos v}{v^2}\left|{\widehat \psi}\!\left(\frac{v}{\kappa(\eta)}\right)\right|^2.
\label{eq:plat-decay-prob-term3}
\end{align}
The large argument behaviour of $\widehat{\psi}$ then implies that as $\eta\to \infty$, the first term in \eqref{eq:plat-decay-prob-term3} $\sim \frac{\tau_p  |E|}{2\pi} |{\widehat \psi}(0)|^2 \eta$, using that ${\widehat \psi}(0)\ne0$ by non-negativity of~$\psi$. 
As $\widehat{\psi}$ is bounded, it follows from the large $v$ falloff of $(1-\cos v)/v^2$ that the second term in \eqref{eq:plat-decay-prob-term3} is bounded as 
$\eta \to \infty$. Therefore,
\begin{equation} \label{eq:plat-deexc-rate}
    \frac{{\mathcal F}_{\chi_\eta}(E)}{\eta} \xrightarrow[\eta\to\infty]{} \frac{\tau_p  |E|}{2\pi} |{\widehat \psi}(0)|^2.
\end{equation}

For the plateau switching family, the excitation probability therefore tends to the non-zero function of $E$ \eqref{eq:plat-exc-prob} in the long interaction duration limit, 
and the excitation rate falls off inversely proportionally to the duration. The de-excitation probability diverges linearly in the interaction duration, and the de-excitation rate approaches $|E|$ times a positive constant.

We summarise. 
For the transition rate, in the long interaction time limit, both the adiabatic switching and the plateau switching reproduce the well-known outcomes that arise by informally factoring out the total interaction time to turn the transition probability into the transition rate~\cite{Birrell1982}. In particular, the \textit{excitation rate\/} vanishes in the long time limit. 
For the transition probability, by contrast, 
we have found here that the \textit{excitation probability\/} depends sensitively on the particular switching family used to implement the long time limit. 
For the adiabatic switching family, the excitation probability becomes zero in the long time limit. 
For the plateau switching family, however, the excitation probability remains nonvanishing in the long time limit, due to residual contributions from the fixed-duration switch-on and switch-off intervals.


\section{UDW and HO detectors always fluctuate}\label{app:sho-udw-nonclassicality}

In this appendix we verify the technical statement that neither the UDW detector with $E\ne0$ nor the HO detector has states, pure or mixed, in which the covariance $\cov\!\left( \mu(\tau),\mu(\tau') \right)$ vanishes. In the terminology of Sec.~\ref{sec:examples-of-systems}, this means that the two-point function $\braket{\mu(\tau)\mu(\tau')}$ always has a fluctuating part. 

Recall that the covariance \eqref{eq:covariance-definition} is defined as 
\begin{align}
\cov(X,Y) := \braket{X Y} - \braket{X} \! \braket{Y} , 
\end{align}
where we suppress the state in which the expectation values are evaluated. We shall show that assuming 
\begin{align}
\cov\bigl(\mu(\tau),\mu(\tau')\bigr) = 0 
\label{eq:classical-cov-condition}
\end{align}
leads to a contradiction. 

\subsection{UDW detector}\label{app:udw-nonclassicality}

For the UDW detector, in the notation of Section~\ref{subsec:udw-detector}, 
\eqref{eq:classical-cov-condition} gives 
\begin{align} \label{eq:UDW_nostates}
    &\mathrm{e}^{-i E(\tau+\tau')} \cov(\sigma^-,\sigma^-) +  \mathrm{e}^{-i E(\tau-\tau')}\cov(\sigma^-,\sigma^+) \notag \\
    &+ \mathrm{e}^{i E (\tau-\tau')}\cov(\sigma^+,\sigma^-)+ \mathrm{e}^{i E (\tau+\tau')}\cov(\sigma^+,\sigma^+) = 0. 
\end{align}
If the detector is nondegenerate, $E\ne0$, 
linear independence of the set $\{\mathrm{e}^{\pm i E (\tau+\tau')},\mathrm{e}^{\pm i E (\tau-\tau')}\}$ implies that $\cov(\sigma^-,\sigma^-) = \cov(\sigma^-,\sigma^+)=\cov(\sigma^+,\sigma^-) = \cov(\sigma^+,\sigma^+)=0. $ Since $(\sigma^-)^2=(\sigma^+)^2=0$, the conditions $\cov(\sigma^-,\sigma^-) =\cov(\sigma^+,\sigma^+)=0$ imply $\braket{\sigma^+}=\braket{\sigma^-} = 0$, and the conditions $\cov(\sigma^-,\sigma^+)=\cov(\sigma^+,\sigma^-)=0$ then imply $\braket{\sigma^+\sigma^-} =\braket{\sigma^-\sigma^+} =0 $. However, we also have
\begin{equation}
    \braket{\sigma^+\sigma^-} + \braket{\sigma^-\sigma^+} = \braket{\{\sigma^+,\sigma^-\}} = 1,
\end{equation}
using $\{\sigma^+,\sigma^-\}=\mathds{1}_D$. This contradiction shows that \eqref{eq:classical-cov-condition} does not hold for any state. 

For the gapless UDW detector, $E=0$, 
the dynamics are frozen, and 
$\mu = \sigma^+ + \sigma^- = \begin{psmallmatrix}
        0 & 1 \\
        1 & 0
\end{psmallmatrix}$. 
It can be verified that 
\eqref{eq:classical-cov-condition} then holds if and only if 
$\rho = \frac12\begin{psmallmatrix}
        1 & \pm1 \\
        \pm1 & 1
\end{psmallmatrix}$, 
which are the two eigenstates of~$\mu$.

\subsection{HO detector}\label{app:ho-nonclassicality}

For the HO detector, in the notation of Section~\ref{subsec:SHO}, \eqref{eq:classical-cov-condition} gives 
\begin{align} \label{eq:QHO_nostates}
    &\mathrm{e}^{-i \Omega(\tau+\tau')} \cov(a,a) +  \mathrm{e}^{-i \Omega(\tau-\tau')}\cov(a,a^\dagger) \notag \\
    &+ \mathrm{e}^{i \Omega (\tau-\tau')}\cov(a^\dagger,a)+ \mathrm{e}^{i \Omega (\tau+\tau')}\cov(a^\dagger,a^\dagger) = 0, 
\end{align}
As $\Omega>0$ by assumption, the 
set $\{\mathrm{e}^{\pm i \Omega (\tau+\tau')},\mathrm{e}^{\pm i \Omega (\tau-\tau')}\}$ is linearly independent, and \eqref{eq:QHO_nostates} implies that $\cov(a,a) = \cov(a,a^\dagger)=\cov(a^\dagger,a) = \cov(a^\dagger,a^\dagger)=0. $ However, we also have
\begin{equation}
    \cov(a,a^\dagger)-\cov(a^\dagger,a) = \braket{[a,a^\dagger]} = 1,
\end{equation}
using $[a,a^\dagger] = \mathds{1}_D$. This contradiction shows that \eqref{eq:classical-cov-condition} does not hold for any state.

\section{Evaluation of \texorpdfstring{$I_{\mu\nu}(\mathsf{x})$ \eqref{eq:Imunu-def}}{I-(mu nu)}}\label{app:derivatives-of-correlator}

In this appendix we verify the result \eqref{eq:masterequation} for the coincidence limit of the differentiated two-point correlator, 
$I_{\mu\nu}(\mathsf{x}) = \lim_{\mathsf{x}'\to\mathsf{x}}\partial_\mu\partial_{\nu'}\braket{\phi(\mathsf{x})\phi(\mathsf{x}')}^{(2)}$~\eqref{eq:Imunu-def}. 

\subsection{Split of \texorpdfstring{$I_{\mu\nu}$}{I-(mu nu)}}

From \eqref{eq:two-point-function-mink} we recall that 
\begin{align}
\braket{\phi(\mathsf{x})\phi(\mathsf{x}')}^{(2)}\! &= \im\biggl[\frac{\chi(\tau_-)\mathrm{e}^{iE\tau_-}}{4\pi^3 f'(\tau_-; \mathsf{x})}\biggl(\dashint_{-\infty}^{\tau_-}\mathrm{d}\tau\frac{\chi(\tau)\mathrm{e}^{-iE\tau}}{f(\tau; \mathsf{x}')}\notag\\
    &+\!i\pi\frac{\chi(\tau_-')\mathrm{e}^{-iE\tau_-'}}{f'(\tau_-'; \mathsf{x}')}\Theta(\tau_- - \tau_-')\biggr)\biggr]\!+ (\mathsf{x} \leftrightarrow \mathsf{x}') , \label{eq:app:two-point-function-mink}
\end{align}
where $\mathsf{x}$ and $\mathsf{x}'$ are timelike separated and sufficiently close to each other. $\dashint$ denotes the Cauchy principal value integral in the integral term in which the denominator has a linear zero and a normal integral in the integral term in which the denominator does not have a zero.  

We split $I_{\mu\nu}(\mathsf{x})$ as
\begin{align}
    I_{\mu\nu}(\mathsf{x}) = \frac{1}{4\pi^3}\lim_{\mathsf{x}\to\mathsf{x}'}\left(C_{\mu\nu}(\mathsf{x}, \mathsf{x}') + S_{\mu\nu}(\mathsf{x}, \mathsf{x}')\right) , 
\end{align}
with 
\begin{subequations}
    \begin{align}
        C_{\mu\nu}(\mathsf{x}, \mathsf{x}') &:= \pi\partial_\mu\partial_{\nu'}\re \! \left[ \frac{\chi(\tau_-)\mathrm{e}^{iE\tau_-}}{f'(\tau_-; \mathsf{x})}\frac{\chi(\tau_-')\mathrm{e}^{-iE\tau_-'}}{f'(\tau_-'; \mathsf{x}')} \right] ,
        \label{eq:Cmunu-def}
        \\
        S_{\mu\nu}(\mathsf{x}, \mathsf{x}') &:= \partial_\mu \partial_{\nu'}\im \! \left[ \frac{\chi(\tau_{-})\mathrm{e}^{ iE\tau_{-} }}{f'(\tau_{-}; \mathsf{x})}\dashint_{-\infty}^{\tau_{-}}\mathrm{d}\tau\frac{\chi(\tau)\mathrm{e}^{ -iE\tau } }{f(\tau; \mathsf{x}')} \right] \notag\\
        &\qquad + (\mathsf{x} \leftrightarrow \mathsf{x}') , \label{eq:smunu}
    \end{align}
\end{subequations}
where $C_{\mu\nu}$ comes from the non-integral terms in~\eqref{eq:app:two-point-function-mink}, combining the terms that involve ${\Theta(\tau_- - \tau_-')}$ and ${\Theta(\tau_-' - \tau_-)}$, 
and $S_{\mu\nu}$ comes from the integral terms.

We consider $C_{\mu\nu}$ and $S_{\mu\nu}$ in turn. 

\subsection{Evaluation of \texorpdfstring{$C_{\mu\nu}$}{C-(mu nu)}}

In $C_{\mu\nu}$~\eqref{eq:Cmunu-def}, evaluating the derivatives and taking the coincidence limit gives 
\begin{align}
    C_{\mu \nu}(\mathsf{x}, \mathsf{x}) & = \pi E^{2} \partial_\mu \tau_{-} \partial_\nu \tau_{-}
    \frac{\chi(\tau_{-})^{2}}{f'(\tau_{-}; \mathsf{x})^{2}} \notag\\
    &\quad+ \pi\partial_{\mu}\left( \frac{\chi(\tau_{-})}{f'(\tau_{-}; \mathsf{x})} \right)\partial_{\nu}\left( \frac{\chi(\tau_{-})}{f'(\tau_{-}; \mathsf{x})} \right), \label{eq:cmunu-initial}
\end{align}
where we have used that the terms linear in $E$ vanish on taking the real part. 
To eliminate the derivatives of~$\tau_-$, we recall that the equation defining $\tau_-$ as a function of $\mathsf{x}$ is $f(\tau_{-}; \mathsf{x}) = 0$, and differentiating this relation gives 
\begin{align}
0 = \frac{\partial}
    {\partial x^\mu}f(\tau_{-}; \mathsf{x}) 
    = \partial_{\mu}\tau_{-}\,f'(\tau_{-}; \mathsf{x}) + f_{\mu}(\tau_{-}; \mathsf{x}) , 
\end{align}
from which we have 
\begin{align}
    \partial_{\mu}\tau_{-} = -\frac{f_{\mu}(\tau_{-}; \mathsf{x})}{f'(\tau_{-}; \mathsf{x})} . \label{eq:partialtauminus}
\end{align}
We further have 
\begin{align}
    \partial_{\mu}\left( \frac{\chi(\tau_{-})}{f'(\tau_{-}; \mathsf{x})} \right)  & = 
    \partial_{\mu}\tau_{-}\partial_{\tau}\left.\left( \frac{\chi(\tau)}{f'(\tau; \mathsf{x})} \right)\right|_{\tau = \tau_-}
    \notag\\
    & \quad - \frac{\chi(\tau_{-})}{f'(\tau_{-}; \mathsf{x})^{2}}f_{\mu}'(\tau_{-}, \mathsf{x}) \notag\\[2mm]
    &= -\frac{1}{f'(\tau_{-}; \mathsf{x})}\partial_{\tau}\left.\left( \frac{\chi(\tau)f_{\mu}(\tau; \mathsf{x})}{f'(\tau; \mathsf{x})}\right) \right|_{\tau = \tau_{-}} , 
\label{eq:inverse-productrule}
\end{align}
where the first term after the first equality comes from the $\mathsf{x}$-dependence in $\tau_-$ and the second term comes from the $\mathsf{x}$-dependence in the second argument of $f'(\tau_-; \mathsf{x})$, and in the second equality we have used \eqref{eq:partialtauminus} and the product rule. 
Defining 
\begin{align}
    \alpha_{\mu}(\tau; \mathsf{x}) := \frac{\chi(\tau)f_\mu(\tau; \mathsf{x})}{f'(\tau; \mathsf{x})} , 
    \label{eq:alpha}
\end{align}
we may therefore write \eqref{eq:cmunu-initial} as 
\begin{align}
    C_{\mu \nu}(\mathsf{x}, \mathsf{x}) = \pi\frac{E^{2}\alpha_{\mu}(\tau_{-}; \mathsf{x})\alpha_{\nu}(\tau_{-}; \mathsf{x}) + \alpha'_{\mu}(\tau_{-}; \mathsf{x})\alpha'_{\nu}(\tau_{-}; \mathsf{x})}{{f'(\tau_{-}; \mathsf{x})}^{2}} .
\label{eq:Cmunu-coincidence}
\end{align}

\subsection{Evaluation of \texorpdfstring{$S_{\mu\nu}$}{S-(mu nu)}}

We now turn to $S_{\mu \nu}$~\eqref{eq:smunu}. In the explicitly displayed term in~\eqref{eq:smunu}, the contribution from $\partial_\mu$ acting on the $\tau_-$ at the upper limit of the integral vanishes, on taking the imaginary part, and similarly in the $(\mathsf{x} \leftrightarrow \mathsf{x}')$ term when the primed derivative acts on the $\tau_-'$ at the upper limit of the integral. We can therefore write
\begin{align}
    S_{\mu \nu}(\mathsf{x}, \mathsf{x}') &= \im\biggl[ \partial_{\mu}\left( \frac{\chi(\tau_{-})\mathrm{e}^{ iE\tau_{-} }}{f'(\tau_{-}; \mathsf{x})} \right)\int_{-\infty}^{\tau_{-}}\mathrm{d}\tau \, \chi(\tau)\notag\\
    &\quad\times\mathrm{e}^{ -iE\tau }\partial_{\nu'}\left(\mathrm{P.V.} \frac{1}{f(\tau; \mathsf{x}')}  \right)  \biggr] + (\mathsf{x} \leftrightarrow \mathsf{x}') .
\label{eq:smunu-differentiation-first}
\end{align}
Under the integral in~\eqref{eq:smunu-differentiation-first}, we use the distributional identity 
\begin{align}
    \partial_{\nu'}\left( \mathrm{P.V.}\frac{1}{f(\tau; \mathsf{x}')} \right) = \frac{f_{\nu'}(\tau; \mathsf{x}')}{f'(\tau; \mathsf{x}')} \partial_{\tau}\left( \mathrm{P.V.} \frac{1}{f(\tau; \mathsf{x}')} \right) , 
\label{eq:princvalid-cited}
\end{align}
which we verify in Appendix~\ref{app:principal-value-identity}. Note that $f'(\tau; \mathsf{x}')$ in the denominator in \eqref{eq:princvalid-cited} is nonvanishing in the integration range in \eqref{eq:smunu-differentiation-first} when $\mathsf{x}$ and  $\mathsf{x}'$ are sufficiently close to each other. For the factor outside the integral in~\eqref{eq:smunu-differentiation-first}, we have 
\begin{align}
    \partial_{\mu}\left( \frac{\chi(\tau_{-})\mathrm{e}^{ iE\tau_{-} }}{f'(\tau_{-}; \mathsf{x})} \right) &= -\frac{1}{f'(\tau_-; \mathsf{x})}\partial_\tau\left.\Bigl(\alpha_\mu(\tau; \mathsf{x})\mathrm{e}^{iE\tau}\Bigr)\right|_{\tau = \tau_-}, 
    \label{eq:alpha-gen}
\end{align}
where $\alpha_\mu$ is given by~\eqref{eq:alpha}, as can be verified by the replacement $\chi(\tau_-) \to \chi(\tau_-)\mathrm{e}^{ iE\tau_-}$ in~\eqref{eq:inverse-productrule}. 
Substituting \eqref{eq:princvalid-cited}
and \eqref{eq:alpha-gen} in~\eqref{eq:smunu-differentiation-first}, we find 
\begin{widetext}
\begin{align}
    S_{\mu \nu}(\mathsf{x}, \mathsf{x}') & = -\frac{1}{f'(\tau_-; \mathsf{x})}\im\biggl[ \partial_\tau\left.\Bigl(\alpha_\mu(\tau; \mathsf{x})\mathrm{e}^{iE\tau}\Bigr)\right|_{\tau = \tau_-} 
    \int_{-\infty}^{\tau_{-}}\mathrm{d}\tau \, \alpha_{\nu'}(\tau; \mathsf{x}')\mathrm{e}^{ -iE\tau }\partial_{\tau}\left( \mathrm{P.V.}\frac{1}{f(\tau; \mathsf{x}')}  \right)  \biggr] 
    +  (\mathsf{x} \leftrightarrow \mathsf{x}'), 
    \notag\\
&= -\frac{1}{f'(\tau_-; \mathsf{x})}\im\Biggl\{ \partial_\tau\left.\Bigl(\alpha_\mu(\tau; \mathsf{x})\mathrm{e}^{iE\tau}\Bigr)\right|_{\tau = \tau_-}\biggl[ 
    \frac{\alpha_{\nu'}(\tau_{-}; \mathsf{x}')\mathrm{e}^{ -iE\tau_{-} }}{f(\tau_{-}; \mathsf{x}')} 
    - \dashint _{-\infty}^{\tau_{-}} \mathrm{d}\tau  \frac{1}{f(\tau; \mathsf{x}')} \partial_{\tau}\biggl( \alpha_{\nu'}(\tau; \mathsf{x}')\mathrm{e}^{ -iE\tau } \biggr) \biggr] \Biggr\} 
    \notag\\
     &\quad
     + (\mathsf{x} \leftrightarrow \mathsf{x}') 
     \notag\\
& = -E \frac{\alpha_{\mu}(\tau_{-}; \mathsf{x})}{f'(\tau_{-}; \mathsf{x})} \frac{\alpha_{\nu'}(\tau_{-}; \mathsf{x}')}{f(\tau_{-}; \mathsf{x}')} \notag\\
        &\quad- \frac{1}{f'(\tau_{-}; \mathsf{x})} \dashint _{-\infty}^{\tau_{-}} \mathrm{d}\tau\frac{\sin(E(\tau - \tau_-))}{f(\tau; \mathsf{x}')}\biggl(E^2\alpha_\mu(\tau_-; \mathsf{x})\alpha_{\nu'}(\tau; \mathsf{x}') + \alpha_\mu'(\tau_-; \mathsf{x})\alpha_{\nu'}'(\tau; \mathsf{x}')\biggr) \notag\\
        &\quad- \frac{E}{f'(\tau_{-}; \mathsf{x})} \dashint _{-\infty}^{\tau_{-}} \mathrm{d}\tau  \frac{\cos(E(\tau - \tau_-))}{f(\tau; \mathsf{x}')}\biggl( \alpha_\mu(\tau_-; \mathsf{x})\alpha_{\nu'}'(\tau; \mathsf{x}') - \alpha_\mu'(\tau_-; \mathsf{x})\alpha_{\nu'}(\tau; \mathsf{x}') \biggr) 
        \notag\\
        &\quad + (\mathsf{x} \leftrightarrow  \mathsf{x}') ,         
        \label{eq:smunu-expanded}
    \end{align}
    \end{widetext}
in the second equality integrating by parts, and in the third equality expanding the derivatives by the product rule and taking the imaginary part. Note that since $\partial_\mu$ and $\partial_{\nu'}$ in \eqref{eq:smunu} act on $\mathsf{x}$ and $\mathsf{x}'$ respectively, and in \eqref{eq:smunu-expanded}
these derivatives are encoded in the index on $\alpha_\mu$ and $\alpha_{\nu'}$ respectively, the interchange $(\mathsf{x} \leftrightarrow \mathsf{x}')$ in \eqref{eq:smunu-expanded} includes the interchange of $\mu$ and~$\nu'$.

We need to take the coincidence limit in~\eqref{eq:smunu-expanded}. 
In the terms with the sine in~\eqref{eq:smunu-expanded}, the coincidence limit is 
\begin{align}
&- \frac{2}{f'(\tau_{-}; \mathsf{x})} \int _{-\infty}^{\tau_{-}} \mathrm{d}\tau\frac{\sin(E(\tau - \tau_-))}{f(\tau; \mathsf{x})}
\notag\\
& \hspace{3ex}
\times \biggl(E^2\alpha_{(\mu}(\tau_-; \mathsf{x})\alpha_{\nu)}(\tau; \mathsf{x}) + \alpha_{(\mu}'(\tau_-; \mathsf{x})\alpha_{\nu)}'(\tau; \mathsf{x})\biggr) , 
\label{eq:Smunu-coincidence-firstpart}
\end{align}
where the parentheses on the indices denote symmetrisation, in the convention $2A_{(\mu\nu)} = A_{\mu\nu} + A_{\nu\mu}$, this symmetrisation has come from the sum of the term shown explicitly in \eqref{eq:smunu-expanded} and its $(\mathsf{x} \leftrightarrow  \mathsf{x}')$ partner, and the linear zero of $\sin(E(\tau - \tau_-))$ at $\tau = \tau_-$ has turned the principal value integral into an ordinary integral in the coincidence limit. 

\vspace{-2cm}

In the terms with the cosine in~\eqref{eq:smunu-expanded}, we change variables by $\tau = \tau_-+s$ in the term shown explicitly, and by $\tau = \tau_-'+s$ in the $(\mathsf{x} \leftrightarrow  \mathsf{x}')$ partner term, where in each case $s\in(-\infty,0)$. Combining the two terms, the integrand is $-E\cos(E s)$ times 
\begin{align}
    &\frac{\alpha_\mu(\tau_-; \mathsf{x})\alpha_{\nu'}'(\tau_- + s; \mathsf{x}')}{f'(\tau_{-}; \mathsf{x}) f(\tau_- + s; \mathsf{x}')} 
    - \frac{\alpha_{\nu'}'(\tau_-'; \mathsf{x}')\alpha_{\mu}(\tau_-' + s; \mathsf{x})}{f'(\tau_{-}'; \mathsf{x}') f(\tau_-' + s; \mathsf{x})} 
    \notag\\
    &\quad 
    + \frac{\alpha_{\nu'}(\tau_-'; \mathsf{x}')\alpha_{\mu}'(\tau_-' + s; \mathsf{x})}{f'(\tau_{-}'; \mathsf{x}') f(\tau_-' + s; \mathsf{x})} 
    - \frac{\alpha_\mu'(\tau_-; \mathsf{x})\alpha_{\nu'}(\tau_- + s; \mathsf{x}')}{f'(\tau_{-}; \mathsf{x}) f(\tau_- + s; \mathsf{x}')} . 
\label{eq:cmunu-cos-integrand}
\end{align}
The coincidence limit is hence 
\begin{align}
    &-\frac{2E}{f'(\tau_-; \mathsf{x})}\int_{-\infty}^{\tau_-}\mathrm{d}\tau \frac{\cos(E(\tau - \tau_-))}{f(\tau; \mathsf{x})} \notag\\
    &\hspace{3ex}\times\biggl(\alpha_{(\mu}(\tau_-; \mathsf{x})\alpha_{\nu)}'(\tau; \mathsf{x}) - \alpha_{(\mu}(\tau; \mathsf{x})\alpha_{\nu)}'(\tau_-; \mathsf{x})\biggr),
\label{eq:Smunu-coincidence-secondpart}
\end{align}
where the regularity of \eqref{eq:cmunu-cos-integrand} in the coincidence limit has turned the principal value integral into an ordinary integral. 

We write the remaining piece in \eqref{eq:smunu-expanded} as $-B_{\mu \nu}(\mathsf{x}, \mathsf{x}')$, where 
\begin{align}
B_{\mu \nu}(\mathsf{x}, \mathsf{x}') &:= E 
\Biggl[  \frac{\alpha_{\mu}(\tau_{-}; \mathsf{x})}{f'(\tau_{-}; \mathsf{x})} \frac{\alpha_{\nu'}(\tau_{-}; \mathsf{x}')}{f(\tau_{-}; \mathsf{x}')} \notag\\
&\hspace{7ex} + \frac{\alpha_{\nu'}(\tau_{-}'; \mathsf{x}')}{f'(\tau_{-}'; \mathsf{x}')} \frac{\alpha_{\mu}(\tau_{-}'; \mathsf{x})}{f(\tau_{-}'; \mathsf{x})} \Biggl]  .
\label{eq:Bmunu-def}
\end{align}
To find the coincidence limit of $B_{\mu \nu}(\mathsf{x}, \mathsf{x}')$, we write 
\begin{align}
\mathsf{x}' = \mathsf{x} + \epsilon , 
\end{align}
where $\epsilon$ is a timelike vector, and we consider the limit in which the direction of $\epsilon$ is fixed and its length tends to zero. We expand $\tau_-'$ in terms of $\mathsf{x}$ and $\epsilon$ as 
\begin{align}
        \tau_{-}' & = \tau_{-} + \epsilon^{\alpha}\partial_{\alpha}\tau_{-} + \frac{\epsilon^{\alpha}\epsilon^{\beta}}{2}\partial_{\alpha\beta}\tau_{-} + O(\epsilon^{3}) . 
        \label{eq:tauminusprime-eps-expansion1}
\end{align}
$\partial_{\alpha}\tau_{-}$ is given by~\eqref{eq:partialtauminus}, and for $\partial_{\alpha\beta}\tau_{-}$ we find 
\begin{align}
\partial_{\alpha\beta}\tau_{-} &  = \partial_{\alpha}\left( -\frac{f_{\beta}(\tau_{-}; \mathsf{x})}{f'(\tau_{-}; \mathsf{x})} \right) \notag\\
     &= -\frac{\partial_\alpha \tau_- f_\beta'(\tau_-; \mathsf{x}) + f_{\alpha\beta}(\tau_-; \mathsf{x})}{f'(\tau_-; \mathsf{x})} \notag\\
     &\quad+ \frac{\partial_\alpha \tau_- f''(\tau_-; \mathsf{x}) + f_{\alpha}'(\tau_-; \mathsf{x})}{{f'(\tau_-; \mathsf{x})}^2}f_\beta(\tau_-; \mathsf{x})
     \notag\\
& = 
        - \frac{f_{\alpha}(\tau_{-}; \mathsf{x}) f_{\beta}(\tau_{-}; \mathsf{x})
        f''(\tau_{-}; \mathsf{x})}{{f'(\tau_{-}; \mathsf{x})}^3} 
        \notag\\
       &\quad
        + 2\frac{f_{(\alpha}(\tau_{-}; \mathsf{x})f'_{\beta)}(\tau_{-}; \mathsf{x})}{{f'(\tau_{-}; \mathsf{x})}^2} 
       - \frac{f_{\alpha\beta}(\tau_{-}; \mathsf{x})}{f'(\tau_{-}; \mathsf{x})} 
        , 
\label{eq:tauminusprime-eps-expansion2}
\end{align}
using \eqref{eq:partialtauminus} in the last equality. By \eqref{eq:tauminusprime-eps-expansion1} and \eqref{eq:tauminusprime-eps-expansion2} we then have 
\begin{align}
    \tau_{-}' &= \tau_{-} - \epsilon^{\alpha}\frac{f_{\alpha}(\tau_{-}; \mathsf{x})}{f'(\tau_{-}; \mathsf{x})}  
    - \frac{(\epsilon^{\alpha}f_{\alpha}(\tau_{-}; \mathsf{x}))^{2} f''(\tau_{-}; \mathsf{x})}{{2f'(\tau_{-}; \mathsf{x})}^3}
    \notag\\
    &\quad
    + \frac{(\epsilon^{\alpha}f_{\alpha}(\tau_{-}; \mathsf{x}))(\epsilon^{\beta}f_{\beta}'(\tau_{-}; \mathsf{x}))}{{f'(\tau_{-}; \mathsf{x})}^2} 
    - \frac{\epsilon^{\alpha}\epsilon^{\beta}f_{\alpha\beta}(\tau_{-}; \mathsf{x})}{2f'(\tau_{-}; \mathsf{x})} 
    \notag\\
    &\quad
    + O(\epsilon^{3}).
    \label{eq:taump-expansion}
\end{align}

For $f(\tau_{-}; \mathsf{x}')$ and $f(\tau_{-}'; \mathsf{x})$, which appear in the denominators in~\eqref{eq:Bmunu-def}, we have 
\begin{subequations}
\label{eq:fprimed-expansion}
\begin{align}
 f(\tau_{-}; \mathsf{x}') & = \epsilon^{\alpha}f_{\alpha}(\tau_{-}; \mathsf{x}) + \frac{\epsilon^{\alpha}\epsilon^{\beta}}{2}f_{\alpha\beta}(\tau_{-}; \mathsf{x}) + O({\epsilon}^{3}) , \\
 f(\tau_{-}'; \mathsf{x}) & = (\tau_-'-\tau_-) f'(\tau_{-}; \mathsf{x}) + 
 \frac{(\tau_-'-\tau_-)^2}{2} f''(\tau_{-}; \mathsf{x})
 \notag\\
 &\quad 
 + O\bigl((\tau_-'-\tau_-)^3\bigr)
 \notag\\
 &= -\epsilon^{\alpha}f_{\alpha}(\tau_{-}; \mathsf{x}) + \frac{(\epsilon^{\alpha}f_{\alpha}(\tau_{-}; \mathsf{x})) (\epsilon^{\beta}f'_{\beta}(\tau_{-}; \mathsf{x}))}{f'(\tau_{-}; \mathsf{x})}
 \notag\\
 &\quad- \frac{\epsilon^{\alpha}\epsilon^{\beta}}{2}f_{\alpha\beta}(\tau_{-}; \mathsf{x}) + O(\epsilon^3) , 
 \label{eq:ftaumpx-expansion}
\end{align}
\end{subequations}
where the last equality in \eqref{eq:ftaumpx-expansion} follows using~\eqref{eq:taump-expansion}. For the reciprocals this implies 
\begin{subequations}
\label{eq:1/f-expansions}
\begin{align}
 \frac{1}{f(\tau_{-}; \mathsf{x}')} & = \frac{1}{\epsilon^{\alpha}f_{\alpha}(\tau_{-}; \mathsf{x})} - \frac{\epsilon^{\alpha}\epsilon^{\beta}f_{\alpha\beta}(\tau_{-}; \mathsf{x})}{2{(\epsilon^{\gamma}f_{\gamma}(\tau_{-}; \mathsf{x}))}^{2}} + O(\epsilon) , 
 \label{eq:1/f-noprime-prime-expansion}\\
 \frac{1}{f(\tau_{-}'; \mathsf{x})} & = -\frac{1}{\epsilon^{\alpha}f_{\alpha}(\tau_{-}; \mathsf{x})} + \frac{\epsilon^{\alpha}\epsilon^{\beta}f_{\alpha\beta}(\tau_{-}; \mathsf{x})}{2{(\epsilon^{\gamma}f_{\gamma}(\tau_{-}; \mathsf{x}))}^{2}} \notag\\
 &\quad- \frac{\epsilon^{\beta}f'_{\beta}(\tau_{-}; \mathsf{x})}{f'(\tau_{-}; \mathsf{x})\epsilon^{\gamma}f_{\gamma}(\tau_{-}; \mathsf{x})} + O(\epsilon) . 
 \label{eq:1/f-prime-noprime-expansion}
\end{align}
\end{subequations}
Substituting \eqref{eq:1/f-expansions} in~\eqref{eq:Bmunu-def}, we find 
\begin{align}
    B_{\mu \nu}(\mathsf{x}, \mathsf{x}') &= \frac{E}{\epsilon^{\alpha}f_{\alpha}(\tau_{-}; \mathsf{x})}\Biggl[ \frac{\alpha_{\mu}(\tau_{-}; \mathsf{x})\alpha_{\nu'}(\tau_{-}; \mathsf{x}')}{f'(\tau_{-}; \mathsf{x})} \notag\\
    &\quad- \frac{\alpha_{\nu'}(\tau_{-}'; \mathsf{x}')\alpha_{\mu}(\tau_{-}'; \mathsf{x})}{f'(\tau_{-}'; \mathsf{x}')} \Biggr] \notag\\
    &\quad- E \frac{\alpha_{\mu}(\tau_{-}'; \mathsf{x})\alpha_{\nu'}(\tau_{-}'; \mathsf{x}')\epsilon^{\beta}f'_{\beta}(\tau_{-}; \mathsf{x})}{{f'(\tau_{-}'; \mathsf{x}')}f'(\tau_{-}; \mathsf{x}) \epsilon^{\gamma}f_{\gamma}(\tau_{-}; \mathsf{x})}  + O(\epsilon)
\label{eq:Bmunu-interm1}
\end{align}
where the contributions from second term in 
\eqref{eq:1/f-noprime-prime-expansion}
and the second term in 
\eqref{eq:1/f-prime-noprime-expansion}
have combined to be~$O(\epsilon)$. 
The square brackets in \eqref{eq:Bmunu-interm1} may be expanded as 
\begin{align}
    &\quad \frac{\alpha_{\nu}(\tau_{-}; \mathsf{x})}{f'(\tau_{-}; \mathsf{x})}\bigl(\alpha_{\mu}(\tau_{-}; \mathsf{x}) - \alpha_{\mu}(\tau_{-}'; \mathsf{x})\bigr) \notag\\
    &\quad + \frac{\alpha_{\mu}(\tau_{-}; \mathsf{x})}{f'(\tau_{-}; \mathsf{x})}\bigl(\alpha_{\nu'}(\tau_{-}; \mathsf{x}') - \alpha_{\nu'}(\tau_{-}'; \mathsf{x}')\bigr) \notag\\
    &\quad + \alpha_{\mu}(\tau_{-}; \mathsf{x})\alpha_{\nu}(\tau_{-}; \mathsf{x})\left( \frac{1}{f'(\tau_{-}; \mathsf{x})} - \frac{1}{f'(\tau_{-}'; \mathsf{x}')} \right) \notag\\
    &\quad+ O (\epsilon^2) \notag\\[2mm]
    &= \frac{2\alpha_{(\nu}'(\tau_-; \mathsf{x})\alpha_{\mu)}(\tau_-; \mathsf{x})}{f'(\tau_-; \mathsf{x})^2}\epsilon^\alpha f_\alpha(\tau_-; \mathsf{x}) \notag\\
    &\quad - \frac{\alpha_\mu(\tau_-; \mathsf{x})\alpha_\nu(\tau_-; \mathsf{x})f''(\tau_-; \mathsf{x})}{{f'(\tau_-; \mathsf{x})}^3}\epsilon^\alpha f_\alpha(\tau_-; \mathsf{x}) \notag\\
    &\quad + \frac{\alpha_\mu(\tau_-; \mathsf{x})\alpha_\nu(\tau_-; \mathsf{x})}{{f'(\tau_-; \mathsf{x})}^2}\epsilon^\alpha f'_\alpha(\tau_-; \mathsf{x}) \notag\\
    &\quad + O(\epsilon^2) ,
\label{eq:Bmunu-brackets-expansion}
\end{align}
using the linear term in \eqref{eq:taump-expansion} in the last step.
The contribution from the last displayed term in \eqref{eq:Bmunu-brackets-expansion} cancels the last displayed term in \eqref{eq:Bmunu-interm1}, and we have 
\begin{align}
 B_{\mu \nu}(\mathsf{x}, \mathsf{x}') & = \frac{E}{{f'(\tau_{-}; \mathsf{x})}^{2}}\Biggl[ 2\alpha_{(\mu}(\tau_{-}; \mathsf{x})\alpha'_{\nu)}(\tau_{-}; \mathsf{x}) \notag\\
 &\qquad- \frac{\alpha_{\mu}(\tau_{-}; \mathsf{x})\alpha_{\nu}(\tau_{-}; \mathsf{x})}{f'(\tau_{-}; \mathsf{x})}f''(\tau_{-}; \mathsf{x}) \Biggr] + O(\epsilon) \notag\\
	 & = \frac{E}{f'(\tau_{-}; \mathsf{x})}\partial_{\tau}\!\left.\left( \frac{\alpha_{\mu}(\tau; \mathsf{x})\alpha_{\nu}(\tau; \mathsf{x})}{f'(\tau; \mathsf{x})} \right)\right|_{\tau = \tau_-} \hspace{-4pt}+ O(\epsilon) , 
\end{align}
with the coincidence limit 
\begin{align}
 B_{\mu \nu}(\mathsf{x}, \mathsf{x}) & = \frac{E}{f'(\tau_{-}; \mathsf{x})}\partial_{\tau}\left.\left( \frac{\alpha_{\mu}(\tau; \mathsf{x})\alpha_{\nu}(\tau; \mathsf{x})}{f'(\tau; \mathsf{x})} \right)\right|_{\tau = \tau_-} , 
\label{eq:Bmunu-coincidence}
\end{align}
and the contribution to the coincidence limit of $S_{\mu\nu}$ \eqref{eq:smunu-expanded} is $-B_{\mu \nu}(\mathsf{x}, \mathsf{x})$. 

\subsection{Result for \texorpdfstring{$I_{\mu\nu}$}{I-(mu nu)-result}}

Combining \eqref{eq:Cmunu-coincidence}, 
\eqref{eq:Smunu-coincidence-firstpart}, 
\eqref{eq:Smunu-coincidence-secondpart} and~\eqref{eq:Bmunu-coincidence}, the final expression for $I_{\mu\nu}$ \eqref{eq:Imunu-def} is 
\begin{align}
    I_{\mu \nu} & = \frac{1}{4\pi^3f'(\tau_{-})} \Biggl[  - E\partial_{\tau}\left.\left( \frac{\alpha_{\mu}(\tau)\alpha_{\nu}(\tau)}{f'(\tau)} \right) \right|_{\tau = \tau_-}   \notag\\
    &\quad + \pi\frac{E^{2}\alpha_{\mu}(\tau_{-})\alpha_{\nu}(\tau_{-}) + \alpha'_{\mu}(\tau_{-})\alpha'_{\nu}(\tau_{-})}{f'(\tau_{-})}\notag\\
    &\quad - 2\int _{-\infty}^{\tau_{-}} \mathrm{d}\tau \biggl(E^{2}   \frac{\sin(E(\tau - \tau_{-}))}{f(\tau)}\alpha_{(\mu}(\tau_{-})\alpha_{\nu)}(\tau) \notag\\
    &\quad + E\cos(E(\tau - \tau_{-})) \frac{\alpha_{(\mu}'(\tau_{-})\alpha_{\nu)}(\tau) - \alpha_{(\mu}'(\tau)\alpha_{\nu)}(\tau_{-})}{f(\tau)}\notag\\
    &\quad  + \frac{\sin(E(\tau - \tau_{-}))}{f(\tau)} \alpha_{(\mu}'(\tau_{-})\alpha_{\nu)}'(\tau) \biggr)  \Biggr], \label{eq:app:masterequation}
\end{align}
where the notation suppresses the spacetime point~$\mathsf{x}$. 

\subsection{Alternative expression for \texorpdfstring{$I_{\mu\nu}$}{I-(mu nu)}}

We shall give for $I_{\mu \nu}$ \eqref{eq:app:masterequation} an alternative expression that will be convenient for handling the Rindler trajectory detector in Appendix~\ref{app:accelerated-emissions}. 

We make the technical assumption that 
$f(\tau)$ is defined for $-\infty < \tau \le \tau_-$, that is, the trajectory is defined at arbitrarily negative proper times in the past. Recalling that $f(\tau_-)=0$, $f'(\tau) > 0$ for $\tau\le0$, and $f(\tau) < -(\tau-\tau_-)^2$ for $\tau<\tau_-$, 
by elementary Lorentz-geometry inequalities, it follows that all the integrals encountered in the analysis below are absolutely convergent. 

We introduce a notation in which 
overtildes on functions of $\tau$ denote evaluation at $\tau=\tau_-$, so that $\tilde\alpha_\mu := \alpha_\mu(\tau_-)$, $\tilde{\alpha}_\mu' := \alpha_\mu'(\tau_-)$, and similarly for $f$ and its derivatives. We also define
\begin{align}
    \beta_\mu(\tau) := \frac{\alpha_\mu(\tau) - \tilde\alpha_\mu}{f(\tau)} , 
\label{eq:betamu-def}
\end{align}
where we recall that $\tau < \tau_-$. $\beta_\mu$ has the limit
$\beta_\mu(\tau) \to \tilde\alpha_\mu'/\tilde f' =: \tilde\beta_\mu$ as $\tau\to\tau_-$, by L'H\^opital, and the limit $\beta_\mu(\tau) \to 0$ as $\tau \to -\infty$, by the technical assumption introduced above. Differentiating \eqref{eq:betamu-def} gives 
\begin{align}
    \beta_\mu'(\tau) = \frac{\alpha_\mu'(\tau) - f'(\tau)\beta_\mu(\tau)}{f(\tau)} . 
\label{eq:betamuprime-formula}
\end{align}

On the third line of~\eqref{eq:app:masterequation}, 
we write the integral involving the sine as 
\begin{align}
    &-2E^2\tilde\alpha_{(\mu}\int_{-\infty}^{\tau_-}\mathrm{d}\tau\frac{\sin(E(\tau - \tau_-))}{f(\tau)}\alpha_{\nu)}(\tau) \notag\\
    &\quad= 2E\tilde\alpha_{(\mu}\tilde\beta_{\nu)} - 2E\tilde\alpha_{(\mu}\int_{-\infty}^{\tau_-}\mathrm{d}\tau \cos(E(\tau - \tau_-))\beta_{\nu)}'(\tau) \notag\\
    &\qquad-2E^2\tilde\alpha_{(\mu}\tilde{\alpha}_{\nu)}\int_{-\infty}^{\tau_-}\mathrm{d}\tau \frac{\sin(E(\tau - \tau_-))}{f(\tau)}, \label{eq:app-tmunu:simplify-third-line}
\end{align}
writing $\alpha_\nu = \tilde\alpha_\nu + f\beta_\nu$ from \eqref{eq:betamu-def} and integrating by parts. Substituting this back in \eqref{eq:app:masterequation} gives 
\begin{align}
    I_{\mu \nu} & = \frac{1}{4\pi^3\tilde f'} \Biggl[  E\frac{\tilde\alpha_\mu\tilde\alpha_\nu\tilde f''}{\tilde{f'}^2} + \pi\frac{E^{2}\tilde\alpha_{\mu}\tilde\alpha_{\nu} + \tilde\alpha'_{\mu}\tilde\alpha'_{\nu}}{\tilde f'}\notag\\
    &\quad - 2E^2\tilde\alpha_{\mu}\tilde\alpha_{\nu}\int _{-\infty}^{\tau_{-}} \mathrm{d}\tau \frac{\sin(E(\tau - \tau_{-}))}{f(\tau)} \notag\\
    &\quad - 2E\tilde\alpha_{(\mu}\int_{-\infty}^{\tau_-}\mathrm{d}\tau \cos(E(\tau - \tau_-))\beta_{\nu)}'(\tau) \notag\\
    &\quad - 2E\int_{-\infty}^{\tau_-} 
    \mathrm{d}\tau \cos(E(\tau - \tau_{-})) 
    \notag\\
    &\hspace{17ex} \times 
    \frac{\tilde\alpha_{(\mu}'\alpha_{\nu)}(\tau) - \tilde\alpha_{(\mu}\alpha_{\nu)}'(\tau)}{f(\tau)}\notag\\
    &\quad -2\tilde\alpha_{(\mu}'\int_{-\infty}^{\tau_-} \mathrm{d}\tau \frac{\sin(E(\tau - \tau_{-}))}{f(\tau)} \alpha_{\nu)}'(\tau) \biggr)  \Biggr], \label{eq:app:masterequation-simplify-1}
\end{align}
after partial cancellations between the first term in \eqref{eq:app:masterequation} and the substitution term in~\eqref{eq:app-tmunu:simplify-third-line}. 

For the integrals involving the cosine in~\eqref{eq:app:masterequation-simplify-1}, we use the rearrangement 
\begin{align}
    &- 2E\tilde\alpha_{(\mu}\int_{-\infty}^{\tau_-}\mathrm{d}\tau \cos(E(\tau - \tau_-))\beta_{\nu)}'(\tau) \notag\\
    &\quad - 2E\int_{-\infty}^{\tau_-} 
    \mathrm{d}\tau \cos(E(\tau - \tau_{-})) \frac{\tilde\alpha_{(\mu}'\alpha_{\nu)}(\tau) - \tilde\alpha_{(\mu}\alpha_{\nu)}'(\tau)}{f(\tau)} \notag\\
    &= -2E\int_{-\infty}^{\tau_-}\mathrm{d}\tau\frac{\cos(E(\tau - \tau_-))}{f(\tau)}\biggl[\tilde\alpha_{(\mu}\alpha_{\nu)}'(\tau) \notag\\
    &\qquad\qquad - f'(\tau)\tilde\alpha_{(\mu}\beta_{\nu)}(\tau) + \tilde\alpha_{(\mu}'\alpha_{\nu)}(\tau) - \tilde\alpha_{(\mu}\alpha_{\nu)}'(\tau) \biggr] \notag\\
    &= -2E\tilde\alpha_{(\mu}\int_{-\infty}^{\tau_-}\mathrm{d}\tau\cos(E(\tau - \tau_-))\frac{\tilde\alpha_{\nu)}' - f'(\tau)\beta_{\nu)}(\tau)}{f(\tau)} \notag\\
    &\quad-2E\tilde\alpha_{(\mu}'\int_{-\infty}^{\tau_-}\mathrm{d}\tau\cos(E(\tau - \tau_-)) \beta_{\nu)}(\tau) , 
\label{eq:app-tmunu:simplify-cosines}
\end{align}
using in the first equality~\eqref{eq:betamuprime-formula}, then spotting a cancellation in the square brackets, and in the last equality splitting the integral by adding and subtracting $\tilde\alpha'_{(\mu}\tilde\alpha_{\nu)}$ in the coefficient of $\cos(E(\tau - \tau_-))/f(\tau)$ and using~\eqref{eq:betamu-def}. 
The integrals created by the split are nonsingular, by the observations below~\eqref{eq:betamu-def}, noting that $\tilde\alpha_{\nu}' - f'(\tau)\beta_{\nu}(\tau)$ vanishes at $\tau=\tau_-$. 

The last term in \eqref{eq:app-tmunu:simplify-cosines} may be integrated by parts to give 
\begin{align}
&2\tilde\alpha_{(\mu}'\int_{-\infty}^{\tau_-}\mathrm{d}\tau
\sin(E(\tau - \tau_-)) 
\beta'_{\nu)}(\tau)
\notag\\
&= 
2\tilde\alpha_{(\mu}'\int_{-\infty}^{\tau_-}\mathrm{d}\tau
\frac{\sin(E(\tau - \tau_-))}{f(\tau)} \alpha_{\nu)}'(\tau)
\notag\\
&\quad 
- 2\tilde\alpha_{(\mu}'\int_{-\infty}^{\tau_-}\mathrm{d}\tau
\frac{\sin(E(\tau - \tau_-))}{f(\tau)} 
f'(\tau)\beta_{\nu)}(\tau) , 
\label{eq:app-tmunu:simplify-second-cosine}
\end{align}
using~\eqref{eq:betamuprime-formula}. Note that the term involving $\alpha_{\nu}'(\tau)$ in \eqref{eq:app-tmunu:simplify-second-cosine} cancels the last term in~\eqref{eq:app:masterequation-simplify-1}. 

Finally, we note that the coincidence limit of $\braket{\phi(\mathsf{x})\phi(\mathsf{x}')}^{(2)}$ \eqref{eq:app:two-point-function-mink} is given by 
\begin{align}
\braket{\phi(\mathsf{x})^2}
& = 
\frac{\chi(\tau_-)}{4\pi^3\tilde f'}
\Biggl( \frac{\pi \chi(\tau_-)}{\tilde f'} 
\notag\\
& \quad 
- 2 \int_{-\infty}^{\tau_-}\mathrm{d}\tau\;\frac{\chi(\tau) \sin(E(\tau - \tau_-))}{f(\tau)} 
\Biggr) , 
\label{eq:vacpol-chi}
\end{align}
using arguments similar to those with~\eqref{eq:Smunu-coincidence-firstpart}. When $\chi(\tau)=1$, so that the detector operates with unit strength for all times, \eqref{eq:vacpol-chi} reduces to 
\begin{align}
\braket{\phi(\mathsf{x})^2}_\infty^{(2)} := \frac{1}{4\pi^3\tilde f'}\left( \frac{\pi}{\tilde f'} - 2 \int_{-\infty}^{\tau_-}\mathrm{d}\tau\;\frac{\sin(E(\tau - \tau_-))}{f(\tau)}  \right) . 
\label{eq:vacpol-infty}
\end{align}
In terms of~\eqref{eq:vacpol-infty}, we may arrange $I_{\mu\nu}$ in the final form 
\begin{align}
    I_{\mu\nu} &= (E^2\tilde\alpha_\mu\tilde\alpha_\nu + \tilde\alpha_\mu'\tilde\alpha_\nu')\braket{\phi^2}_\infty^{(2)} + \frac{E\tilde\alpha_\mu\tilde\alpha_\nu \tilde{f}''}{4\pi^3\tilde{f}'^3} \notag\\
    &\quad-\frac{E\tilde\alpha_{(\mu}}{2\pi^3\tilde{f}'}\int_{-\infty}^{\tau_-}\mathrm{d}\tau\cos(E(\tau - \tau_-))\frac{\tilde\alpha_{\nu)}' - f'(\tau)\beta_{\nu)}(\tau)}{f(\tau)} \notag\\
    &\quad+\frac{\tilde\alpha_{(\mu}'}{2\pi^3\tilde f'}\int_{-\infty}^{\tau_-}\mathrm{d}\tau\sin(E(\tau - \tau_-))\frac{\tilde\alpha_{\nu)}' - f'(\tau)\beta_{\nu)}(\tau)}{f(\tau)},  \label{eq:rearranged-master-equation}
\end{align}
suppressing the dependence on $\mathsf{x}$ in $\braket{\phi^2}_\infty^{(2)}$. 

The advantage of \eqref{eq:rearranged-master-equation} over \eqref{eq:app:masterequation} is that the integrals explicit in \eqref{eq:rearranged-master-equation} have only one spacetime index and the integral in $\braket{\phi(\mathsf{x})^2}_\infty^{(2)}$
\eqref{eq:vacpol-infty} has no spacetime indices, giving fewer integrals to compute. We shall use \eqref{eq:rearranged-master-equation} for the detector on a Rindler trajectory in Appendix~\ref{app:accelerated-emissions}.

\section{Principal value integral identity}\label{app:principal-value-identity}

In this appendix we verify the distributional identity~\eqref{eq:princvalid-cited}. 

The identity \eqref{eq:princvalid-cited} is a special case of the identity 
\begin{align}
    \partial_y \left\{\mathrm{P.V.}\left( \frac{1}{f(\tau, y)} \right)\right\} = \frac{f_y(\tau, y)}{f_\tau(\tau, y)} \partial_\tau\left\{ \mathrm{P.V.}\left( \frac{1}{f(\tau, y)} \right) \right\} ,
    \label{eq:app-distr-identity}
\end{align}
where $f(\tau, y)$ is a smooth function of two real variables, with a linear zero in the first argument at $\tau = \tau_0(y)$, determined implicitly by $f(\tau_0(y), y) = 0$, such that $\tau_0(y)$ is smooth in~$y$, and has no other zeros. 
$f_\tau$~and $f_y$ denote the derivatives of $f$ with respect to the first and second argument, respectively, and we assume $f_\tau$ is nowhere vanishing. We consider each side of \eqref{eq:app-distr-identity} as a distribution in $\tau$, with $y$ as a parameter, and P.V. denotes the principal value integral in~$\tau$. 

We now proceed to verify~\eqref{eq:app-distr-identity}. 

Let $\phi \in C^\infty_0$ be a test function. 
Let $g(y)$ denote the action of the distribution on the left-hand side of \eqref{eq:app-distr-identity} on~$\phi$, 
\begin{align}
    g(y) & := \partial_{y}\left\{ \mathrm{P.V.}\int_{-\infty}^{\infty} \mathrm{d}\tau\frac{ \phi(\tau)}{f(\tau, y)}  \right\} 
    \notag \\ 
    & = \partial_{y}\Biggl\{ \lim_{\epsilon\to 0}\biggl(\int_{-\infty}^{\tau_0 - \epsilon}\mathrm{d}\tau\frac{ \phi(\tau)}{f(\tau, y)} 
    + \int_{\tau_0 + \epsilon}^{\infty} \mathrm{d}\tau\frac{ \phi(\tau)}{f(\tau, y)} \biggr) \Biggr\} , 
    \label{eq:gy-initialsplit}
\end{align}
where the second equality uses the definition of the principal value integral, and the notation suppresses the $y$-dependence of~$\tau_0$. Making the substitution $\tau = \tau_0 + z$ in the first integral and $\tau = \tau_0 - z$ in the second integral, combining the two terms, and taking the $\epsilon\to0$ limit, we have 
\begin{align}
    g(y) & =  \partial_{y}\int_{0}^{\infty} \mathrm{d}z \left( \frac{\phi(\tau_0 + z)}{f(\tau_0 + z, y)} + \frac{\phi(\tau_0 - z)}{f(\tau_0 - z, y)} \right) , 
    \label{eq:appPV:substitute-to-function}
\end{align}
where each term in the integrand has a $1/z$ singularity at $z=0$ but the sum of the two terms is finite. 

Bringing $\partial_y$ under the integral in \eqref{eq:appPV:substitute-to-function}, and recalling the $y$-dependence of~$\tau_0$, gives
\begin{align}
    g(y) & = \int_{0}^{\infty} \mathrm{d}z \;\Biggl( \frac{\tau_{0}'\phi'(\tau_0 + z)}{f(\tau_0 + z, y)} + \frac{\tau_{0}'\phi'(\tau_0 - z)}{f(\tau_0 - z, y)} \notag\\
    &\qquad- \frac{\phi(\tau_0 + z)\tau_{0}'f_{\tau}(\tau_0 + z, y)}{f(\tau_0 + z, y)^{2}} \notag\\
    &\qquad- \frac{\phi(\tau_0 + z)f_{y}(\tau_0 + z, y)}{f(\tau_0 + z, y)^{2}} \notag\\
    &\qquad- \frac{\phi(\tau_0 - z)\tau_{0}'f_{\tau}(\tau_0 - z, y)}{f(\tau_0 - z, y)^{2}} \notag\\
    &\qquad- \frac{\phi(\tau_0 - z)f_{y}(\tau_0 - z, y)}{f(\tau_0 - z, y)^{2}}\Biggr) , 
    \label{eq:appPV:diff-wrt-y}
\end{align}
where the prime denotes the derivative with respect to the argument. The individual terms in the integral are singular at $z=0$ but the sum of the terms is finite, as can be verified using 
\begin{align}
    \tau_0' = -\frac{f_y(\tau_0, y)}{f_\tau(\tau_0, y)} , 
    \label{eq:appPV:tau-prime}
\end{align}
which follows by differentiating the definition $f(\tau_0(y),y) = 0$. 

We regroup \eqref{eq:appPV:diff-wrt-y} as 
\begin{align}
    g(y) & = \int_{0}^{\infty} \mathrm{d}z \;\Biggl( \frac{\tau_{0}'\phi'(\tau_0 + z)}{f(\tau_0 + z, y)} + \frac{\tau_{0}'\phi'(\tau_0 - z)}{f(\tau_0 - z, y)} \notag\\
    &+ \phi(\tau_0 + z)\left( \tau_{0}' + \frac{f_{y}(\tau_0 + z, y)}{f_{\tau}(\tau_0 + z, y)} \right) \partial_{z}\frac{1}{f(\tau_0 + z, y)} \notag\\
    &- \phi(\tau_0 - z)\left( \tau_{0}' + \frac{f_{y}(\tau_0 - z, y)}{f_{\tau}(\tau_0 - z, y)} \right)  \partial_{z} \frac{1}{f(\tau_0 - z, y)} \Biggr) . 
    \label{eq:gy-before-intbyparts}
\end{align}
Integrating the last two lines in \eqref{eq:gy-before-intbyparts} by parts gives 
\begin{align}
    g(y) & = -\lim_{ z \to 0 } \Biggl[ \frac{\phi(\tau_0 + z)}{f(\tau_0 + z, y)}\left( \tau_{0}' + \frac{f_{y}(\tau_0 + z, y)}{f_{\tau}(\tau_0 + z, y)} \right) \notag\\
    &\qquad\qquad- \frac{\phi(\tau_0 - z)}{f(\tau_0 - z, y)}\left( \tau_{0}' + \frac{f_{y}(\tau_0 - z, y)}{f_{\tau}(\tau_0 - z, y)} \right) \Biggr]  \notag\\
    &+ \int_{0}^{\infty} \mathrm{d}z \;\Biggl( \frac{\tau_{0}'\phi'(\tau_0 + z)}{f(\tau_0 + z, y)} + \frac{\tau_{0}'\phi'(\tau_0 - z)}{f(\tau_0 - z, y)} \notag\\
    &- \partial_{z}\!\left[ \phi(\tau_0 + z)\!\left( \tau_{0}' + \frac{f_{y}(\tau_0 + z, y)}{f_{\tau}(\tau_0 + z, y)} \right) \right] \!\frac{1}{f(\tau_0 + z, y)} \notag\\
    &+ \partial_{z}\!\left[ \phi(\tau_0 - z)\!\left( \tau_{0}' + \frac{f_{y}(\tau_0 - z, y)}{f_{\tau}(\tau_0 - z, y)} \right) \right] \!\frac{1}{f(\tau_0 - z, y)} \Biggr) .
    \label{eq:gy-after-intbyparts}
\end{align}
The $z\to0$ substitution term in \eqref{eq:gy-after-intbyparts} vanishes, using~\eqref{eq:appPV:tau-prime}. 
In the integral term in \eqref{eq:gy-after-intbyparts}, the terms linear in $\tau_0'$ 
cancel pairwise, leaving  
\begin{align}
    g(y) & = - \int_{0}^{\infty} \mathrm{d}z\Biggl( \partial_{z}\hspace{-1mm}\left[ \frac{f_{y}(\tau_0 + z, y)\phi(\tau_0 + z)}{f_{\tau}(\tau_0 + z, y)} \right]\hspace{-1mm} \frac{1}{f(\tau_0 + z, y)} \notag\\
    &\qquad- \partial_{z}\left[ \frac{f_{y}(\tau_0 - z, y)\phi(\tau_0 - z)}{f_{\tau}(\tau_0 - z, y)} \right] \frac{1}{f(\tau_0 - z, y)} \Biggr) . 
    \label{eq:gy-after-substzero}
\end{align}
Following the steps from \eqref{eq:gy-initialsplit} to \eqref{eq:appPV:substitute-to-function} in reverse order, \eqref{eq:gy-after-substzero} becomes 
\begin{align}
    g(y)& = -\mathrm{P.V.}\int_{-\infty}^{\infty} \mathrm{d}\tau  \frac{1}{f(\tau, y)}\partial_{\tau}\left[ \frac{f_{y}(\tau, y)\phi(\tau)}{f_{\tau}(\tau, y)} \right] .
    \label{eq:gy-back-to-tau}
\end{align}
Finally, integrating \eqref{eq:gy-back-to-tau} by parts gives 
\begin{align}
    g(y) = \int_{-\infty}^{\infty} \mathrm{d}\tau \, \phi(\tau)\,\frac{f_{y}(\tau, y)}{f_{\tau}(\tau, y)} \partial_{\tau} \left( \mathrm{P.V.} \frac{1}{f(\tau; y)} \right) , 
\end{align}
which is the action of the distribution on the right-hand side of \eqref{eq:app-distr-identity} on~$\phi$. 

This completes the verification of~\eqref{eq:app-distr-identity}.

\section{Positivity of energy density for an inertial UDW detector}\label{app:T00pos}
In this appendix, we show that the renormalised energy density $\braket{T_{tt}}^{(2)}$~\eqref{eq:inertial-T_(tt)}, which is a function of just $E$ and~$r$, is positive for all $E\in \mathbb{R}$ and $r>0$. We start by writing~\eqref{eq:inertial-T_(tt)} as
\begin{equation}
     \braket{T_{tt}}^{(2)} = \frac{1}{16\pi^3r^4} F(Er),
\end{equation}
where $F:\mathbb{R}\to \mathbb{R}$ is defined by
\begin{equation} \label{eq:F(x)def_T00pos}
    F(q):= 
    \begin{cases}
        \displaystyle{\pi(1+2q^2) - \!\int_{0}^\infty \!\mathrm{d} t\, \frac{\sin(t)+|q|\cos(t)}{t+2|q|}}, &  q<0  \\[2ex]
        \displaystyle{\frac{\pi}{2}}, & q=0  \\[2ex]
        \displaystyle{\int_{0}^\infty \mathrm{d} t\, \frac{\sin(t)+|q|\cos(t)}{t+2|q|}}, & q>0. 
    \end{cases}
\end{equation}
It is thus sufficient to show that $F(q)>0$ for all $q\in\mathbb{R}$.

For $q\ne0$, 6.7.13 and 6.7.14 in \cite{DLMF} give the identities 
\begin{subequations}
    \label{eq:sin_cos_int_identity}
    \begin{align}
        \int_{0}^\infty \mathrm{d} t\, \frac{\sin(t)}{t+2|q|} &= \int_0^\infty \mathrm{d} y\, \mathrm{e}^{-2|q|y}\frac{1}{1+y^2}, \label{eq:sin(t)_int_identity}\\
         \int_{0}^\infty \mathrm{d} t\, \frac{\cos(t)}{t+2|q|} &= \int_0^\infty \mathrm{d} y\, \mathrm{e}^{-2|q|y}\frac{y}{1+y^2}.  \label{eq:cos(t)_int_identity}
    \end{align}
\end{subequations}
For $q\ne0$, substituting  
\eqref{eq:sin_cos_int_identity} into \eqref{eq:F(x)def_T00pos} gives the formula 
\begin{align}
    F(q) &= \pi (1 + 2 q^2) \Theta(-q) \notag \\
    &\hspace{0.5cm} + \sgn(q) \int_{0}^\infty \mathrm{d} y \, \mathrm{e}^{-2|q| y} \left(\frac{1+|q|y}{1+y^2}\right). \label{eq:F(x)def_T00pos2}
\end{align}

$F(q)$ is everywhere continuous. For $q\ne0$, continuity follows using dominated convergence in the second term in~\eqref{eq:F(x)def_T00pos2}. 
For $q=0$, continuity follows because the 
$q\to0_\pm$ limits in the second term in \eqref{eq:F(x)def_T00pos2} are equal to $\pm \pi/2$, by dominated convergence, and recalling from \eqref{eq:F(x)def_T00pos} that $F(0)=\pi/2$. 

For $q\ne0$, differentiating \eqref{eq:F(x)def_T00pos2} gives 
\begin{align}
    F'(q) &= \begin{cases}
       \displaystyle{ - G(q)} & q>0  
       \\[1ex]
        \displaystyle{-4\pi |q|  - G(q)} & q<0 , 
    \end{cases}
\label{eq:Fprime(q)_appA}
\end{align}
where 
\begin{align}
     G(q) &:=  \int_{0}^\infty \mathrm{d} y \, \mathrm{e}^{-2|q| y}  \frac{y(1+2|q|y)}{1+y^2}. 
     \label{eq:G(q)_appA}
\end{align}
The integrand in \eqref{eq:G(q)_appA} is positive for $q\neq 0$ and $y>0$, and thus $G(q)>0$ for $q\neq 0$. Therefore, $F'(q)<0$ for $q\ne0$. As $F(q)$ is continuous at $q=0$, $F(q)$ is a strictly decreasing function for all $q\in \mathbb{R}$. 

Finally, a dominated convergence argument in the second term in \eqref{eq:F(x)def_T00pos2} shows that $F(q) \to 0$ as $q\to \infty$. 
Therefore, $F(q)>0$ for all $q\in \mathbb{R}$.

As an aside, we note that although $F(q)$ is continuous at $q=0$,
$F(q)$ is not differentiable at $q=0$, and 
${F'(q) \to -\infty}$ as $q\to 0$. 
To see this, we write $G(q)$ \eqref{eq:G(q)_appA} as 
\begin{align}
    G(q) &= 1 - 2|q|\Big( \Ci(2|q|)\sin(2|q|) - \si(2|q|)\cos(2|q|) \Big) \notag \\
    &\hspace{0.4cm}-\Ci(2|q|)\cos(2|q|) - \si(2|q|) \sin(2|q|),
\end{align}
using 3.356.1 and 3.356.2 in~\cite{Gradshteyn2007}, 
where $\si$ and $\Ci$ are the sine and cosine integrals~\cite{DLMF}. 
By the small argument asymptotics of $\si$ and $\Ci$ (6.6.5 and 6.6.6 in~\cite{DLMF}), we have 
\begin{equation}
    G(q) = -\log(2|q|) + 1 - \gamma_E + O(|q|) 
\end{equation}
as $q\to0$, where $\gamma_E$ is the Euler-Mascheroni constant. This shows that $G(q) \to \infty$ as $q\to0$. Hence $F'(q) \to -\infty$ as $q\to0$, by \eqref{eq:Fprime(q)_appA} and~\eqref{eq:G(q)_appA}. By l'H\^opital's rule we then have $\bigl(F(q)-F(0)\bigr)/q \to -\infty$ as $q\to0$, and thus $F'(0)$ does not exist. 

\section{SET for a detector on the Rindler trajectory}\label{app:accelerated-emissions}

In this appendix we calculate the SET for a detector on the Rindler trajectory \eqref{eq:linear-acc-trajectory} in the long interaction time limit, $\chi(\tau)=1$. We use  
\eqref{eq:rearranged-master-equation}, which we verify to remain well defined in this limit.

\subsection{Null coordinates}

Recall that in the Minkowski coordinates $(t, x, y, z)$ the Rindler trajectory \eqref{eq:linear-acc-trajectory} is given by  
\begin{align}
    \mathsf{Z}(\tau) = \left( \frac{1}{a}\sinh(a\tau), 0, 0, \frac{1}{a}\cosh(a\tau) \right), 
\end{align}
where the positive constant $a$ is the proper acceleration. In terms of the null coordinates $u = t - z$ and $v = t + z$, the squared geodesic distance \eqref{eq:f-func-def} is 
\begin{align}
    f(\tau; \mathsf{x}) &= u a^{-1}\mathrm{e}^{a\tau} - v a^{-1}\mathrm{e}^{-a\tau} - uv + x^2 + y^2 + a^{-2} , 
\label{eq:f-func-rindlertraj}
\end{align}
and we recall that $\mathsf{x}$ is by assumption in the $v>0$ half of Minkowski space. 
In terms of the dimensionless coordinates $U = au \in \mathbb{R}$, $V = av >0$ and 
$R = a \sqrt{x^2 + y^2} \ge 0$, we then have 
\begin{subequations}
\label{eq:f-fprime-fprimeprime-rindler}
    \begin{align}
        a^2f(\tau; \mathsf{x}) & = U\mathrm{e}^{a\tau} - V\mathrm{e}^{-a\tau} + K,  \label{eq:asqrf-rindler}\\
        af'(\tau; \mathsf{x}) &= U\mathrm{e}^{a\tau} + V\mathrm{e}^{-a\tau},  
        \label{eq:afprime-rindler}\\
        f''(\tau; \mathsf{x}) &= a^2f(\tau; \mathsf{x})-K , 
    \end{align}
\end{subequations}
where $K := -UV + R^2 + 1$. 

Solving $f(\tau_-; \mathsf{x})=0$ for $\tau_-$ from \eqref{eq:asqrf-rindler} gives 
\begin{align}
\mathrm{e}^{-a\tau_-} = 
\frac{K}{2V} + \sqrt{\frac{K^2}{4 V^2} + \frac{U}{V}} , 
\label{eq:tauminus-rindlersol}
\end{align}
and \eqref{eq:afprime-rindler} then gives  
\begin{align}
    a \tilde f' = \sqrt{K^2 + 4 U V } . 
\label{eq:fprime-rindler-exact}
\end{align}
We also note that \eqref{eq:f-fprime-fprimeprime-rindler} at $\tau = \tau_-$ yields the identities 
\begin{align}
    2 U \mathrm{e}^{a\tau_-} &= a\tilde f' - K 
    = a\tilde f' + \tilde f'' , 
\label{eq:rindler-aux1}
\end{align}
which will be useful in what follows. 

\subsection{Simplifying \texorpdfstring{$\braket{T_{\mu\nu}}^{(2)}$}{stress energy}}

In the expression \eqref{eq:rearranged-master-equation} for~$I_{\mu\nu}$, 
the key simplification for the Rindler trajectory is the identity 
\begin{align}
    \frac{\tilde\alpha_\mu' - f'(\tau)\beta_\mu(\tau)}{f(\tau)} &= a\frac{U\tilde\alpha_\mu - A_\mu}{\tilde f'}\partial_\tau\left( \frac{2\mathrm{e}^{a\tau_-}}{V\mathrm{e}^{-a\tau} + U\mathrm{e}^{a\tau_-}} \right), 
\label{eq:rindlertraj-identity}
\end{align}
where 
\begin{align}
    A_t &= \frac{V-U}{2V},  & A_z &= -\frac{V+U}{2V},& A_x = A_y &= 0, 
\label{eq:Amu}
\end{align}
which can be verified by a direct calculation, using 
\begin{align}
K = - U\mathrm{e}^{a\tau_-} + V\mathrm{e}^{-a\tau_-} . 
\label{eq:K-ito-tauminus}
\end{align}
In the cosine term in~\eqref{eq:rearranged-master-equation}, using \eqref{eq:rindlertraj-identity} gives 
\begin{align}
    &\int_{-\infty}^{\tau_-}\mathrm{d}\tau\cos(E(\tau - \tau_-))\frac{\tilde\alpha_\mu' - f'(\tau)\beta_\mu(\tau)}{f(\tau)} \notag\\
    &= 2\mathrm{e}^{a\tau_-}\frac{U\tilde\alpha_\mu - A_\mu}{\tilde f'}\left[ \frac{1}{\tilde f'} + aE\int_{-\infty}^{\tau_-}\!\mathrm{d}\tau \frac{\sin(E(\tau - \tau_-))}{V\mathrm{e}^{-a\tau} + U\mathrm{e}^{a\tau_-}} \right], 
\end{align}
integrating by parts and using \eqref{eq:afprime-rindler} at $\tau=\tau_-$. Proceeding similarly with the sine term in~\eqref{eq:rearranged-master-equation}, we find 
\begin{align}
    I_{\mu\nu} &= (E^2\tilde\alpha_\mu\tilde\alpha_\nu + \tilde\alpha_\mu'\tilde\alpha_\nu')\braket{\phi^2}_\infty^{(2)} + \frac{E\tilde\alpha_\mu\tilde\alpha_\nu \tilde f''}{4\pi^3\tilde{f}'^3} \notag\\
    &\quad + a \mathrm{e}^{a\tau_-}\!\frac{E^2\tilde\alpha_{(\mu}(A_{\nu)} - U\tilde\alpha_{\nu)})}{\pi^3\tilde f'^2}\!\int_{-\infty}^{\tau_-}\!\!\mathrm{d}\tau\frac{\sin(E(\tau - \tau_-))}{V\mathrm{e}^{-a\tau} + U\mathrm{e}^{a\tau_-}} \notag\\
    &\quad + a \mathrm{e}^{a\tau_-}\frac{E\tilde\alpha_{(\mu}'(A_{\nu)} - U\tilde\alpha_{\nu)})}{\pi^3\tilde f'^2}\int_{-\infty}^{\tau_-}\!\!\mathrm{d}\tau \frac{\cos(E(\tau - \tau_-))}{V\mathrm{e}^{-a\tau} + U\mathrm{e}^{a\tau_-}} \notag\\
    &\quad + \frac{E\tilde\alpha_{(\mu}(A_{\nu)} - U\tilde\alpha_{\nu)})}{\pi^3\tilde f'^3}\mathrm{e}^{a\tau_-}. \label{eq:app:imunu-rindler-simplified-1}
\end{align}
For $\braket{\phi^2}_\infty^{(2)}$~\eqref{eq:vacpol-infty}, we have 
\begin{align}
\braket{\phi^2}_\infty^{(2)} 
    &= \frac{1}{4\pi^3\tilde f'}\left[ \frac{\pi}{\tilde f'} - 2\int_{-\infty}^{\tau_-}\mathrm{d}\tau \frac{\sin(E(\tau - \tau_-))}{f(\tau)} \right] \notag\\
    &= \frac{1}{4\pi^3\tilde f'^2}\Biggl[\pi + 2a\int_{-\infty}^{\tau_-}\mathrm{d}\tau\frac{\sin(E(\tau-\tau_-))}{\mathrm{e}^{-a(\tau-\tau_-)} - 1} \notag\\
    &\quad+ 2aU\mathrm{e}^{a\tau_-}\int_{-\infty}^{\tau_-}\mathrm{d}\tau \frac{\sin(E(\tau - \tau_-))}{V\mathrm{e}^{-a\tau} + U\mathrm{e}^{a\tau_-}}\Biggr]
    \notag\\
    &= \frac{1}{4\pi^3\tilde f'^2}\Biggl[\frac{2\pi}{1 - \mathrm{e}^{2\pi E/a}} + \frac{a}{E} \notag\\
    &\quad+ 2aU\mathrm{e}^{a\tau_-}\int_{-\infty}^{\tau_-}\mathrm{d}\tau\frac{\sin(E(\tau - \tau_-))}{V\mathrm{e}^{-a\tau} + U\mathrm{e}^{a\tau_-}}\Biggr], 
    \label{eq:phisqinfty-rindler}
\end{align}
first splitting $1/f(\tau)$ by partial fractions using~\eqref{eq:asqrf-rindler}
with~\eqref{eq:K-ito-tauminus}, 
and then using \cite[Eq.~3.911.2]{Gradshteyn2007} to evaluate the first integral. 
Substituting \eqref{eq:phisqinfty-rindler} into 
$E^2\tilde\alpha_\mu\tilde\alpha_\nu\braket{\phi^2}_\infty^{(2)}$ in 
\eqref{eq:app:imunu-rindler-simplified-1}, combining the terms with a sine under the integral, and integrating these terms by parts, we find, using~\eqref{eq:rindler-aux1}, 
\begin{align}
    I_{\mu\nu} &= \tilde\alpha_\mu'\tilde\alpha_\nu'\braket{\phi^2}_\infty^{(2)} + \frac{\tilde\alpha_\mu\tilde\alpha_\nu}{2\pi^2\tilde f'^2}\frac{E^2}{1 - \mathrm{e}^{2\pi E/a}} \notag\\
    &\quad + a \mathrm{e}^{a\tau_-}\frac{E\tilde\alpha_{(\mu}'(A_{\nu)} - U\tilde\alpha_{\nu)})}{\pi^3\tilde f'^2}\int_{-\infty}^{\tau_-}\!\!\mathrm{d}\tau\frac{\cos(E(\tau - \tau_-))}{V\mathrm{e}^{-a\tau} + U\mathrm{e}^{a\tau_-}} \notag\\
    &\quad + a^2V\mathrm{e}^{a\tau_-}\frac{E\tilde\alpha_{(\mu}(2A_{\nu)} - U\tilde\alpha_{\nu)})}{2\pi^3\tilde f'^2}\notag\\
    &\qquad\qquad\times\int_{-\infty}^{\tau_-}\mathrm{d}\tau\frac{\cos(E(\tau - \tau_-))\mathrm{e}^{-a\tau}}{{(V\mathrm{e}^{-a\tau} + U\mathrm{e}^{a\tau_-})}^2}
    \label{eq:imunu-rindler-final} .
\end{align}

Recall from \eqref{eq:Tmunu-from-masterequation} that $\braket{T_{\mu\nu}(\mathsf{x})}^{(2)} = I_{\mu\nu}(\mathsf{x}) - \frac12 g_{\mu\nu} g^{\sigma\rho}I_{\sigma\rho}(\mathsf{x})$. To simplify $g^{\sigma\rho}I_{\sigma\rho}(\mathsf{x})$, 
we use the identifies 
\begin{subequations}
\label{eq:alpha-alpha-tilde-identities}
 \begin{align}
    \tilde\alpha^\mu\tilde\alpha_{\mu} &= 0 , 
    \label{eq:alphamu-alphamu-tilde}\\
    \tilde\alpha^\mu\tilde\alpha_\mu' &= \frac{2}{\tilde f'} , 
    \label{eq:alphamuprime-alphamu-tilde}\\
(\tilde\alpha')^\mu(\tilde\alpha')_\mu &= -4\frac{1 + \tilde f''}{\tilde f'^2} . 
\label{eq:alphamuprime-alphamuprime-tilde}
\end{align}
\end{subequations}
\eqref{eq:alphamu-alphamu-tilde} follows from \eqref{eq:alpha} with $\chi(\tau)=1$, using $f(\tau) = (\mathsf{Z}(\tau) - \mathsf{x})^2$ and $f_\mu(\tau) = -2(\mathsf{Z}(\tau) - \mathsf{x})_\mu$ to obtain 
\begin{align}
    \alpha^\mu(\tau)\alpha_\mu(\tau) = \frac{4f(\tau)}{f'(\tau)^2} , 
\label{eq:alphamu-alphamu}
\end{align}
and evaluating \eqref{eq:alphamu-alphamu} at $\tau = \tau_-$. 
Differentiating \eqref{eq:alphamu-alphamu} 
and setting $\tau = \tau_-$ gives~\eqref{eq:alphamuprime-alphamu-tilde}. \eqref{eq:alphamuprime-alphamuprime-tilde}~is verified similarly, using 
$f'(\tau) = 2 \dot{\mathsf{Z}}^\mu(\tau)
(\mathsf{Z}(\tau) - \mathsf{x})_\mu$
and 
$f'_\mu(\tau) = -2 \dot {\mathsf{Z}}_\mu(\tau)$. 
Note that \eqref{eq:alpha-alpha-tilde-identities} can be thought of as consequences of the identities satisfied by Synge's world function~\cite{Synge1960,Poisson2004}. 
We also use the formulas 
\begin{subequations}
    \begin{align}
        \tilde\alpha_\mu A^\mu &= \frac{1}{V} , \\
        \tilde\alpha_\mu' A^\mu &= 0 , 
    \end{align}
\end{subequations}
which hold for the Rindler trajectory. 
Combining these observations, we find 
\begin{align}
    \braket{T_{\mu\nu}}^{(2)} &= B_{\mu\nu}\braket{\phi^2}_\infty^{(2)} + \frac{\tilde\alpha_\mu\tilde\alpha_\nu}{2\pi^2\tilde f'^2}\frac{E^2}{1 - \mathrm{e}^{2\pi E/a}} \notag\\
    &\quad + a^2  V\mathrm{e}^{a\tau_-}\frac{EC_{\mu\nu}}{2\pi^3\tilde f'^2}\int_{-\infty}^{\tau_-}\mathrm{d}\tau\frac{\cos(E(\tau - \tau_-))\mathrm{e}^{-a\tau}}{{(V\mathrm{e}^{-a\tau} + U\mathrm{e}^{a\tau_-})}^2} \notag\\
    &\quad 
    + a \mathrm{e}^{a\tau_-}\frac{ED_{\mu\nu}}{\pi^3\tilde f'^2}\int_{-\infty}^{\tau_-}\mathrm{d}\tau\frac{\cos(E(\tau - \tau_-))}{V\mathrm{e}^{-a\tau} + U\mathrm{e}^{a\tau_-}} , 
\label{eq:Tmunu2-app-final}
\end{align}
where
\begin{subequations}
\label{eq:B-C-D-formulas}
    \begin{align}
        B_{\mu\nu} &= \tilde\alpha_\mu'\tilde\alpha_\nu' 
        + 2 g_{\mu\nu}\frac{1 + \tilde f''}{\tilde f'^2} ,\\
        C_{\mu\nu} &= \tilde\alpha_{(\mu}\left(2A_{\nu)} - U\tilde\alpha_{\nu)}\right) - \frac{g_{\mu\nu}}{V} ,\\
        D_{\mu\nu} &= \tilde\alpha_{(\mu}'\left(A_{\nu)} - U\tilde\alpha_{\nu)}\right) + U\frac{g_{\mu\nu}}{\tilde f'} . 
    \end{align}
\end{subequations}

\subsection{\texorpdfstring{$\braket{T_{tt}}^{(2)}$}{Minkowski energy density}}\label{app:subsec:T-tt}

For $\braket{T_{tt}}^{(2)}$, 
we compute 
\begin{subequations}
    \label{eq:alpha-t}
    \begin{align}
        \tilde\alpha_t &= \frac{V - U}{2UV} - \frac{(V + U)(K + 2UV)}{2aUV\tilde f'} ,  
        \label{eq:alpha-t-tilde-noprime}\\
        \tilde\alpha_t' &= a\frac{(V + U)(2 - K)}{K^2 + 4UV} , 
    \end{align}
\end{subequations}
where \eqref{eq:rindler-aux1} shows that \eqref{eq:alpha-t-tilde-noprime} is regular at $U=0$. 
Using~\eqref{eq:Amu}, we then find 
\begin{subequations}
    \begin{align}
        B_{tt} &= a^2\frac{(U^2 + V^2)(1 + UV - R^2)^2 + 2K^2R^2}{{(K^2 + 4UV)}^2} , \\
        C_{tt} &= -\frac{(U + V)^2}{V(K^2 + 4UV)}R^2 , \\
        D_{tt} &= \frac{(U^2 + V^2)((UV + 1)^2 - R^4) }{2a^2V\tilde f'^3} - U\frac{2KR^2}{a^2\tilde f'^3}.
    \end{align}
\end{subequations}

Specialising to $R=0$, 
\eqref{eq:tauminus-rindlersol}
and 
\eqref{eq:fprime-rindler-exact}
give 
\begin{subequations}
\begin{align}
e^{-a\tau_-} &= 
\begin{cases}
V^{-1} & \text{for\ } S=1,
\\
-U & \text{for\ } S=-1,
\end{cases}
\\
a \tilde f' &= |UV+1| , 
\end{align}
\end{subequations}
where 
$S := \sgn(UV + 1)$. Note that $S=1$ ($S=-1$) for points to the left (right) of the detector's trajectory, respectively. We then find 
\begin{align}
    \braket{T_{tt}}^{(2)}_{R=0} &= a^2\frac{U^2 + V^2}{{(UV + 1)}^2}\braket{\phi^2}_\infty^{(2)} \notag\\
    &\quad + a^4\frac{E(UV)^{S-1}(U^2 + V^2)}{2\pi^3{|UV + 1|}^3}\int_{0}^{\infty}\mathrm{d}\tau  \frac{\cos(E\tau)}{\mathrm{e}^{a\tau} + (UV)^S} \notag\\
    &\quad + \frac{1}{2\pi^2\zeta^2{(UV + 1)}^2}\frac{E^2}{1 - \mathrm{e}^{2\pi E/a}}, 
\end{align}
where 
\begin{align}
    \zeta := \begin{cases}
        V/a = v & \text{for}\ S = 1 ,\\
        U/a = u & \text{for}\ S = -1, 
    \end{cases}
\end{align}
after the change of variables $\tau \to \tau_--\tau$ in the integral. 

\subsection{\texorpdfstring{$\braket{T_{zz}}^{(2)}$}{Pressure in the z direction}}\label{app:subsec:T-zz}

For $\braket{T_{zz}}^{(2)}$, we compute
\begin{subequations}
    \label{eq:alpha-z}
    \begin{align}
        \tilde\alpha_z &= -\frac{V + U}{2UV} + \frac{(V - U)(2UV + K)}{2aUV\tilde{f}'} , 
        \label{eq:alpha-z-tilde-noprime}\\
        \tilde\alpha_z' &= a\frac{(V - U)(K - 2)}{K^2 + 4UV} , 
    \end{align}
\end{subequations}
where \eqref{eq:rindler-aux1} shows that \eqref{eq:alpha-z-tilde-noprime} is regular at $U=0$. Using~\eqref{eq:Amu}, we then find 
\begin{subequations}
    \begin{align}
        B_{zz} &= a^2\frac{(V^2 + U^2)(UV + 1 - R^2)^2 - 2R^2K^2}{{(K^2 + 4UV)}^2} , \\
        C_{zz} &= -\frac{(U - V)^2}{V{(K^2 + 4UV)}}R^2 , \\
        D_{zz} &= \frac{(U^2 + V^2)((UV + 1)^2 - R^4)}{2a^2V\tilde f'^3} + U\frac{2KR^2}{a^2\tilde f'^3} .
    \end{align}
\end{subequations}

Specialising to $R=0$, we find that $\tilde\alpha^2_t = \tilde\alpha^2_z$, $B_{zz} = B_{tt}$, $C_{zz} = C_{tt}=0$, and $D_{zz} = D_{tt}$. 
This shows that 
\begin{align}
    \braket{T_{tt}}^{(2)}_{R=0} = \braket{T_{zz}}^{(2)}_{R=0} , 
\end{align}
that is, $\braket{T_{tt}}^{(2)}$ and $\braket{T_{zz}}^{(2)}$ are equal in the plane of the detector's trajectory.

\subsection{\texorpdfstring{$\braket{T_{tz}}^{(2)}$}{Minkowski energy flux in the z direction}}

For $\braket{T_{tz}}^{(2)}$, it follows from \eqref{eq:alpha-t}, \eqref{eq:alpha-z} and \eqref{eq:Amu} that  
\begin{subequations}
    \begin{align}
        B_{tz} &= a^2\frac{(U^2 - V^2)(K - 2)^2}{{(K^2 + 4UV)}^2} , \\
        C_{tz} &= \frac{V^2 - U^2}{V(K^2 + 4UV)}R^2 , \\
        D_{tz} &= \frac{(V^2 - U^2)(K - 2)(K + 2UV)}{2Va^2\tilde{f}'^3} . 
    \end{align}
\end{subequations}

Specialising to $R=0$, we find 
\begin{align}
    \braket{T_{tz}}^{(2)}_{R=0} &= a^2\frac{U^2 - V^2}{{(UV + 1)}^2}\braket{\phi^2}_\infty^{(2)} \notag\\
    &\quad + a^4 \frac{E(UV)^{S - 1}(U^2 - V^2)}{2\pi^3{|UV + 1|}^3} 
    \notag\\
    & \hspace{6ex} \times \int_{0}^{\infty}\mathrm{d}\tau  \frac{\cos(E\tau)}{\mathrm{e}^{a\tau} + (UV)^S} \notag\\
    &\quad + \frac{S}{2\pi^2\zeta^2{(UV + 1)}^2}\frac{E^2}{1 - \mathrm{e}^{2\pi E/a}} .
\end{align}

\subsection{\texorpdfstring{$\braket{T_{\eta b}}^{(2)}$: Rindler frame energy flux}{Energy flux in Rindler coordinates}}\label{appsec:rindlerflux-transverse}

In the notation of Section~\ref{subsec:rindler-observers}, we consider the Rindler frame energy flux in the Rindler wedge that contains the detector's worldline, not just in the plane of the detector's trajectory but everywhere in the wedge. In the Rindler coordinates $(\eta, \xi, x, y)$ \eqref{eq:rindler-coords-def}, the metric is given by~\eqref{eq:rindler-metric}. The energy flux is the negative of $\braket{T_{\eta b}}^{(2)}$, where $b \in \{\xi,x,y \}$. 

Expressing \eqref{eq:f-func-rindlertraj} in the Rindler coordinates shows that $\alpha_\eta = -a^{-1}$, which implies $\tilde\alpha_\eta = -a^{-1}$ and $\tilde\alpha'_\eta = 0$. 
\eqref{eq:rindler-coords-def}~and \eqref{eq:Amu} give 
$A_\eta  = -U/a$ and $A_b = 0$. Substituting this in \eqref{eq:B-C-D-formulas} and using $g_{\eta b}=0$ shows that 
\begin{align}
    B_{\eta b} = C_{\eta b} = D_{\eta b} = 0.
\end{align}
In the expression \eqref{eq:Tmunu2-app-final} for $\braket{T_{\eta b}}^{(2)}$, all terms hence vanish except the second one. We find 
\begin{align}
    \braket{T_{\eta b}}^{(2)} &= -\frac{\tilde f_b}{2\pi^2a\tilde f'^3}\frac{E^2}{1 - \mathrm{e}^{2\pi E/a}} , 
    \label{eq:app-fluxvector}
\end{align}
where 
\begin{align}
    \tilde f' = \sqrt{a^2 {(\xi^2+x^2+y^2 + a^{-2})}^2 - 4 \xi^2}  
\label{eq:fprimexi-exact}
\end{align}
and 
\begin{subequations}
\label{eq:ftildeb}
    \begin{align}
        \tilde f_\xi &= \frac{\xi^2 - x^2 - y^2 - a^{-2}}{\xi} , 
\label{eq:fxi-exact}\\
        \tilde f_x &= 2x , \\
        \tilde f_y &= 2y . 
    \end{align}
\end{subequations}
\eqref{eq:fprimexi-exact} has come from~\eqref{eq:fprime-rindler-exact}, and \eqref{eq:ftildeb} has come by differentiating \eqref{eq:f-func-rindlertraj} and using~\eqref{eq:tauminus-rindlersol}. 

Now, close to the detector's worldline, $\xi = a^{-1}$ and $x=y=0$, 
\eqref{eq:fprimexi-exact} and \eqref{eq:fxi-exact} reduce to 
\begin{subequations}
\begin{align}
    \tilde f' &\approx 2\sqrt{(\xi - a^{-1})^2 + x^2 + y^2} ,
\label{eq:fprimetilde-neartraj}
    \\
\tilde f_\xi & \approx 2(\xi - a^{-1}) . 
\end{align}
\end{subequations}
The term shown in \eqref{eq:fprimetilde-neartraj} equals twice the proper distance to the detector's worldline on a hypersurface of constant~$\eta$. Close to the detector's worldline, the Rindler energy flux is hence isotropic, pointing away from the detector for $E<0$ and towards the detector for $E>0$, and it has an inverse square fall-off in the proper distance to the detector. 

As a consistency check, we may verify from \eqref{eq:app-fluxvector} with  \eqref{eq:fprimexi-exact} and
\eqref{eq:ftildeb} by a direct calculation that $\nabla_a T^a{}_\eta = 0$. 

\bibliographystyle{apsrev4-2}
\bibliography{References}

\end{document}